\documentclass[12pt,lettersize]{article}

\usepackage{amssymb,amscd}
\usepackage{mathrsfs}
\usepackage{amsmath,amsthm}
\usepackage[hypertex]{hyperref}
\usepackage{array,longtable}
\usepackage[dvips]{graphicx}
\usepackage[dvips]{color}
\usepackage{wrapfig}
\usepackage{bbding}
\usepackage{pifont}
\usepackage{bbm}

\newtheorem{thm}{Theorem}[section]
\newtheorem{lemma}{Lemma}[section]
\newtheorem{cor}{Corollary}[section]
\newtheorem{rem}{Remark}[section]

\newcommand{\Ds}{\mathscr{D}}

\newcommand{\Rh}{\mathbb R}
\newcommand{\Ch}{\mathbb C}
\newcommand{\Zh}{\mathbb Z}

\newcommand{\Lc}{\mathcal{L}}
\newcommand{\Nc}{\mathcal{N}}

\newcommand{\Hc}{\mathcal{H}}

\newcommand{\Oc}{\mathcal{O}}

\newcommand{\pd}{\partial}

\newcommand{\vol}{\mathop{\mathrm{vol}}\nolimits}

\newcommand{\CS}{\mathop{\textsl{CS}\,}\nolimits}

\newcommand{\Lb}{\boldsymbol{\mathcal{L}}}

\renewcommand{\Im}{\mathop{\mathrm{Im}}\nolimits}
\renewcommand{\Re}{\mathop{\mathrm{Re}}\nolimits}

\newcommand{\tr}{\mathop{\mathrm{tr}}\nolimits}

\newcommand{\hcat}{\mathcal{H}}

\newcommand{\chcat}{\check\hcat}

\renewcommand{\theequation}{\arabic{section}.\arabic{equation}}

\begin{document}
\title{
\begin{flushright}
{\small NSF-KITP-05-120}
\\
\vspace{-6mm}
{\small Imperial/TP/06/DMB/01}
\\
\vspace{-6mm}
{\small hep-th/0605038}
\end{flushright}
\vspace{1cm} \textbf{Holographic  Action for the Self-Dual Field}}
\author{
{Dmitriy M.~Belov$\,{}^a$\; and \;Gregory
W.~Moore$\,{}^b$} \vspace{6mm}
\\
\emph{\normalsize ${}^a$ The Blackett Laboratory, Imperial College London}
\vspace{-2mm}
\\
\emph{\normalsize Prince Consort Road, London SW7 2AZ, UK}
\vspace{5mm}
\\
\emph{\normalsize ${}^b$ Department of Physics, Rutgers University}
\vspace{-2mm}
\\
\emph{\normalsize136 Frelinghuysen Road, Piscataway, NJ 08854, USA}
}

\date {~}

\maketitle
\thispagestyle{empty}

\begin{abstract}
We revisit the construction of self-dual field theory in $4\ell+2$
dimensions using Chern-Simons theory in $4\ell+3$ dimensions,
building on the work of Witten. Careful quantization of the
Chern-Simons theory reveals all the topological subtleties
associated with the self-dual partition function, including the
generalization of the choice of spin structure needed to define the
theory. We write the partition function for arbitrary torsion
background charge, and in the presence of sources. We show how this
approach leads to the formulation of an action principle for the
self-dual field.
\end{abstract}

\vspace{1cm} $~~$ May 2006

\clearpage

\tableofcontents
\clearpage

\clearpage
\section{Introduction}
\label{sec:intro}

The problem of formulating a quantum self-dual field is an important
part of the formulation of string theory and supergravity. It is
very subtle. It was pointed out some time ago by Marcus and Schwarz
\cite{6} that there is no simple Poincar\'e invariant action
principle for the self-dual gauge field. Since then, much has been
written about the action and the quantization of the self-dual
field. For an incomplete sampling of the literature see
\cite{7}-\cite{29}.  Nevertheless, we believe the last word has not
yet been said on this problem. The main point of the present paper
is to describe a new approach to the formulation of an action
principle for self-dual fields. Our main motivation for writing the
paper is that we needed to solve this problem more thoroughly than
heretofore before writing a corresponding action for RR fields in
type IIA and IIB supergravity. The action for RR fields will be
described in a separate publication.

The action principle is described below in the introduction and in
section~\ref{sec:eom}.  In a word, the action is a period matrix defined by a
canonical complex structure and a {\it choice} of Lagrangian
decomposition of fieldspace.  We arrive at this principle following
the lead of Witten \cite{Witten:1996hc,Witten:1999vg}, who
stressed that the best way to formulate a self-dual theory is to
rely on a Chern-Simons theory in one higher dimension. This is the
``holographic approach'' of the title. One advantage of this
approach is that it takes proper account of topological aspects
ignored in other discussions. These are not --- as is often stated ---
minor topological subtleties, but can lead to qualitative physical
effects. Even in the simplest example of a chiral scalar in $1+1$
dimensions, that chiral scalar is equivalent to a chiral Weyl
fermion. Accordingly,  one cannot formulate the theory without
making a choice of spin structure. We will explain how the spin
structure is generalized and how the theory depends upon it. The
holographic approach has other advantages: It is the correct way to
capture the subtle half-integer shifts in Dirac quantization laws
for the fieldstrength.  It is also a good way to approach the
question of the metric dependence of the self-dual partition
function --- a subject of some relevance to stabilization of string
theory moduli. Not much is known about this metric dependence, and
we take some initial steps towards understanding it.

Our route to the action proceeds by careful construction of  the
self-dual partition function. The self-dual partition function has
been already discussed by Witten in
\cite{Witten:1996hc,Witten:1999vg} and by Hopkins and Singer in
\cite{Hopkins:2002rd} (see also
\cite{Henningson:1999dm,Dolan:1998qk,Gerasimov:2004yx,Moore:2004jv}).
Nevertheless, there were some technical points in these papers which
we found confusing and we hope our work will add some useful
clarification. In particular, we hope that our paper will make clear
the physical relevance of the main theorem of Hopkins and Singer in
\cite{Hopkins:2002rd}. The basic principle we employ for writing the
action has in fact already been used in \cite{Diaconescu:2000wy},
but the discussion of \cite{Diaconescu:2000wy}  was restricted to
the harmonic sector of the fields. Here we have generalized it to
the full infinite-dimensional fieldspace and broadened the context
in a way we believe will be useful.

\subsection{Main result}

Let us now describe our results in more technical detail. Consider a
$4\ell+2$-dimensional space-time manifold $M$ equipped with a
Lorentzian metric of signature $-+\dots +$. The Hodge $*$ squares to
$+1$ on the middle dimensional forms $\Omega^{2\ell+1}(M)$, making
it possible to impose a self-duality constraint on a field strength
$\mathcal{F}\in\Omega^{2\ell+1}(M)$:
\begin{equation}
*_g\mathcal{F}^+=\mathcal{F}^+.
\label{SD}
\end{equation}
When we impose \eqref{SD}  the Bianchi identity and equation of motion
coincide
\begin{equation}
d\mathcal{F}^+=0. \label{eom}
\end{equation}
A classical field theory describing the self-dual particle is
completely specified by these two equations. The quantum theory,
however, is problematic. As we have noted, folklore states there is
no straightforward Lorentz covariant action. Moreover, an important
aspect of the quantum theory is Dirac quantization. In the string
theory literature many authors attempt to impose a Dirac
quantization condition of the form
\begin{equation}
\mathcal{F}^+\in \Omega^{2\ell+1}_{\Zh}(M), \label{trialq}
\end{equation}
i.e. $\mathcal{F}^+$ is a closed form with integral periods. However
this quantization condition is \textit{incompatible} with the
self-duality constraint \eqref{SD} since the self-duality condition
\eqref{SD} varies continuously with the metric $g$. As we will see,
both of these difficulties are nicely overcome by the holographic
approach.

For technical reasons (e.g. use of the Hodge theorem) it is much more
convenient to work on a manifold with Riemannian metric. Let
$(X,g_E)$ be a compact Riemannian $4\ell+2$-dimensional manifold.
Now the Hodge $*_E$ squares to $-1$ on the space of $2\ell+1$ forms.
So the self-dual form becomes \textit{imaginary anti-self-dual}:
\begin{equation}
*_E R^+=-i R^+.
\end{equation}

In section 2 below we will explain the key insight that the
partition function of an imaginary anti self-dual field on a
$4\ell+2$-manifold $X$ should be viewed as a \textit{wave function}
of an abelian spin  Chern-Simons theory on a $4\ell+3$-manifold $Y$ where
$X$ is a component of $\partial Y$. The Chern-Simons field plays the
role of a current coupling to the self-dual field. The wavefunction
as a function of the Chern-Simons field $A\in \Omega^{2\ell+1}(X)$
is the self-dual partition function as a function of an external
current.

The following heuristic argument should make this connection to
Chern-Simons theory quite plausible \cite{Maldacena:2001ss}. The
Chern-Simons action $\mathrm{CS}\sim \int_Y A \wedge dA$, so on  a manifold
with boundary  $\delta\mathrm{CS} = \int_X \delta A \wedge A$. To
get a well-posed boundary value problem we set $A=*A\vert_X$. But in
Chern-Simons theory the gauge modes in the bulk, $Y$, become
dynamical fields (``edge states'') on the boundary $X$.  In this
case the gauge freedom is $A \to A+R$, $R\in
\Omega^{2\ell+1}_{\Zh}(Y)$, so we have a dynamical field with
$R =* R$ on the boundary.

When we go beyond this heuristic level we find  that the
Chern-Simons theory depends on an integer level $k$. \footnote{A
slight generalization, described in section 2, shows that there is
in fact a theory for any pair of  integers $p,q$. The level is then
given by $k=pq$. In another kind of generalization, one can assume
that $\mathcal{F}$ takes values in a vector space equipped with an
involution $I^2=\pm 1$. Such generalizations naturally arise in
compactifications of self-dual theories. This is related to the
generalization where $k$ can be taken to be an integral matrix. A
thorough study in the three-dimensional case of the latter
generalization can be found in \cite{Belov:2005ze}. } Even defining
the Chern-Simons term at the fundamental level $k=1$ turns out to be
very subtle indeed, and this leads to the most difficult aspects in
the work of Hopkins, Singer and Witten. In the case of the self-dual
scalar in two dimensions the corresponding Chern-Simons theory in
three dimensions is at ``half-integer level,'' (this corresponds to
$k=1$ in our normalization) and is known as spin Chern-Simons
theory. In this case, in addition to a level, one must also specify
a spin structure even to define the Chern-Simons term. In the higher
dimensional case there is an analogous choice generalizing the
choice of spin structure.  We must stress the word ``generalizing'';
in physical applications we do {\it not} want to restrict attention
to spin manifolds for $\ell>0$. Since the term ``generalized spin
structure'' is already in use for something entirely different
\cite{Avis:1979de}, we will refer to our generalized spin structure
as a \textit{QRIF}  --- for reasons to be explained below.

Of course, the difficulties in defining the $k=1$ Chern-Simons term
only arise in the presence of topologically nontrivial fields. In
section 3 we describe how to formulate fieldspace in a way that
properly accounts for topology. The space of gauge equivalence
classes of fields is formulated in terms of differential cohomology.
A trickier aspect is how to describe gauge potentials, and here we
take a somewhat pragmatic approach. At the cost of some mathematical
naturality, we gain in physical insight. We then review some aspects
of the Hopkins-Singer theory in section 4, but an understanding of
this theory is not strictly necessary in order to follow the rest of
the paper: we will make an end-run around their key theorem, to be
described presently.

In section~\ref{sec:quant} we turn to the real technical work, the quantization
of the Chern-Simons theory. Once we have understood how to formulate
the action, the quantization of this free topological theory follows
the standard pattern. The physical space of states is the space of
wavefunctions satisfying the Gauss law.   For level $k=1$ it turns
out that the Chern-Simons theory has a 1-dimensional Hilbert space.
As we vary the external current and the metric, both of which couple
to the self-dual field, the partition function thus becomes a
covariantly constant  section of a line bundle with connection.
Therefore, up to a constant, the construction of the partition
function is thus the construction of this line bundle with
connection.

\paragraph{Partition function.}
Let us sketch briefly the construction of the partition function. (A
much more precise discussion is the subject of sections 5 and 6). It
is important for the whole construction that the space
$V_{\Rh}=\Omega^{2\ell+1}(X, \Rh)$ is a real symplectic vector space
with the symplectic form
\begin{equation}
\omega(u,v)=\int_X u\wedge v. \label{omega}
\end{equation}
To get a wave function we need to choose a polarization on the phase
space. This can be obtained by choosing the Hodge complex structure $J=-*_E$
on $V_{\Rh}$. Using this complex
structure we decompose the space of forms as
\begin{center}
\includegraphics[width=260pt]{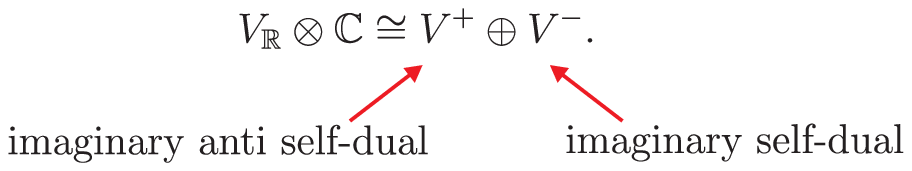}
\end{center}
To any real vector $R\in V_{\Rh}$ we associate  $R^{\pm}$ in the complex vector
space $V^\pm$ by
\begin{equation}
R^\pm=\frac12(R\pm i*_E R).
\end{equation}
We may now quantize using holomorphic polarization.  Holomorphy and
gauge invariance fix the dependence on the Chern-Simons field to be
essentially a ``theta function'' in infinite dimensions. The precise
formula is given in Theorem~\ref{thm:1} and Corollary~\ref{cor61}. When restricted
to the harmonic fields the wavefunction is essentially a theta
function on the finite-dimensional torus
$\mathscr{H}^{2\ell+1}(X)/\mathscr{H}^{2\ell+1}_{\Zh}(X)$ where
$\mathscr{H}^{2\ell+1}(X)$ is the space of harmonic $(2\ell+1)$-forms
and $\mathscr{H}^{2\ell+1}_{\Zh}(X)$ is the lattice of harmonic forms with
integral periods.
 Denoting by $a$ the Chern-Simons field, the
partition function in the harmonic sector takes the form
\begin{equation}
\mathcal{Z}\left[\begin{smallmatrix} \varepsilon_1
\\
\varepsilon_2
\end{smallmatrix}
\right](a^+,a^-)=\mathcal{N}_g \,e^{i\pi\int_X a^-\wedge a^+}
\vartheta\left[\begin{smallmatrix} \varepsilon_1
\\
\varepsilon_2
\end{smallmatrix}
\right](a^+|\tau)
\label{zeetheta}
\end{equation}
where $\vartheta\left[\begin{smallmatrix} \varepsilon_1
\\
\varepsilon_2
\end{smallmatrix}
\right]$ is a theta function with characteristics $\varepsilon_1$
and $\varepsilon_2$ (see Theorem~\ref{thm62} or appendix B for a definition),
$\mathcal{N}_g$ is a normalization factor which captures the metric
dependence, and $\tau$ is a complex period matrix. It is completely
determined by the metric and a choice of Lagrangian decomposition of
the lattice of harmonic forms with integer periods:
$\bar{\Gamma}_1^h\oplus \bar{\Gamma}_2^h=\mathscr{H}^{2\ell+1}_{\Zh}(X)$.
Equation \eqref{zeetheta} is only a caricature.  See Theorem~\ref{thm62} for
the precise result. In particular it   hides some important
subtleties to which we will soon return.  But before that, in the
next two subsections, we will reconsider the two main problems with
the quantum theory mentioned above in light of \eqref{zeetheta}.

\subsection{Action and classical equation of motion}

In section 7 we describe an action principle for the self-dual
field. We give a brief summary of that action here.  The relation of
the partition function to a theta function suggests the proper way
to formulate the action. For simplicity we put $a=0$ and
$\varepsilon=0$ in \eqref{zeetheta}. From the definition of the
theta function as an infinite sum we learn that the   period matrix
can be viewed as the on-shell action in the harmonic sector of the
theory:
\begin{equation}
S_E(R)=i\pi\tau(R^+)
\label{claction}
\end{equation}
where $R^+=\frac12(R+i*_E R)$ and $R\in \bar{\Gamma}_1^h$.
\footnote{Here we ignore possible half-integer shifts in the Dirac
quantization law. See below.} Now we need to extend equation
\eqref{claction} to the vector space $V_{\Rh}:=\Omega^{2\ell+1}(X)$
of all $2\ell+1$-forms.

\paragraph{Euclidean action.}
To be able to write the Euclidean action in a simple and
workable form we need to choose an  orthogonal coordinate system
$V_{\Rh}=V_2\oplus V_2^{\perp}$ where
$V_2$ is a Lagrangian subspace
\begin{wrapfigure}{l}{150pt}
\includegraphics[width=145pt]{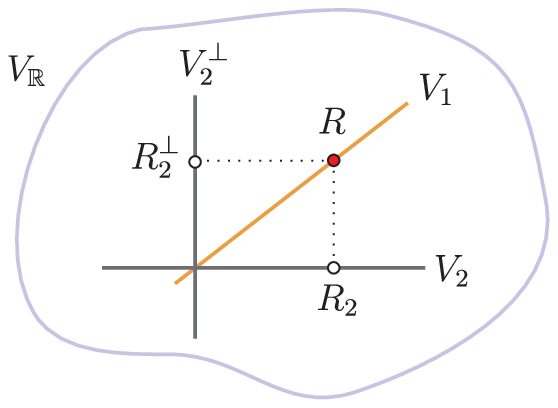}
\end{wrapfigure}
and $V_2^{\perp}=*_E V_2$ is its orthogonal complement with respect to
the Hodge metric. From the positivity of
the Riemannian metric $g_E$
it follows that $V_2\cap *_EV_2=\{0\}$, and thus this orthogonal
decomposition is also a Lagrangian decomposition. So any
form $v\in V_{\Rh}$ can be
uniquely written in the
form
$v=v_2+v_2^{\perp}$ for some $v_2\in V_2$
and $v_2^{\perp}\in V_2^{\perp}$.

A choice of Lagrangian subspace $V_2$ defines a Lagrangian subspace
$\Gamma_2\subset H^{2\ell+1}_{DR}(X)$ in the DeRham cohomology.
Next, we choose an arbitrary \textit{complementary} Lagrangian
subspace $\Gamma_1 \subset H^{2\ell+1}_{DR}(X)$,
 and define a Lagrangian subspace $V_1\subset
V_{\Rh}$ by
\begin{equation}
V_1=\{R\in \Omega^{2\ell+1}_d(X)\,|\,[R]_{DR}\in \Gamma_1\}.
\end{equation}
 Here $\Omega^{2\ell+1}_d(X)$ is the space of all closed
$2\ell+1$-forms. Note that the subspaces $V_1$ and $V_2$ are not
complementary: $V_{12}:=V_1\cap V_2 =\{\text{exact forms in }V_2\}$.
This fact will result in an extra gauge invariance of the action.

In the orthogonal coordinates $R\in V_1$
can be uniquely written as
\begin{equation}
R=R_2+R_2^{\perp}.
\end{equation}
Since $R$ is constrained to be in the Lagrangian subspace $V_1$ the
coordinates $R_2$ and $R_2^{\perp}$ are \textit{not} independent. In
these coordinates the Euclidian action \eqref{claction} for the
imaginary anti-self dual field takes the simple form
\begin{equation}
S_E(R):=\pi\int_X \bigl(R_2^{\perp}\wedge *_E R_2^{\perp}-iR_2\wedge R_2^{\perp}\bigr).
\label{claction.simple}
\end{equation}
Note that by construction the action vanishes on elements from
$V_2$. Since $V_1$ has a nontrivial intersection with $V_2$ the
action \eqref{claction.simple} has an extra gauge symmetry: for any
$R\in V_1$ and $v_{12}\in V_1\cap V_2$ we have
$S_E(R+v_{12})=S_E(R)$.

It is important that the action  \eqref{claction.simple} {\it
depends} on the choice of Lagrangian decomposition, while a properly
normalized partition function does not. This comes as follows: the
theta function and the normalization factor $\Nc_g$ in
\eqref{zeetheta} transform in metaplectic representations of the
group $Sp(b_{2\ell+1},\Zh)$ of weight $\frac12$ and $-\frac12$
respectively. Thus modulo the phase factor the partition function
\eqref{zeetheta} does not depend on the choice of Lagrangian
decomposition. This is the source of much of the difficulty people
have had in writing an action principle for the self-dual field.

\paragraph{Lorentzian action.}
The action on a Lorentzian manifold $(M,g)$ can be obtained from
\eqref{claction.simple} by  Wick rotation:
\begin{equation}
S_L(R):=\pi\int_M \bigl(R_2^{\perp}\wedge *R_2^{\perp} + R_2\wedge R_2^{\perp}
\bigr).
\label{claction.m}
\end{equation}
However for this expression to be meaningful we need to require
that the Lagrangian subspace $V_2$ be such that
\begin{equation}
V_2\cap *V_2 =\{0\}.
\end{equation}
(Otherwise $V_2\oplus *V_2$ does not define an orthogonal
coordinate system on $\Omega^{2\ell+1}(M)$.)

The variation of the action \eqref{claction.m}
with respect to $R\mapsto R+d\delta c$
where $\delta c\in\Omega^{2\ell}_{cpt}(M)$ is
\begin{equation}
\delta S_L(R)=-2\pi\int_M \delta c
\wedge d \mathcal{F}^+(R)
\label{intro:dS}
\end{equation}
where $\mathcal{F}^+(R):=R_2^{\perp}+*R_2^{\perp}$.
%
%

As a check on our action, we show in section~\ref{sec:metric} that
the variation of the action with respect to the metric yields the
standard stress-energy tensor for the self-dual field
$\mathcal{F}^+$:
\begin{equation}
\delta_g S_L(R)=
\frac{\pi}{2}\int_M
(\delta g^{-1} g)^{\mu}{}_{\nu}\,
\mathcal{F}^+\wedge *(dx^{\nu}\wedge i({\tfrac{\pd}{\pd x^{\mu}}})\mathcal{F}^+).
\end{equation}
Note that although the action \eqref{claction.m}
depends on $R\in V_1$ its metric variation
depends only on $\mathcal{F}^+(R)$.
Note also that this result is independent of Lagrangian
decomposition. Moreover, as noted in footnote 10 of
\cite{Witten:1996hc} it offers a promising way to describe the
important metric-dependent factor $\mathcal{N}_g$ in the partition
function \eqref{zeetheta}. We make some preliminary remarks on this topic in section~\ref{subsec:norm}.
%
%
One relatively straightforward thing we explain is the norm-square
of $\mathcal{N}_g$. See section~\ref{subsec:norm}.  This follows
simply from normalizing the Chern-Simons wavefunction and
strengthens the idea, common in AdS/CFT, that the boundary partition
function should be identified with the \textit{normalized} bulk
wavefunction.

\paragraph{Examples.}
In section~\ref{sec:examples} we consider the  example of a
self-dual field on a product manifold $M=\Rh\times N$ where $N$ is a
compact Riemannian $4\ell+1$-manifold. On a product manifold there
is a natural choice of Lagrangian subspaces:
\begin{equation*}
V_2=\Omega^1(\Rh)\otimes\Omega^{2\ell}(N),\quad
\Gamma_2\cong H^{2\ell}_{DR}(N)\quad\text{and}\quad \Gamma_1\cong H^{2\ell+1}_{DR}(N).
\end{equation*}
With this choice  the action \eqref{claction.m}   for $\ell=0$ gives
an action for the chiral scalar which has appeared
previously \cite{Floreanini:1987as} for $M=\Rh^2$.  For
$\ell\geqslant 1$ it gives the Henneaux-Teitelboim action
\cite{Henneaux:1988gg}. Note, however,  that our interpretation of
the action differs from the one given by Henneaux and Teitelboim.
See section~\ref{sec:other} where we compare the two constructions.
We stress that these are just special cases of our general action
which can be formulated on arbitrary manifolds with arbitrary
metrics.

\subsection{Dirac quantization}

Let us now return to the second conundrum surrounding
\eqref{trialq}. We see from \eqref{intro:dS} that we should
distinguish between $R$ and the self-dual flux $\mathcal{F}^+(R)$. Thus
the way out is to understand the quantization condition in a broader
sense: there is an abelian group $\mathscr{F}^+(g,V_1,V_2)$
with nontrivial connected components inside the space
\begin{figure}[!h]
\centering
\includegraphics[width=300pt]{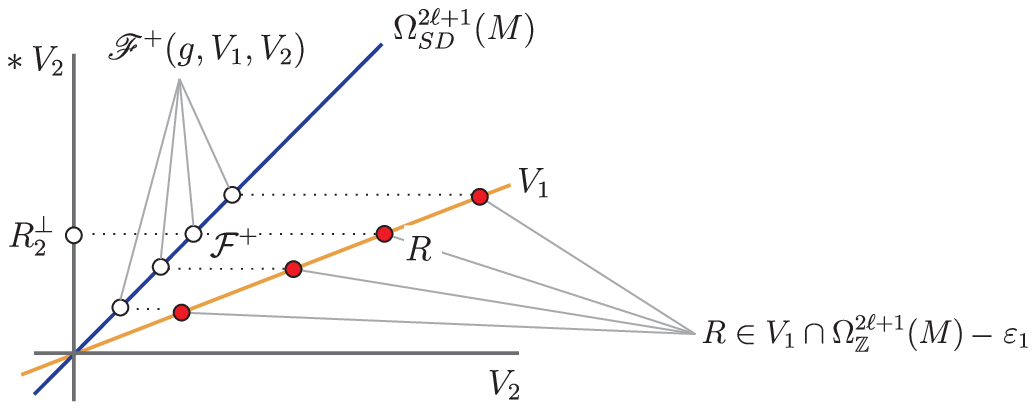}
\end{figure}
of   \textit{closed} self-dual forms $\Omega^{2\ell+1}_{SD}(M)$, and
the classical self-dual field $\mathcal{F}^+(R)$ takes values only
in this group: $\mathcal{F}^+(R):=R_2^{\perp}+*R_2^{\perp}$ where
$R=R_2^{\perp}+R_2$ and $[R]_{DR}\in\Gamma_1-[\varepsilon_1]$.

There is a further subtlety in Dirac quantization involving
half-integral shifts related to the characteristics. In order to
understand these we must return to one of the subtleties suppressed
in our discussion thus far of \eqref{zeetheta}. We mentioned above
that one must make extra choices even to define a level $k=1$
Chern-Simons term. These choices enter the theory through the
formulation of the Gauss law. In order to write a Gauss law one must
choose a $U(1)$-valued function $\Omega:H^{2\ell+1}(X;\Zh)\to U(1)$
which satisfies the cocycle condition
\begin{equation}
\Omega(a_1+a_2) = \Omega(a_1) \Omega(a_2) (-1)^{\int_X a_1 \cup
a_2}.
  \label{eqOmega2}
\end{equation}
In other words, to specify a Chern-Simons theory one must not only
choose a level $k$, but also an $\Omega$-function.\footnote{Here is
the ``end-run'' around the Hopkins-Singer theorem mentioned above.
Rather than working directly with their Chern-Simons term we instead
formulate the theory in terms of $\Omega$, following
\cite{Witten:1996hc,Witten:1999vg}.}  As mentioned above, we will
refer to this as a \textit{QRIF}.
 \footnote{QRIF stands for ``Quadratic Refinement of the
 Intersection Form.'' Note that a
choice of QRIF on a $4\ell+2$-manifold $X$ is closely related to a
choice of spin structure on the finite dimensional torus
$\Omega^{2\ell+1}_d(X)/\Omega^{2\ell+1}_{\Zh}(X)$.}
For a pure self dual field the  $\Omega$ functions is $\Zh_2$-valued. In this
case the  set of such functions is a principal homogeneous space
(torsor) for the 2-torsion points in $H^{2\ell+1}(X;\Rh/\Zh)$. The
choice of $\Omega$ then generalizes the choice of spin structure for
the self-dual scalar in two dimensions. The half-integer shifts in
the quantization law depends on the choice of $\Omega$. Once a
choice of $\Omega$ has been made, the level $1$ theta function is
uniquely fixed. However, in order to write a formula such as
\eqref{zeetheta}, and thus deduce the shift $\varepsilon_1$ in the
quantization law of $R$, we must also choose a Lagrangian
decomposition. Thus, the shift in the quantization law depends on
both the QRIF and the Lagrangian decomposition.

The $\Omega$ function has certainly appeared in previous treatments
of self-dual fields. It was first used in the   case of the chiral
scalar ($\ell=0$) in
\cite{Alvarez-Gaume:1986mi,Alvarez-Gaume:1987vm}, in which case it
is related to the mod two index of the Dirac operator. In that
treatment the factor was introduced in order to obtain a
factorization of a sum over instanton  sectors of the nonchiral
field into a square of a \textit{single} theta function.
\footnote{The original computation of
\cite{Alvarez-Gaume:1986mi,Alvarez-Gaume:1987vm} can be generalized
to higher dimensions. The fact that the insertion of $\Omega$
changes the splitting of the nonchiral sum over instantons is still
overlooked or misunderstood even by quite reputable authors, so we
present the details in appendix E.} In
 \cite{Witten:1996hc,Witten:1999vg} Witten showed that it
plays a central role in the theory of self-dual forms in general.
His discussion was based on the general theory of theta functions.
He used $\Omega$ to define precisely the holonomies of a line bundle
with connection over the intermediate Jacobian. An analogous
function   also plays an important role in the theory of RR fields
\cite{Moore:1999gb,Freed:2000tt,Diaconescu:2000wy}. One novelty in
the present paper is that we show how    $\Omega$ arises directly in
formulating the Gauss law of the Chern-Simons theory.

There is a further subtlety associated with the $\Omega$ function.
As explained in section~\ref{subsec:Omega}, associated with $\Omega$ is a
(torsion) characteristic class $\mu \in H^{2\ell+2}(X,\Zh)$. The
Gauss law for the Chern-Simons theory (in the absence of Wilson line
sources) says that $F =0$, where $F$ is the fieldstrength of the
Chern-Simons field. However, it also says that the topological
component $a(\check{ A} )\in H^{2\ell+2}(X;\Zh)$ of the Chern-Simons field is constrained
by the tadpole constraint:
\begin{equation}
k a(\check{ A}) + k\mu =0.
\end{equation}
The class $\mu$ has the physical interpretation of being the
Poincar\'e dual of a brane wrapping a torsion cycle. In section~\ref{subsec:Omega}
we explain that when we choose $\Omega$ we choose whether branes
wrap certain $2$-torsion cycles in $X$. The set of choices of $\Omega$
for {\it fixed} background charge $\mu$ is a torsor for
$\bar{H}^{2\ell+1}(X;\Zh)/2\bar{H}^{2\ell+1}(X;\Zh)$. This defines a
set of $2^{\,b_{2\ell+1}}$ distinct partition functions,
generalizing the set of partition functions of the self-dual scalar
on a Riemann surface.

It is important to note that in the $M$-theory five-brane the analog
of $\Omega$ is $U(1)$ valued, not just $\Zh_2$-valued. Moreover the
$\Omega$ function is not a choice, but rather is determined by the
topology and geometry of the embedded 5-brane and will in general
vary continuously with the metric \cite{Witten:1999vg,Moore:2004jv}.

\section{Self-dual field on a $4\ell+2$ dimensional manifold}
\label{sec:selfdual} \setcounter{equation}{0} In
\cite{Witten:1996hc} Witten proposed a recipe for constructing a
partition function of  the chiral $2\ell$-form potential on a
$4\ell+2$ dimensional Riemannian manifold $X$. In this section we
review his basic construction.  (We will be slightly more general,
introducing a theory depending on two integers $p,q$.) Consider the
Euclidean action for a topologically trivial $2\ell$-form gauge
field $C\in\Omega^{2\ell}(X)$ with coupling constant $g^2$:
\begin{equation}
S=\frac{1}{2g^2}\int_X dC\wedge *_EdC.
\label{Sc}
\end{equation}

We introduce now a topologically trivial $U(1)$ external gauge field $A\in\Omega^{2\ell+1}(X)$
with gauge transformation
law $\delta A=d\lambda$ and $\delta C=q\lambda$, so that the field $C$ has charge
$q\in\Zh$ under this $U(1)$. Consider the Lagrangian
\begin{equation}
e^{-S(C,A)}=\exp\left[
-\frac{1}{2g^2}\int_X (dC-qA)\wedge *_E(dC-qA)+i\pi p\int_X A\wedge dC
\right]
\label{lagr}
\end{equation}
where $p$ is an integer. For simplicity we will assume
that $p$ and $q$ are relatively prime integers.\footnote{If
$\mathrm{gcd}(p,q)=m\ne 1$ we can rescale the external gauge field
$A\mapsto A'=mA$ and consider the action \eqref{lagr} as function of
the rescaled gauge field $A'$. The normalization of $A$ is fixed, so
this is really a different theory. In particular,
the factorization in appendix~\ref{app:sum} is more subtle.}
To understand the effect of the
topological interaction it is useful to rewrite this Lagrangian in
complex coordinates. Thus, setting $R=dC$ we have $R=R^++R^-$ and
$A=A^++A^-$ where
\begin{equation}
R^{\pm}=\frac12(R\pm i*_E R)
\end{equation}
and similarly for $A^{\pm}$. Here $A$ and $C$ are real forms and
thus $(A^-)^*=A^{+}$ and $(R^-)^*=R^+$. In this notation we obtain
\begin{equation*}
e^{-S(C,A)}=\exp\left[
\frac{i}{g^2}\int_X \Bigl\{R^-R^++q^2 A^-A^+
+(\pi g^2 p + q)\,A^+R^- +(\pi g^2 p  -q)\,A^-R^+
\Bigr\}
\right].
\end{equation*}
Now one sees that at special values of the coupling constant $g_{p/q}^{-2}=\pi p/q$
only the $A^-$ part of the gauge field $A$
couples to the $C$ field (here we assume that $p/q$ is positive).
At this coupling constant the action has the form
\begin{equation}
e^{-S(C,A)}=\exp\left[
i\pi\int_X \Bigl\{\frac{p}{q}\,R^-R^++pq\, A^-A^+
+2q\,A^+(R^+)^*
\Bigr\}
\right].
\label{Sca}
\end{equation}
It follows that the holomorphic dependence of the partition function
\begin{equation}
\mathcal{Z}(A)=\int_{\text{top. trivial.}} \Ds C\,e^{-S(C,A)}
\label{ZA}
\end{equation}
on $A^-$ represents the coupling to the self-dual degree of freedom.
Therefore, let us  introduce the Hodge complex structure $J=-*_E$ on
the space of complexified gauge fields in which $A^+$ is holomorphic
and $A^-$ is antiholomorphic. The covariant derivatives are
\begin{equation}
D^{+}=\int_X\delta A^+\,\Bigl[\frac{\delta}{\delta A^+}-i\pi k\, A^-\Bigr]
\quad\text{and}\quad
D^{-}=\int_X\delta A^-\,\Bigl[\frac{\delta}{\delta A^-}-i\pi k\, A^+\Bigr]
\label{connectionA}
\end{equation}
where $k=pq$.
The partition function obeys the holomorphic equation
\begin{equation}
D^{-}\mathcal{Z}(A)=0.
\end{equation}
This easily follows since the Lagrangian \eqref{Sca}
satisfies this equation.
Since $[D^-,D^-]=0$ the connection \eqref{connectionA}
defines a holomorphic
line bundle $\mathcal{L}^{\otimes k}$ over the space of complexified gauge fields $A$.

The partition function is a holomorphic section of
$\mathcal{L}^{\otimes k}$. The fact that the partition function is
not a function but a section of a line bundle is related to the fact
that the action \eqref{lagr} is not gauge invariant. If $X$ is
a closed manifold then under  the gauge transformation $\delta C=q \lambda$, $\delta
A=d\lambda$ it transforms as
\begin{equation}
\delta S=i\pi k \int_X c\wedge F\quad\Rightarrow\quad
\mathcal{Z}(A+d\lambda)=\mathcal{Z}(A)\,e^{i\pi k \int_X \lambda\wedge F}
\end{equation}
where $F=dA$ is the curvature of $A$. Thus the partition function
obeys the non standard gauge-invariance:
\begin{equation*}
\bigl[d D_A-i\pi k\,F\bigr]\mathcal{Z}(A)=0
\end{equation*}
where $D=D^++D^-=\delta-i\pi k\int_X\delta A\wedge A$
with $\delta$ being the differential on the space of gauge fields.
The connection $D$ has a nonzero curvature
\begin{equation}
D^2=-2\pi i k \omega\quad\text{where}\quad
\omega=\frac12\int_X\delta A\wedge \delta A.
\label{omega1}
\end{equation}

\paragraph{Complexification of the gauge group.}
The fact that the partition function is a \textit{holomorphic}
section of $\mathcal{L}^{\otimes k}$ allows us to complexify the
gauge group. Recall that originally the partition function
$\mathcal{Z}(A)$ was a function of a real gauge field $A$. By
writing $A=A^++A^-$ we realized that it depends holomorphically on
the complex field $A^+$. This   means that $A^+$ and $A^-$ can be
considered as independent complex variables so $(A^-)^* \not= A^+$.
This in turn allows us to complexify the gauge group. Originally,
the gauge transformations were given by a real form
$c\in\Omega^{2\ell}(X)$: $A\mapsto A+dc$. Complexification of the
gauge group means that now we have two complex gauge parameters
$c^+$ and $c^-$, and gauge transformations
\begin{equation*}
A^+\mapsto A^++\frac12(dc^++ i*dc^+)\quad
\text{and}\quad
A^-\mapsto A^-+\frac12(dc^-- i*dc^-).
\end{equation*}
Notice that the field strength $F=dA^++dA^-$ is not invariant under
the complex gauge transformation:
\begin{equation*}
F\mapsto F+ \frac{1}{2 i}\,d*d(c^+-c^-).
\end{equation*}
Evidently, by a complex gauge transformation we can restrict a
topologically trivial gauge field $A$ to be flat, $dA=0$.

To proceed further we need to modify the partition function
\eqref{ZA} to include a sum over topological sectors. This step is
quite  nontrivial, and requires conceptual changes.  We postpone the
details of the construction to the next section. The partition
function takes the schematic form
\begin{equation}
\mathcal{Z}_{p,q}(A):=\sum_{a\in H^{2\ell+1}(X;\Zh)} (\Omega(a))^k\,
\int_{\text{fixed top. sector}}\Ds C_a \,e^{-S(R,A)} \label{p.func}
\end{equation}
where $\Omega:H^{2\ell+1}(X;\Zh)\to \{\pm 1\}$ is the crucial phase factor described in the
introduction and discussed in detail in section~\ref{subsec:Omega}.

\paragraph{Partition function as a holomorphic section of a line bundle.}
The space of   topologically trivial flat gauge fields is a torus:
\begin{equation}
\mathcal{W}^{2\ell+1}(X)
=\Omega^{2\ell+1}_{d}(X,\Rh)/\Omega^{2\ell+1}_{\Zh}(X)
\end{equation}
which is a quotient of the space of  closed fields,
$\Omega^{2\ell+1}_d(X)$, by the group of  large gauge
transformations $A\mapsto A+R$ where $R$ is a closed $2\ell+1$-form
with integral periods. Thus the partition function is a holomorphic
section of the line bundle $\Lb^{\otimes k}$ over the
complex torus $\mathcal{W}^{2\ell+1}_{\Ch}(X)$ which is obtained
from the real torus $\mathcal{W}^{2\ell+1}(X)$ by using the Hodge
complex structure $J$. Thus $\dim _{\Rh} \mathcal{W}_{\Ch} =
b_{2\ell+1}$.  The line bundle $\Lb\to
\mathcal{W}^{2\ell+1}(X)$ has a nonzero first Chern class
$c_1(\Lb)=k[\omega]_{DR}$. The symplectic form $\omega$ is
of type $(1,1)$ in the Hodge complex structure $J$. From the Kodaira
vanishing theorem and the index of $\bar{\pd}$-operator it follows
that
\begin{equation}
\dim H^0(\mathcal{W}_{\Ch}^{2\ell+1},\mathcal{\Lb}^{\otimes k})=
\int_{\mathcal{W}^{2\ell+1}}e^{kc_1(\mathcal{\Lb})}\mathrm{Td}(T\mathcal{W}^{2\ell+1})
=\int_{\mathcal{W}^{2\ell+1}}e^{k\omega}=k^g
\label{dimL}
\end{equation}
where $g=\frac12\dim H^{2\ell+1}_{DR}(X)$.

It was argued in \cite{Witten:1996hc} that for $k=1$ this
construction describes the partition function of a self-dual field.
{}From \eqref{dimL} it follows that the line bundle $\Lb$
has \textit{unique} holomorphic section. This holomorphic section is
the partition function for a self-dual field.

Therefore to construct a partition function for a self-dual particle
we need to
\begin{enumerate}
\item construct a line bundle $\Lb$ over the torus $\mathcal{W}^{2\ell+1}(X)$ equipped
with norm and hermitian connection $\nabla$ with curvature $-2\pi i \omega$.

\item choose the Hodge complex structure on the torus $\mathcal{W}^{2\ell+1}(X)$.
Using the connection $\nabla^{0,1}$ we can define holomorphic
sections of $\Lb$.
\end{enumerate}

A natural geometrical way of constructing the line bundle and connection on it
is to use Chern-Simons theory in one dimension higher.

\paragraph{Relation to Chern-Simons theory.}
A lot of information about
the line bundle $\mathcal{L}^{\otimes k}$ is encoded in the topological term
\begin{equation}
e^{i\pi p\int_X A\wedge dC}.
\label{topterm}
\end{equation}
Recall that this exponential is not gauge invariant: under the gauge transformation
$\delta A=d\lambda$ and $\delta C=q\lambda$ it transforms by
\begin{equation}
e^{i\pi p\int_X A\wedge dC}\mapsto
\exp\left[i\pi qp\int_X \lambda\wedge F\right]\,e^{i\pi p\int_X A\wedge dC}
\label{gtr}
\end{equation}
This extra phase coming from the gauge transformation looks like the
boundary term of a level
 $k=qp$
``spin'' abelian Chern-Simons theory
 in one dimension higher. Indeed, let $Y$ be a $(4\ell+3)$-manifold
with boundary $X$.
Consider the following topological action
for a topologically trivial gauge field $A\in\Omega^{2\ell+1}(Y)$
\begin{equation}
e^{2\pi i k\CS_Y(A)}=e^{i\pi k\int_Y A\wedge dA}.
\end{equation}
This Lagrangian is not gauge invariant on a manifold with boundary.
Under the gauge transformation $A\mapsto A+d\lambda$ it shifts  by the
boundary term \eqref{gtr}. The Chern-Simons functional on a manifold
$Y$ with boundary $X$ is most naturally considered as a section of
the line bundle $\mathcal{L}_{\CS}$ over the space of gauge fields
$A_X$ on the boundary $X$. Our simple calculation shows that
$\mathcal{L}_{\CS}$ and $\mathcal{L}$ are isomorphic line bundles.

 Up to now we were able to identify $k=qp$ with the level of the
 ``spin''
abelian Chern-Simons in one dimensional higher. But how do the
separate factors  $p$ and $q$ appear in the construction? One should
think of the self-dual partition functions $\mathcal{Z}^+_{pq}$ as
conformal blocks. There are several ways to construct a correlation
function by gluing the conformal blocks: different gluings
corresponds to different factorizations of the level $k$ into
relatively prime factors $k=pq$.

Now we need to add source terms to the self-dual partition function
(see also Appendix~A in \cite{Maldacena:2001ss} and
section 8.2 of Witten's lectures in \cite{qftstr}).
We start again from the action \eqref{Sc}.
Clearly the gauge
invariant coupling of the $C$-field is
\begin{equation}
e^{2\pi i \oint_{\Sigma} C}
\label{Ccopuling1}
\end{equation}
\begin{wrapfigure}{l}{130pt}
\vspace{-5mm}
\includegraphics[width=110pt]{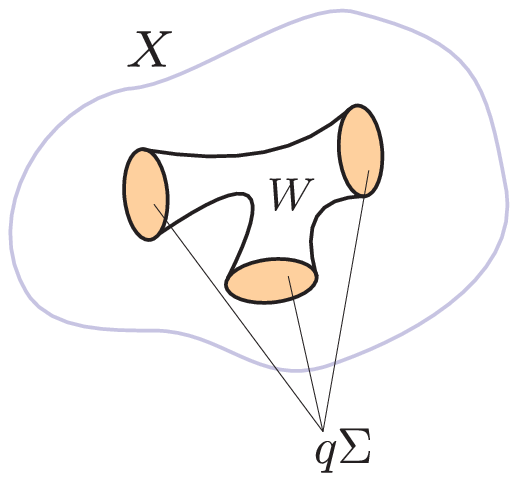}
\vspace{5mm}
\end{wrapfigure}
where $\Sigma$ is a closed $2\ell$-cycle in $X$. The cycle $\Sigma$
is not necessarily connected, but might have several connected components.
Since $X$ is compact we are forced to assume that $[\Sigma]=0$ in
the homology (the total charge on a compact manifold must be zero).
Now we need to generalize this coupling to the gauged theory
\eqref{lagr}. The coupling \eqref{Ccopuling1} is not gauge
invariant: under a gauge transformation $\delta C=q\lambda$ it is
multiplied by the factor
\begin{equation}
\exp\left[2\pi i q\oint_{\Sigma}\lambda\right].
\label{ff}
\end{equation}
It is natural to think of the coupling \eqref{Ccopuling1} as a
section of a line bundle over the space of $\Sigma$'s. More
precisely we choose a cobordism $W$ of $q$ copies of $\Sigma$: $\pd
W=q\Sigma$ and consider the coupling of the $A$ field
\begin{equation}
\exp\left[2\pi i \int_{W}A\right].
\label{Ccopuling2}
\end{equation}
This expression is not gauge invariant under $\delta A=d\lambda$ but
multiplies by factor \eqref{ff}. Thus the couplings
\eqref{Ccopuling1} and \eqref{Ccopuling2} are sections of isomorphic
line bundles. To interpret the coupling \eqref{Ccopuling2} in the
Chern-Simons theory on $Y$ we just need to push out $W$ from $X$ to
$Y$ while keeping the boundary components of $W$ on $X$ so that the
embedding $(q\Sigma,W)\hookrightarrow (X,Y)$ is a neat map.

\paragraph{Summary.}
In this section we argued that the partition function of imaginary
self-dual field $C$, or more general CFT at coupling
$g^{-2}_{p/q}=p/q$, can be obtained by quantizing level $k=qp$ spin
abelian Chern-Simons theory in one dimension higher. The coupling of
the field $C$ to the external sources can be obtained by considering
Wilson surfaces in the Chern-Simons theory. For a topologically
trivial gauge field $A$ the theory is of the form
\begin{equation}
\exp\left[i\pi k\int_Y A\wedge dA+2\pi i \int_W A\right]
\label{CStoptriv}
\end{equation}
where $\pd W=q \Sigma$. To proceed further we need to generalize
this action to topologically nontrivial gauge fields $A$. This is
the subject of the next two sections.

\section{Field space and gauge transformations}
\setcounter{equation}{0}

To proceed further we need to generalize the above construction to
allow topologically nontrivial gauge fields $C$ and $A$. The set of
gauge-inequivalent fields   is an infinite dimensional abelian group
$\check{H}^{2\ell+2}(Y)$, known as a Cheeger-Simons cohomology
group. For an explanation of this see the  pedagogical introduction
to Cheeger-Simons cohomology  in section~2 of \cite{FMS}. This group
can be described by two exact sequences:
\begin{itemize}
\item[\ding{46}] Field strength exact sequence
\begin{equation}
0\to \mathop{\underbrace{H^{2\ell+1}(Y;\Rh/\Zh)}}_{\text{flat
fields}}\to\check{H}^{2\ell+2}(Y) \stackrel{F}{\longrightarrow}
\Omega^{2\ell+2}_{\Zh}(Y) \to 0. \label{fieldstr}
\end{equation}
Every differential character $\check{A}$ has a field strength
$F(\check{A})$ which is a closed $(2\ell+2)$-form with integral
periods.

\item[\ding{46}] Characteristic class exact sequence
\begin{equation}
0\to
\mathop{\underbrace{\Omega^{2\ell+1}(Y)/\Omega^{2\ell+1}_{\Zh}(Y)}}_{
\text{topologically trivial}} \to
\check{H}^{2\ell+2}(Y)\mathop{\longrightarrow}^{a}
H^{2\ell+2}(Y;\Zh)\to 0 \label{charclass}
\end{equation}
Every differential character $\check{A}$ has a characteristic class
$a(\check{A})$ which is an element of integral cohomology
$H^{2\ell+2}(Y;\Zh)$.
\end{itemize}
The field strength and characteristic class are compatible in the
sense that  the reduction $\bar{a}$ of the characteristic class
modulo torsion must coincide with the DeRham cohomology class
$[F]_{DR}$ defined by the field strength: $\bar{a}=[F]_{DR}$.
Putting together the two sequences we can visualize the infinite
dimensional abelian group $\check{H}^{2\ell+2}(Y)$ as
\begin{figure}[!!h]
\centering
\includegraphics[width=330pt]{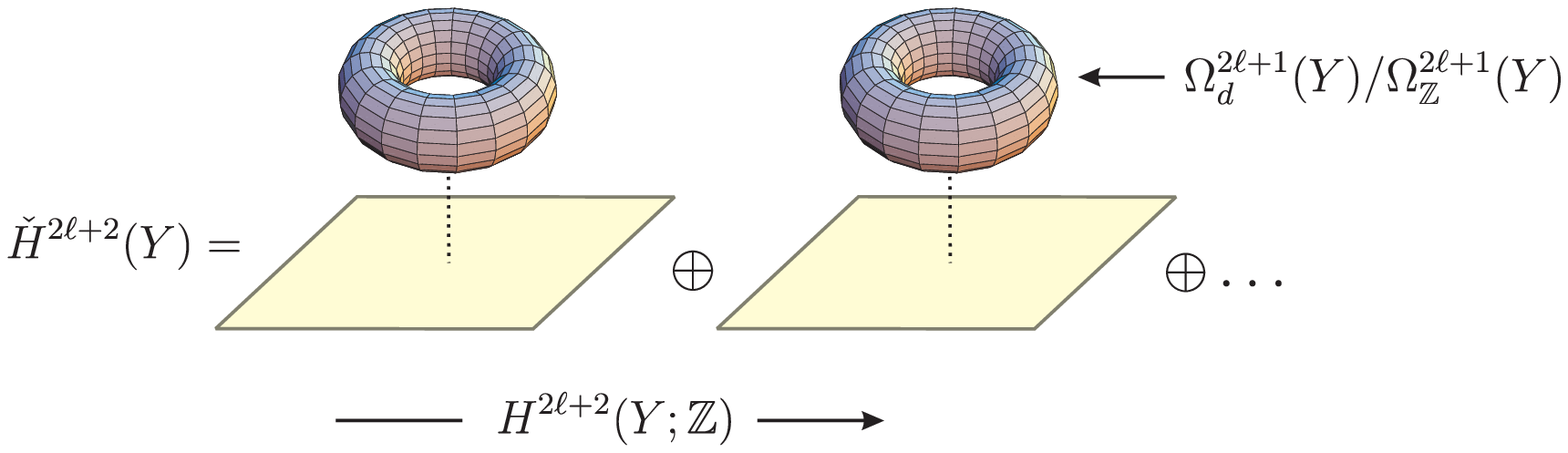}
\end{figure}

\noindent The group $\check{H}^{2\ell+2}(Y)$ consists of many
connected components labeled by the characteristic class $a\in
H^{2\ell+2}(Y;\Zh)$. Each component is a torus fibration over a
vector space. The fibres are finite dimensional tori
$\mathcal{W}^{2\ell+1}(Y)
=\Omega^{2\ell+1}_d(Y)/\Omega^{2\ell+1}_{\Zh}(Y)$ represented by
  topologically trivial flat gauge fields.

There is a product and integration on characters. The product
$[\check{A_1}] \cdot [\check{A_2}]$ induces a graded ring structure
on $\check{H}^*(Y)$, and the integration $\int^{\check{H}}: \check
H^{n+1}(X) \to \Rh/\Zh$ for an $n$-dimensional manifold $Y$.

In terms of differential cohomology classes the action of the
previous section is generalized to be
\begin{equation}
e^{-S(\check{C},\check{A})}=\exp\left[ -\frac{\pi p}{2q}\int_X
(F(\check{C})-q\check{A})\wedge *_E(F(\check{C})-q\check{A}) +i\pi
p\int_X^{\check{H}} \check{A}\cdot \check{C} \right]. \label{lagr2}
\end{equation}

As in Yang-Mills theory, locality forces one to work with gauge
potentials, rather than gauge isomorphism classes of fields. In
generalized abelian gauge theories the proper framework   is to find
a groupoid whose set of isomorphism classes is the set of gauge
equivalence classes. The objects in the category are the gauge
potentials and the ``gauge transformations'' are the morphisms
between objects. One such groupoid was constructed by Hopkins and
Singer, and is known as the groupoid of differential cocycles,
denoted by $\check{\Hc}^{2\ell+2}(Y)$. (We quote the  definitions in
appendix~\ref{sec:appA}). The notation $\check{A}$ is very
intuitive:  it reminds us that locally $\check{A}$ is described by a
differential $(2\ell+1)$-form, but  the $\check{}$   reminds us that
it is not globally well defined.

Unfortunately the category of differential cocycles constructed by
Hopkins and Singer involves non-differentiable objects such as
real-valued cocycles and is somewhat  alien to the intuition of
physicists. At the cost of mathematical naturality we will instead
postulate that there exists an equivalent category
$\check{\mathscr{H}}^{2\ell+2}(X)$ which is closer to the  way we think
about these objects in physics. We would like our category to be a
groupoid obtained by the action of a gauge group on a set of
objects.\footnote{Recall that given any set $S$ and group $G$ acting
on $S$ one can form the category $S/\!\!/G$ whose objects are points
in $S$ and whose morphisms are group actions $s \to s'= g\cdot s$. }

The gauge group, from which we get the  morphisms of the category
$\check{\mathscr{H}}^{2\ell+2}(X)$ is, by hypothesis,  the group
$\check{H}^{2\ell+1}(X)$. The simplest way to motivate this
hypothesis is to consider the action of the gauge group on
\textit{open} Wilson surfaces on $\Sigma$ with nonempty boundary.
The gauge transformation law should be:
\begin{equation}
\chi_{  \check{A}}(\Sigma) \rightarrow \check{\chi}(\pd\Sigma)\,
\chi_{\check{A}}(\Sigma)
\end{equation}
and thus a gauge transformation is precisely given by an element
$\check{\chi} \in \check{H}^{2\ell+1}(X)$.

Now, the set of objects of our category  forms a space,
$\mathscr{C}(X)$. Connected components   are labeled by
$H^{2\ell+2}(X;\Zh)$. \textit{We assume that each component can be taken
to be a  torsor for $\Omega^{2\ell+1}(X)$.}  At the cost of
naturality, we may choose a basepoint $\check{A}_\bullet$, and write
$\check{A} = \check{A}_\bullet + a$, with $a\in
\Omega^{2\ell+1}(X)$. Since $\check\chi(\pd\Sigma) = \exp[2\pi i
\int_{\Sigma} F(\check{\chi}) ] $, it follows that the gauge
transformations are given by
\begin{equation}
g_{\check{C}}\check{A}_X=\check{A}_X+F(\check{C}).
\end{equation}

 Notice that flat characters $H^{2\ell}(X,\Rh/\Zh)$ act trivially
on the space of gauge fields  $\check{\mathscr{H}}^{2\ell+2}(X)$,
therefore the group of automorphisms of any object is
$\mathrm{Aut}(\check{A})=H^{2\ell}(X;\Rh/\Zh)$.

\clearpage
\section{Defining Spin abelian Chern-Simons term in $4\ell+3$ dimensions}
\setcounter{equation}{0} The purpose of this section is to review
the Hopkins-Singer definition of a ``spin'' abelian Chern-Simons
term on a $(4\ell+3)$-dimensional manifold $Y$. We use word ``spin''
loosely here. It does \textit{not} necessarily mean that the
manifold $Y$ admits a spin structure, although this is an important
special case.

\subsection{Chern-Simons functional}

First we define a Chern-Simons action on a \textit{closed} manifold
$Y$.  For motivation let us begin by assuming that   $Y$ is a
boundary of a $(4\ell+4)$-manifold $Z$ and that the differential
character $\check{A}\in \check{\Hc}^{2\ell+2}(Y)$ extends to a
differential character $\check{A}_Z$ defined on $Z$ \footnote{In
general there are obstructions to the existence of both $Z$ and the
extension $\check{A}_{Z}$. In a moment we will define the
Chern-Simons functional without appealing to an extension.}. In this
case we can define the spin Chern-Simons action by
\begin{equation}
e^{2\pi i k \CS_{\pd Z}(\check{A})}=
\exp\left[i\pi k\int_Z F(\check{A}_Z)\wedge F(\check{A}_Z)\right].
\label{CS1}
\end{equation}
This expression does not depend on the extension provided that the
integral $k \oint_Z F\wedge F$ over any closed $(4\ell+4)$-manifold is
an  \textit{even} integer. This is not always true unless $k$ is an
even integer. However from the theory of Wu-classes it follows that
\begin{equation}
a\cup a=\nu_{2\ell+2}\cup a\mod 2
\end{equation}
where $a=a(\check{A}_Z)$ is a characteristic class of the
differential cocycle $\check{A}_Z$, and $\nu_{2\ell+2}(Z)$ is the
Wu-class of degree $2\ell+2$ on $Z$ \footnote{For some pedagogical
material on the Wu class see \cite{MS,Stong}}. Thus if the Wu-class
$\nu_{2\ell+2}(Z)$ vanishes equation \eqref{CS1} indeed defines a
topological action. The total Wu class $\nu$ is related to the
Stiefel-Whitney class $w$ by the Steenrod square operation
$w=\mathrm{Sq}\, \nu$. Using this one easily finds the first few
nonzero Wu classes for an orientable manifold
\begin{equation}
\nu_2=w_2,\quad \nu_4=w_4+w_2^2,
\quad \nu_6=w_2w_4+w_3^2.
\end{equation}
Actually on an $n$-dimensional manifold all Wu classes $\nu_i$ for
$i>[n/2]$ vanish \cite{Stong}. Thus in particular
$\nu_{2\ell+2}(Y)=0$ for \textit{any} oriented $(4\ell+3)$-manifold
$Y$. However it does not necessarily vanish on the extending
$(4\ell+4)$-manifold $Z$. Thus the requirement $\nu_{2\ell+2}(Z)=0$ is
a restriction on a choice of $Z$.

Even in the case when the Wu-class $\nu_{2\ell+2}(Z)$ does not
vanish, we can define a topological action. To this end we need to
choose \textit{an integral lift} $\bar{\lambda}$ of the Wu-class
$\nu_{2\ell+2}$. By an integral lift of the Wu-class we mean an
integral cocycle $\bar{\lambda}\in Z^{2\ell+2}(Z;\Zh)$ such that
$\nu_{2\ell+2}=[\bar{\lambda}]\mod 2$. Clearly the choice of
integral lift is not unique: we can add to $\bar{\lambda}$ any
integral cocycle multiplied by $2$. Now we can define a Chern-Simons
action by
\begin{equation}
e^{2\pi i k\CS_{\pd Z}(\check{A})}= \exp\left[2\pi i k\frac12\int_Z
F(\check{A}_Z)\wedge (F(\check{A}_Z)-\bar{\lambda}_Z) \right].
\label{csdefext}
\end{equation}

Now we would like to write a formula that does not make use of
extensions. Equation \eqref{csdefext} motivates us to choose a
refinement of an integral Wu class, namely  a \textit{differential
integral Wu class} $\check{\lambda}$. A differential integral Wu
class is an element of a category $\check{\Hc}^{2\ell+2}_{\nu}(Y)$
which is a torsor for $\check{\Hc}^{2\ell+2}(Y)$. The objects  are
differential cocycles such that $\nu_{2\ell+2}=a(\check{\lambda})
\mod 2$. Two choices of differential integral Wu class differ by
$\check{\lambda}_1 = \check{\lambda}_2 + 2 \check{A}$ for some
differential cocycle $\check{A}$.  For further details see
appendix~\ref{sec:appA}. The fact that \eqref{csdefext} is
well-defined suggests that $\int_{Y}\check{A}\cdot
(\check{A}-\check{\lambda}) \in \Rh/\Zh$ can be divided by two in a
well-defined way. It is exactly at this point that a choice of QRIF
enters the theory and provides an unambiguous definition of
$\frac12\int_{Y}\check{A}\cdot (\check{A}-\check{\lambda})\in
\Rh/\Zh$.

In fact, what Hopkins and Singer take as the basis for their
Chern-Simons term is
\begin{equation}
\exp\left[2\pi i k\frac18 \int_Y(\check{\lambda}\cdot
\check{\lambda}-\check{L}_{4\ell+4}) \right] \label{hirzexpr}
\end{equation}
where  $\check{L}_{4\ell+4}$ is a differential cocycle refining the
degree $4\ell+4$ component of the Hirzebruch
$L$-polynomial\footnote{ Recall that the Hirzebruch polynomial is
$$
L=\prod_i {x_i \over \tanh x_i} =
1-\frac{p_1}{3}+\frac{7p_2-2p_1^2}{90}+\dots.
$$}.
Again, having chosen a QRIF the division by $8$ is well-defined.
This at first seems unrelated to our Chern-Simons term, but if we
set $\check{\lambda} = \check{\lambda}_0 - 2 \check{A}$ then
\eqref{hirzexpr} becomes
\begin{equation}
e^{2\pi i k\CS_{Y,\check{\lambda}_0}(\check{A})}:= \exp\left[2\pi i
k\frac12\int_{Y}\check{A}\cdot (\check{A}-\check{\lambda}_0) +2\pi i
k\frac18 \int_Y(\check{\lambda}_0\cdot
\check{\lambda}_0-\check{L}_{4\ell+4}) \right] \label{CSdef2}
\end{equation}

In this paper we are mostly concerned with the $A$-dependence of the
Chern-Simons term, but we expect that when one takes into account
metric dependence it will be very useful to include the second term.
The Chern-Simons functional depends on a choice of
$\check{\lambda}$. Its dependence is given by the following simple
formulas:
\begin{subequations}
\begin{equation}
\CS_{Y,\check{\lambda}-2\check{B}}(\check{A})
=\CS_{Y,\check{\lambda}}(\check{A}+\check{B}) \mod 1;
\end{equation}
\begin{equation}
\CS(\check{A}+\check{B})-\CS(\check{A})-\CS(\check{B})+\CS(\check{0})
=\int_Y \check{A}\cdot\check{B}
\mod 1.
\end{equation}
\end{subequations}
Thus the Chern-Simons functional is a quadratic refinement of the
bilinear form on the objects of $\check{\Hc}^{2\ell+2}(Y)$.

\paragraph{Variational formula.}
Suppose we are given a family $Z$ of $4\ell+3$-manifolds over the
interval $[0,1]$.
\begin{wrapfigure}{l}{110pt}
\includegraphics[width=100pt]{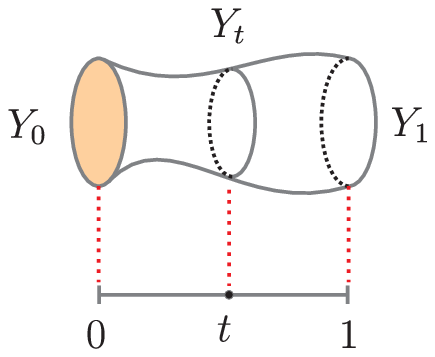}
\end{wrapfigure}
We denote by $Y_t$ the fibre of this family over point $t$. Let
$\check{A}\in \check{\Hc}^{2\ell+2}(Z)$ be a differential cocycle
and let $\check{\lambda}\in\check{\Hc}^{2\ell+2}_{\nu_Z}(Z)$ be a
differential integral Wu class on the total space of the family $Z$.
Then one can show that the following formula is true
\begin{multline}
\CS_{Y_1,\check{\lambda}_1}(\check{A}_1)
-\CS_{Y_0,\check{\lambda}_0}(\check{A}_0) =\frac12\int_{Z}
F(\check{A})\wedge F(\check{A}-\check{\lambda})
\\
+\frac18\int_Z
[F(\check{\lambda})\wedge F(\check{\lambda})-F(\check{L}_{4\ell+4})]
\label{variat} \mod 1.
\end{multline}

\paragraph{Chern-Simons functional on a manifold with boundary.}
On a manifold $Y$ with boundary $X$ the
Chern-Simons functional is naturally defined
as a section of a line bundle $\mathcal{L}_{\CS}$
over the space of gauge inequivalent fields on the boundary $\check{H}^{2\ell+2}(X)$.

The section is constructed as follows:
if $Y$ is \textit{any} $(4\ell+3)$-manifold with boundary $X$ over
which the differential cocycles $\check{A}_X$ and  $\check{\lambda}_X$ extend
 then there is a section $\Psi_{Y}(\check{A})$
of $\mathcal{L}_{\CS}$.
Now given
two possible extensions $(Y_1,\check{A}_1)$ and $(Y_2,\check{A}_2)$
we must specify the gluing function between them. Let $\check{A}_{12}$
be a differential character on $Y_1\cup_X \bar{Y}_2$ obtained by combining two extensions
$\check{A}_1$ and $\check{A}_2$.
Here $\bar{Y}_2$ denotes manifold $Y_2$ with opposite orientation.
The relation between the sections is
\begin{equation}
\Psi_{Y_1}(\check{A}_1)=e^{2\pi i\CS_{Y_1\cup_X \bar{Y}_2,\,\check{\lambda}}(\check{A}_{12})}\,
\Psi_{Y_2}(\check{A}_2).
\end{equation}

\paragraph{Differential cocycles on chains.}
To define the action for topologically nontrivial gauge field
\begin{wrapfigure}{l}{90pt}
\includegraphics[width=85pt]{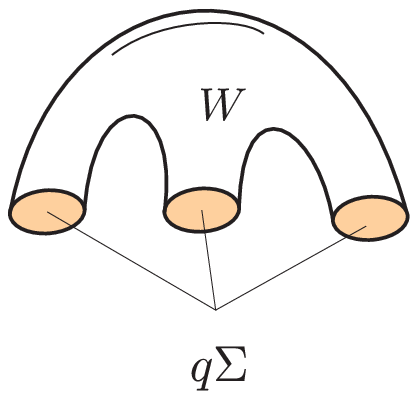}
\end{wrapfigure}
we also need to define a holonomy of $\check{A}$ over a chain
$W\in Z_{2\ell+1}(Y;\Zh)$ with boundary $q\Sigma$.
The boundary $\Sigma$ is not necessarily connected but might
have several connected components. The holonomy of $\check{A}$ over
a chain with boundary is most naturally considered
as a section of a line bundle $\mathcal{L}_{q\Sigma}$
over the space of restrictions of gauge field $A$ to $\Sigma$.

The section is constructed as follows:
if $W$ is \textit{any} $(2\ell+1)$-chain with the boundary $q\Sigma$ over
which the differential cocycles $\check{A}_X$ and  $\check{\lambda}_X$ extend
then there is a section $\mathrm{Hol}_W(\check{A})$
of $\mathcal{L}_{q\Sigma}$.
Now given two possible extensions $W_1$ and $W_2$
we must specify the gluing function between them:
\begin{equation}
\mathrm{Hol}_{W_1}(\check{A})=e^{2\pi i \check{A}(W_1\cup_{q\Sigma}\bar{W}_2)}\,
\mathrm{Hol}_{W_2}(\check{A}).
\label{Hol-line}
\end{equation}

\paragraph{Chern-Simons functional as a quadratic function.}
In the previous section we saw that Chern-Simons functional
appears in three different kinds:
as a number on a closed $4\ell+4$ manifold;
as a map to $\Rh/\Zh$ on a closed $4\ell+3$ manifold;
and as a line bundle with connection on a closed $4\ell+2$ manifold.

Let $E/S$ be a family of manifolds of relative dimension
$4\ell+4-i$, with $i\leqslant 2$. In \cite{Hopkins:2002rd}  Hopkins
and  Singer constructed a Chern-Simons functor from the category of
differential integral Wu structures on $E$ to a category
$\check{\Hc}^i(S)$. So depending on $i$ they constructed: for $i=0$
an integral valued function, for $i=1$ a function with values in
$\Rh/\Zh$ and for $i=2$ a line bundle with connection. Moreover,
these constructions satisfy natural compatibility conditions. We
described this construction in a slightly different language in the
previous paragraphs.

\clearpage
\section{Quantization of the spin abelian Chern-Simons theory}
\label{sec:quant} \setcounter{equation}{0} In this section we
consider Hamiltonian quantization of the spin abelian Chern-Simons
theory on a direct product space $Y=\Rh\times X$. In general,
there   are two ways to quantize Chern-Simons theory: one can first
impose the equation of motion  classically and then quantize the
space  of solutions of this equation, alternatively one can first
quantize the space of all gauge fields and then impose the equation
of motion as an operator constraint. In this paper we mostly follow
the second approach, although our ultimate goal is to construct
wavefunctions on the gauge invariant phase space.

Consider the following topological field theory
\begin{equation}
e^{iS}=e^{2\pi i k \CS_{Y,\check{\lambda}}(\check{A})}e^{-2\pi i\check{A}(W)}
\label{top_action}
\end{equation}
where $W=\Rh\times q\Sigma$ and $q\Sigma$ is a $2\ell$-dimensional
cycle on $X$. The cycle $\Sigma$ is not necessarily connected but
can have several connected components. Since $\nu_{2\ell+2}=0$ on $Y$ one can
shift $\check{A}$ to choose $\check{\lambda}$ to be flat:
$F(\check{\lambda})=0$. We will always make this choice.

Using the variational formula \eqref{variat} one obtains
the familiar equation of motion
\begin{equation}
k F(\check{A})=\delta_{W}
\label{eqr}
\end{equation}
where $\delta_{W}$ is a $(2\ell+2)$-form delta-function supported on the cycle $W$:
$\oint_W a=\int_Y a\wedge \delta_W$.

\paragraph{Hamiltonian.} Recall that the set of objects $\mathscr{C}(Y)$ in
the category $\check{\mathscr{H}}^{2\ell+2}(Y)$ consists of many
connected components labeled by the elements of $H^{2\ell+2}(Y;\Zh)$. Each
component is a torsor for $\Omega^{2\ell+1}(Y)$. Thus we can
parameterize any gauge field $\check{A}$ with a fixed characteristic
class $a(\check{A})$ by choosing a reference gauge field
$\check{A}_{\bullet}$, then $\check{A}=\check{A}_{\bullet} +a$ where
$a$ is a globally well defined $(2\ell+1)$-form. We hope that there
will be no confusion between the differential form $a$ and
characteristic class $a(\check{A})$. The latter will always come
with an argument. Using the variational formula \eqref{variat} one
finds the local action:
\begin{equation}
S_{loc}(\check{A}_{\bullet}+a)=k\int_Y(2\pi a\wedge F_{\bullet}+\pi a\wedge da)-2\pi\oint_{W}a.
\end{equation}

Decomposing
\begin{equation}
a=a_X+dt\wedge a_0\quad\text{and}\quad F_{\bullet}=F_{\bullet,X}
+dt\wedge F_{\bullet,0}
\end{equation}
where all forms on the right hand side are $t$-dependent
forms on $X$ one finds the local Lagrangian
\begin{equation}
L_{loc}(\check{A}_{\bullet}+a)
=2\pi k\int_X \bigl(\tfrac12\,\dot{a}_X\wedge a_X+F_{\bullet,0}\wedge a_X\bigr)
+2\pi \int_X a_0\wedge (k F_X-q\delta_{\Sigma}).
\end{equation}
Here $F_X=F_{\bullet,X}+da$ and $\delta_{\Sigma}$ is a $2\ell+2$-form delta function supported on $\Sigma$:
$\oint_{\Sigma} a_0=:\int_{X}a_0\wedge \delta_{\Sigma}$.

\paragraph{Phase space.}
The symplectic form is $2\pi k\, \omega$ where $\omega$ is defined
in \eqref{omega1}. The phase space $P_{q\Sigma}$ is a union of
components of $\check{H}^{2\ell+2}(X)$ given by the differential
characters with the characteristic class $ka(\check{A})+k\mu
=PD[q\Sigma]$ and (after imposing the classical equation of motion)
is isomorphic to a disjoint union of tori modelled on
$\mathcal{W}^{2\ell+1}(X)=\Omega^{2\ell+1}_d(X)/\Omega^{2\ell+1}_{\Zh}(X)$.
\begin{equation}
P_{q\Sigma}=\{\check{A}_X\in \check{H}^{2\ell+2}(X)\,|\,
ka(\check{A}_X)+k\mu=
PD[q\Sigma]\quad\text{and}\quad
k F(\check{A}_X)=q\delta_{\Sigma}
\}.
\end{equation}

\begin{figure}
\centering
\includegraphics[width=260pt]{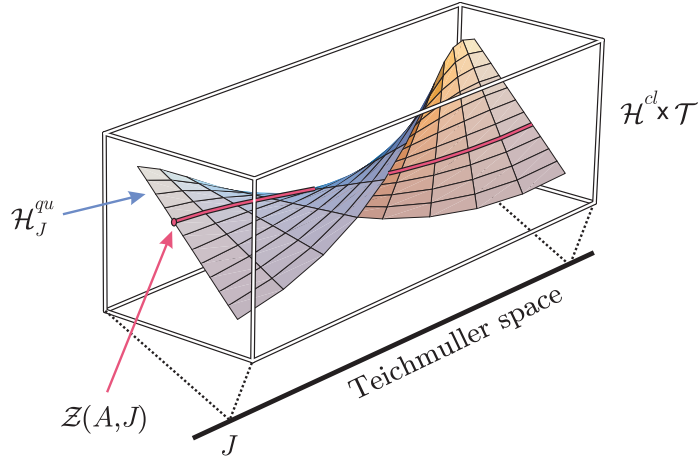}
\caption{The quantum Hilbert space $\Hc^{qu}$
is defined as a subbundle inside the trivial bundle $\Hc^{cl}\times \mathcal{T}$.
The fiber $\Hc^{qu}_J$ over the complex structure $J$ is defined by
$L^2$-normalizable $J$-holomorphic sections of $\Lb\to P_{q\Sigma}$.}
\label{fig:Hqu}
\end{figure}

To quantize Chern-Simons theory we use  geometric quantization.
Recall that geometric quantization consists of three parts:
\begin{enumerate}
\item A choice of prequantum line bundle $\Lb\to P_{q\Sigma}$ over the phase space $P_{q\Sigma}$.
It must be equipped with norm and hermitian connection with
curvature $-2\pi i k\,\omega$.
Then the  prequantum Hilbert space $\mathcal{H}^{cl}=L^2(P_{q\Sigma},\Lb)$
is the space of $L^2$-normalizable sections of $\Lb$.
Note that prequantum line bundle is not unique: one can
take $\Lb\to\Lb\otimes S$ where $S$ is a flat unitary
line bundle defined by an element in $H^1(P_{q\Sigma},\Rh/\Zh)$.

\item Polarization. If $P_{q\Sigma}$ is a K\"ahler manifold then
there is a natural choice of polarization given by
the compatible complex structure $J$. The quantum Hilbert
space $\Hc^{qu}_J=H^0_{L^2}(\Lb)$ is a subspace of $\mathcal{H}^{cl}$
given by holomorphic sections of $\Lb$.

\item The choice of polarization should not be important.
The space of quantum  Hilbert spaces $\Hc^{qu}_J$ form a sub-bundle
in $\Hc^{cl}\times \mathcal{T}$ where $\mathcal{T}$ is a
Teichm\"uller space of complex structures on $P_{q\Sigma}$ (see
Figure~\ref{fig:Hqu}). The fact that the quantization is independent
of the choice of polarization is made precise by equipping this
subbundle with  a  \textit{ projectively flat} connection.
\end{enumerate}

We will approach this by  imposing the Gauss law on
wavefunctions on an infinite dimensional space.

\paragraph{Prequantum line bundle and connection.}
The phase space of the Chern-Simons theory is
\begin{equation*}
\mathscr{C}:=\mathrm{Obj}(\check{\mathscr{H}}^{2\ell+2}(X))
\end{equation*}
the space of gauge fields.
The topological action \eqref{top_action} defines a natural line
bundle
\begin{equation}
\mathcal{L}=\mathcal{L}_{\CS}^{\otimes k}\otimes
\Lc_{Hol}
\to \mathscr{C}\times Z_{2\ell}^{N}.
\end{equation}
We assume that a cocycle $\Sigma\in Z_{2\ell}^{N}$
consists of $N$ connected components, $\Sigma=n_1\Sigma_1+\dots+n_N\Sigma_N$.
Here $\mathcal{L}_{\CS}$
is the line bundle corresponding to the level $1$ Chern-Simons
functional on a manifold with boundary, and $\Lc_{Hol}$
is a line bundle on which the gauge group $\check{H}^{2\ell+1}(X)$
acts as follows:
\begin{equation}
(g_{\check{C}}\Psi)(\Sigma)=e^{2\pi i q\check{C}(\Sigma)}\,\Psi(\Sigma).
\end{equation}

The line bundle $\mathcal{L}_{\CS}^{\otimes k}$ has a natural connection defined by
the level $k$ Chern-Simons phase. Consider a
path $\check{A}_X(t)$ in the space of differential
characters $\mathscr{C}$ where $t\in[0,1]$ is the coordinate on the
path. One can think of $\check{A}_X(t)$ as of a differential character
from $\check{Z}^{2\ell+2}([0,1]\times X)$.
The parallel transport is defined by
\begin{equation}
\mathscr{U}(\{\check{A}_X(t)\}):=e^{2\pi i k\CS_{[0,1]\times X}(\check{A}_X(t))}
\in\mathrm{Hom}(
\Lc_{\CS}^{\otimes k}|_{\check{A}_X(0)},\Lc_{\CS}^{\otimes k}|_{\check{A}_X(1)}).
\label{connection}
\end{equation}
The tangent vector to the path $\{\check{A}_X(t)\}$
is $\phi\in\Omega^{2\ell+1}(X)$. The curvature of the connection
\eqref{connection} can be computed from the variational formula
\eqref{variat}:
\begin{equation}
\Omega_{\check{A}_X}(
\phi_1,\phi_2)=-2\pi i k\,
\omega(\phi_1,\phi_2)
\quad\text{where}\quad
\omega(\phi_1,\phi_2):=\int_X\phi_1\wedge \phi_2.
\end{equation}

The holonomy function allows one to define
a parallel transport of sections of $\mathcal{L}_{Hol}$
along a path  $\{\check{A}_X\}\times W_t$ ($t\in[0,1]$)
in the space of $2\ell$-cycles:
\begin{equation}
\mathscr{U}(\{\check{A}_X\}\times W_t):=e^{-2\pi i q\check{A}_X(W)}
\in\mathrm{Hom}(
\Lc_{Hol}|_{(\check{A}_X,W_0)},\Lc_{Hol}|_{(\check{A}_X,W_1)}).
\label{connection:2}
\end{equation}
The tangent vector to the path $W_t$ is a vector field $\eta$
defined in the vicinity of $W$ and producing an infinitesimal
deformation of the cycle $\Sigma$. The curvature of the connection
\eqref{connection:2} at the point $(\check{A},q \Sigma)$ is
\begin{equation}
\Omega_{(\check{A}_X,\Sigma)}(\eta_1,\eta_2)
=2\pi i q \int_{\Sigma}i_{\eta_1}i_{\eta_2}F(\check{A}_{X}).
\end{equation}

In most equations below we will assume that the $2\ell$-cycle
$\Sigma$ is fixed. So to simplify notations we denote a path
$\check{A}_X(t)\times\{\Sigma\}$ in the space of differential
characters for the fixed $2\ell$-cycle $\Sigma$ by $\check{A}_X(t)$.

\begin{wrapfigure}{l}{95pt}
\vspace{-3mm}
\includegraphics[width=85pt]{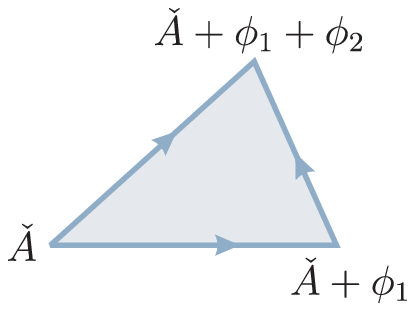}
\end{wrapfigure}
 Now for any $\phi\in\Omega^{2k+1}(X)$ we
introduce a straightline
path $p_{\check{A};\phi}(t)=\check{A}+t\phi$
in the space of
differential characters $\mathscr{C}$.
Using the variational formula \eqref{variat} one finds
\begin{equation}
\mathscr{U}(p_{\check{A}+\phi_1;\phi_2})
\mathscr{U}(p_{\check{A};\phi_1})
=e^{-i\pi k\omega(\phi_1,\phi_2)}\,\mathscr{U}(p_{\check{A};\phi_1+\phi_2})
.
\end{equation}

Now we need to lift action of the gauge group (defined in section 3
above) to the line bundle $\mathcal{L}$. The difference between a
group lift and parallel transport is a cocycle. That is, we can
\begin{wrapfigure}{l}{150pt}
\includegraphics[width=140pt]{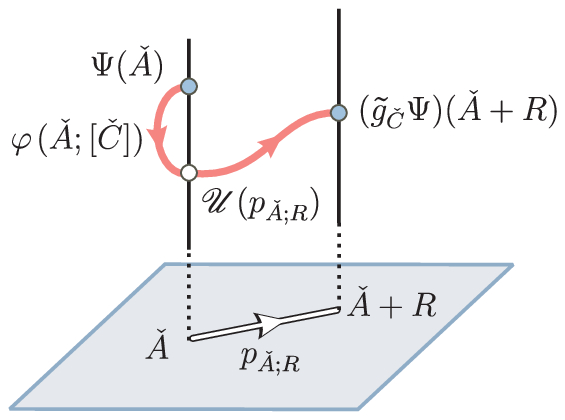}
\end{wrapfigure}
define the group lift by
\begin{equation}
(\tilde{g}_{\check{C}}\Psi)(g_{\check{C}}\check{A}_X;\Sigma)
:=\varphi(\check{A}_X;[\check{C}])e^{-2\pi iq\check{C}(\Sigma)}
\mathscr{U}(p_{\check{A}_X;R(\check{C})})
\Psi(\check{A}_X;\Sigma)
\label{Gaugetr}
\end{equation}
provided  $\varphi$ is a  phase satisfying the cocycle condition:
\begin{multline}
\varphi(g_{\check{C}_1}\check{A}_X;[\check{C}_2])
\varphi(\check{A}_X;[\check{C}_1])
\\
=\varphi(\check{A}_X;[\check{C}_1]+[\check{C}_2]) \,e^{-i\pi k\,
\omega(R(\check{C}_1),R(\check{C}_2))}. \label{cocycle}
\end{multline}

We will impose an operator constraint on the wave function --
the   Gauss law -- which says
$(\tilde{g}_{\check{C}}\Psi)(g_{\check{C}}\check{A};\Sigma)=
\Psi(g_{\check{C}}\check{A};\Sigma)$ (for more details see
section~\ref{subseq:QGL}).

\subsection{Construction of the cocycle via a Chern-Simons term }
One way to construct a cocycle proceeds using a construction going
back to Witten and described in detail in  \cite{Diaconescu:2003bm}.
It makes use of the Chern-Simons term as constructed by
Hopkins-Singer and described in the previous section. Since the
Hopkins-Singer definition is not explicit enough for our purposes,
in the next section we will take a different route to the cocycle.
However, the Chern-Simons definition of the cocycle provides an
important motivation for the construction we use.


Recall that a Chern-Simons functional on $4\ell+2$-manifold
$X$ defines a line bundle with connection. The holonomy
of this connection around the loops in $\check{H}^{2\ell+2}(X)$
is a natural candidate for the  cocycle $\varphi$ \cite{Witten:1999vg}.
This holonomy can be calculated as follows \cite{Witten:1999vg,Diaconescu:2003bm}:
construct a differential cocycle on closed $(4\ell+3)$-manifold
$S^1\times X$:
\begin{equation*}
\check{A}_X+\check{t}\cdot\check{C}
\end{equation*}
where $[\check{t}]\in \check{H}^1(S^1)$ is the canonical
character associated with $S^1\cong U(1)$. This character has
a field strength $F(\check{A})+dt\wedge R(\check{C})$
and characteristic class $a+[dt]\cup a(\check{C})$.
The holonomy on cycles of type $\{t\}\times W_{2\ell+1}$ is
\begin{equation*}
e^{2\pi i \check{A}(W_{2\ell+1})}\,e^{2\pi i t\oint_{W_{2\ell+1}}R(\check{C})},
\end{equation*}
and $e^{2\pi i \check{C}(\Sigma_{2\ell})}$ on cycles of type
$S^1\times\Sigma_{2\ell}$. Now using this twisted differential
character we can define
\begin{equation}
\varphi(\check{A}_X;[\check{C}]):=
e^{2\pi i k\CS_{S^1\times X}(\check{A}_X+\check{t}\cdot \check{C})}.
\label{cocycle_def}
\end{equation}

Two remarks are in order. First, the half-integer level Chern-Simons
term was defined by Hopkins and Singer using a different category
from what we are using, but it only depends on the isomorphism class
of the objects, and hence it can be applied here. Second, the
Chern-Simons term depends on a choice of QRIF. That includes in
particular a choice of spin structure on the $S^1$, which should be
the bounding spin structure $S^1_-$,  for reasons we next discuss.

A standard cobordism argument shows that the functional
\eqref{cocycle_def} satisfies the cocycle relation \eqref{cocycle}.
Indeed, consider a differential character on the
$(4\ell+3)$-dimensional spin manifold $(S^1_-\times X)\cup
(S^1_-\times X)\cup (S^1_-\times X)^P$ restricting to
$\check{A}_X+\check{t}\cdot \check{C}_1$,
$\check{A}_X+R(\check{C}_1)+\check{t}\cdot \check{C}_2$ and
$\check{A}_X+\check{t}\cdot (\check{C}_1+\check{C}_2)$
\begin{wrapfigure}{l}{95pt}
\includegraphics[width=85pt]{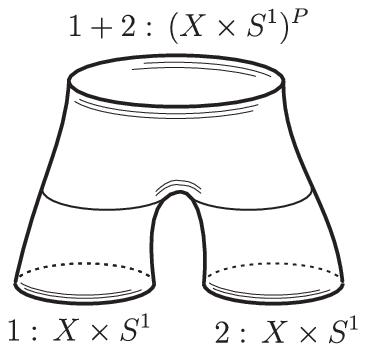}
\end{wrapfigure}
on the three components ($X^P$ means a change of orientation on a
manifold $X$). Then we choose the extending ``generalized spin''
$(4\ell+4)$-manifold to be $Z =\Delta\times X$ where $\Delta$ is a
pair of pants bounding the three circles with spin structure
restricting to $S^1_-$ on the three components. To be explicit we
can choose $\Delta$ to be the simplex $\{(t_1, t_2)\,:\,
 0 \leqslant t_2 \leqslant t_1 \leqslant 1\}$
with identifications $t_i \sim t_i + 1$. We extend the
differential character $\check{A}(t_1, t_2) :=
\check{A}_X + \check{t}_1\cdot \check{C}_1 + \check{t}_2\cdot \check{C}_2$
which clearly restricts to the required
characters on the boundary. The field strength of this character is
$F_{Z}=F(\check{A}_X)+dt_1\wedge R(\check{C}_1)+dt_2\wedge R(\check{C}_2)$.
We therefore can use the variational formula \eqref{variat} to show
\begin{equation*}
\varphi(g_{\check{C}_1}\check{A}_X);[\check{C}_2])
\varphi(\check{A}_X;[\check{C}_1])
\varphi^*(\check{A}_X;[\check{C}_1]+[\check{C}_2])
=e^{i\pi k\int_Z F_Z\wedge F_Z}=e^{-i\pi k\,\omega(R(\check{C}_1),R(\check{C_2}))}.
\end{equation*}

Using properties of the multiplication of differential characters one
can rewrite the cocycle \eqref{cocycle_def} as
\begin{equation}
\varphi(\check{A}_X;[\check{C}])=
\Omega([\check{C}])^k\,e^{2\pi i k\int_X [\check{C}]\cdot [\check{A}_X]}
.
\end{equation}
{}From the properties of the Chern-Simons functional we find that
$\Omega([\check{C}])$ is a locally constant function of
$[\check{C}]$. Therefore it only depends on $a([\check{C}])$. Since
there is no difficulty in defining the integral level Chern-Simons
term (i.e. $k$ divisible by two),   $\Omega$ must take values $\{\pm
1\}$. Finally, from the cocycle condition we derive:
\begin{equation}
\Omega([\check{C}_1]+[\check{C}_2])=\Omega([\check{C}_1])\,
\Omega([\check{C}_2])(-1)^{\int_X a(\check{C}_1)
\cup a(\check{C}_2)}
\label{eqOmega1}
\end{equation}

\subsection{A direct construction of the cocycle}
\label{subsec:Omega}

The definition of the half-integral level Chern-Simons term ($k=1$
in our notations) in \cite{Hopkins:2002rd} is very subtle,
especially in its dependence on certain choices. Therefore, we take
a different view here. Our viewpoint is closer to that of Witten's
in \cite{Witten:1999vg}.

Using the Chern-Simons definition as motivation we construct the
cocycle by setting
\begin{equation}
\varphi(\check{A}_X;[\check{C}])= \Omega^k(a(\check{C}))\,e^{2\pi i
k\int_X [\check{C}]\cdot [\check{A}_X]}.
\label{newdefomega}
\end{equation}
but now, we seek to \textit{define} the cocycle by \textit{choosing}
a function $\Omega: H^{2\ell+1}(X;\Zh) \to U(1)$ such that
\begin{equation}
\Omega(a_1+a_2) = \Omega(a_1) \Omega(a_2) (-1)^{\int_X a_1 \cup
a_2}.
  \label{eqOmega2}
\end{equation}
Any such function $\Omega$ can be used to construct a Chern-Simons
theory. A choice of $\Omega$ is a choice of theory.

What are the possible choices of $\Omega$? Two solutions of
\eqref{eqOmega2} differ by a homomorphism from $H^{2\ell+1}(X;\Zh)$
to $\Rh/\Zh$. By Poincar\'e duality it follows that any two
solutions $\Omega_1$ and $\Omega_2$ are related by $\Omega_2(a) = \Omega_1(a)\, e^{i \pi \int
a \cup \varepsilon} $
where $\varepsilon\in H^{2\ell+1}(X,\Rh/\Zh)$. If we want $\Omega$
to take values $\pm 1$ then $\varepsilon$ is $2$-torsion,
\begin{wrapfigure}{l}{110pt}
\vspace{-2mm}
\includegraphics[width=100pt]{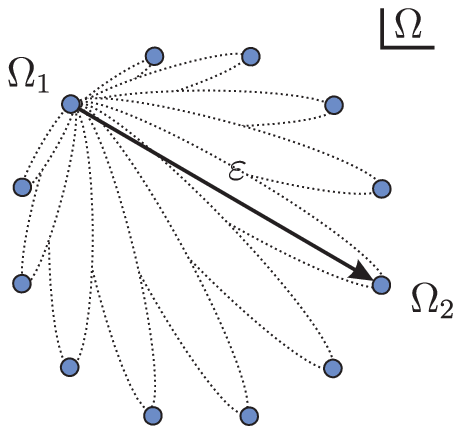}
\end{wrapfigure}
 i.e. $2
\varepsilon=0$.

Now, associated to $\Omega$ is an important invariant. Note that
since the  bilinear form $\int_X a\cup b$ vanishes on   torsion
classes, $\Omega$ is a homomorphism from $H^{2\ell+1}_{tors}(X;\Zh)$
to $\Rh/\Zh$. Since there is a perfect pairing on torsion classes it
follows that there is a $\mu \in H^{2\ell+2}_{tors}(X;\Zh)$ such
that
\begin{equation}
\Omega(a_T) = e^{2\pi i T( a_T, \mu ) } = e^{2\pi i \int_X \alpha \cup
\mu } \label{mu}
\end{equation}
for all torsion classes $a_T$. In the second equality we have written
out the definition of the torsion pairing $T( a, \mu )$, namely, if
$a_T = \beta(\alpha)$ where $\beta$ is the Bockstein map then we can
express it as a cup product. If we choose $\Omega$ to be
$\Zh_2$-valued then $\mu$ is $2$-torsion. Note that if $\Omega_2(a)
= \Omega_1(a) e^{i \pi \int a \cup \varepsilon} $ then $\mu_2 = \mu_1 +
\beta(\varepsilon)$.

Thus, the set of $\Zh_2$-valued solutions $\Omega$ is a torsor for
the group of $2$-torsion points $G=(H^{2\ell+1}(X,\Rh/\Zh))_2$. The
set of solutions with a fixed value of $\mu$ is a torsor for the
$2$-torsion points in the identify component $G_0 =
\mathcal{W}^{2\ell+1}_2(X)$. The group $G_0$  is isomorphic to $
\bar{H}^{2\ell+1}(X;\Zh) /2\bar{H}^{2\ell+1}(X;\Zh)$
where $\bar{H}^{2\ell+1}(X;\Zh)$ denotes the reduction of
the cohomology group modulo torsion.

It remains to establish the existence of a solution to
\eqref{eqOmega2}. To do this we choose a Lagrangian decomposition
$\bar H^{2\ell+1}(X;\Zh) = \bar{\Gamma}_1 \oplus \bar{\Gamma}_2$ and then define
\begin{equation}
\Omega_{\Gamma_1\oplus\Gamma_2}(a):= e^{i \pi \int_X  \bar a^1 \wedge \bar a^2 }
\label{baseomega}
\end{equation}
where $\bar a^1 \in \bar{\Gamma}_1,\, \bar a^2 \in \bar{\Gamma}_2$.  One easily
checks that this is a cocycle. Moreover, it clearly has $\mu=0$.

For $\mu=0$ the $\Zh_2$-valued function $\Omega$ can be related to quadratic refinements of the cup product.
The Gauss-Milgram sum
 formula
 allows one to define a mod $2$ invariant of $\Omega$
(the Arf invariant)
\begin{equation}
\mathrm{arf}(\Omega):=|G|^{-1/2}
\sum_{x\in G}
\Omega(x).
\label{arf}
\end{equation}
{}From equation \eqref{eqOmega2} it follows that $\mathrm{arf}(\Omega)$ takes values in
$\{\pm 1\}$. Thus there are two types of $\Omega$'s
depending on the value of the Arf invariant:
\begin{equation*}
\text{\textbf{even} solutions: } \mathrm{arf}(\Omega)=+1
\quad\text{and}\quad\text{\textbf{odd} solutions: }\mathrm{arf}(\Omega)=-1.
\end{equation*}

Let $\{\Omega_s\}$ be  the  finite set of all solutions with $\mu=0$. It is a
torsor for $G_0$. Denote $q(s):=\mathrm{arf}(\Omega_s)$, now using
\eqref{eqOmega2} one easily obtains:
\begin{equation}
q(s+x)=q(s)\,\Omega_s(x),
\quad\text{or}\quad \Omega_s(x)=\frac{q(s+x)}{q(s)}.
\end{equation}
This shows that $\Omega(x)$ is a ratio of two $\mathrm{mod}\,2$
invariants. Moreover, $q$ is a quadratic refinement of the
intersection pairing:
\begin{equation}
{q(s+x+y)\, q(s) \over q(s+x)\, q(s+y)} = (-1)^{\int x \cup y }
\end{equation}
As mentioned in the introduction,  will refer to a choice of
solution $\Omega$ as to a choice of QRIF. Note  that for $\ell=0$
the set $\{s\}$ is set of spin structures, and $q(s)$ is the mod $2$
index of the Dirac operator corresponding to spin structure $s$, so
a choice of QRIF generalizes the choice of spin structure.

\subsection{Quantum Gauss law}
\label{subseq:QGL} The wave function must define a section of the
line bundle $\Lb$ over a component of the space of gauge
inequivalent fields $\check{H}^{2\ell+2}(X)$ satisfying the tadpole
constraint. In the previous section we constructed a line bundle
$\mathcal{L}$ over the objects $\mathscr{C}$ of the category
$\check{\mathscr{H}}^{2\ell+2}(X)$. A
section $\Psi$ of $\mathcal{L}\to \mathscr{C}$ descends to a section
of $\Lb$ iff it satisfies the Gauss law constraint
\begin{equation}
\Psi(g_{\check{C}}\check{A},q\Sigma)=\varphi(\check{A};[\check{C}])\,
e^{-2\pi i q\check{C}(\Sigma)}
\mathscr{U}(p_{\check{A};R(\check{C})})\Psi(\check{A},q\Sigma).
\label{Glaw}
\end{equation}

\paragraph{Tadpole constraint.}
We have constructed a line bundle with connection over the space of
gauge potentials. Now we would like it to descend to a line bundle
with connection over the isomorphism classes of fields. The
condition for this is the ``tadpole condition,'' which in our
context is the condition that the automorphism group of an object
acts trivially on the fiber of the line bundle. (Here we are
following the general line of reasoning of
\cite{Diaconescu:2003bm}.) This amounts to the condition
\begin{equation*}
\varphi(\check{A};[\check{C}])\,e^{-2\pi i q\check{C}(\Sigma)}=1
\end{equation*}
 on flat characters
$[\check{C}]\in H^{2\ell}(X;\Rh/\Zh)$. Combining \eqref{newdefomega}
with \eqref{mu} we obtain the condition:
\begin{equation}
k\mu + k a(\check{A})=PD[q\Sigma] \label{tadpole}
\end{equation}
where $PD[\Sigma]$ is the Poincar\'{e} dual to the homology class $[\Sigma]\in
H_{2\ell}(X;\Zh)$.

What is the physical interpretation of $\mu$?  If we view $PD[q\Sigma]
-k\mu$ together then we   see
  that we can interpret $\mu$ as the class of a
background brane wrapping a two-torsion cycle. For example, in the
case of the  $5$-brane partition function, nonzero $\mu$ means that
the background contains  a string wrapping a cycle Poincar\'{e} dual
to $\mu$.

Thus, from the physical point of view we should first choose $\mu$.
This partially specifies the background --- telling us the homology
class of possible torsion branes.  Having fixed that background, the
set of possible partition functions is a torsor for $G_0$. These are
the possible partition functions generalizing the well-known set of
partition functions of a self-dual scalar on a Riemann surface.

Finally, let us remark that if   $k>1$ there might be several
solutions to \eqref{tadpole}, e.g., if there is $k$-torsion. In this
case the partition function becomes a section of a line bundle
$\Lb^{\otimes k}$ over several connected components.

The cocycle $\varphi$ looks particularly simple for
differential characters satisfying the tadpole constraint \eqref{tadpole}.
We choose a flat differential cocycle $[\check{\mu}]\in \check{H}^{2\ell+2}(X)$
with (torsion) characteristic class $a(\check{\mu})=\mu$.
If $\Omega([\check{C}])$ is $\Zh_2$-valued then one
can choose $\check{\mu}$ to be $2$-torsion.
Now we can rewrite the phase in  \eqref{Glaw} in the form
\begin{equation*}
\varphi(\check{A};[\check{C}])\,e^{-2\pi i q\check{C}(\Sigma)}
=
\bigl[\Omega(a(\check{C}))e^{-2\pi i \int_X [\check{C}]\cdot [\check{\mu}]}\bigr]^{k}\,
e^{2\pi i k
\int_X [\check{C}]\cdot ([\check{A}]+[\check{\mu}]-\frac{q}{k}[\check{\delta}(\Sigma)])}.
\end{equation*}
Notice that the first term defines a QRIF with zero characteristic class
which we denote by $\Omega_0([R(\check{C})])$. It depends only on the
DeRham cohomology class of the curvature $R(\check{C})$ of the differential cocycle $[\check{C}]$.
For $\check{A}$ satisfying the tadpole constraint \eqref{tadpole}
the term $[\check{A}]+[\check{\mu}]-\frac{q}{k}[\check{\delta}(\Sigma)]$
in the exponential is a topologically trivial differential character.
Thus it can be represented by a $2\ell+1$ form $\sigma(\check{A},\Sigma)$
which satisfies two properties
\begin{equation}
d\sigma(\check{A},\Sigma)=F(\check{A})-\frac{q}{k}\,\delta(\Sigma)
\quad\text{and}\quad
\sigma(\check{A}+a,\Sigma)=\sigma(\check{A},\Sigma)+a
\quad\forall a\in\Omega^{2\ell+1}(X).
\label{sigmadef}
\end{equation}
So finally the Gauss law \eqref{Glaw} can be written as
\begin{equation}
\Psi(g_{\check{C}}\check{A},q\Sigma)=\Omega_0([R])^k\,e^{-2\pi ik
\,\omega(R,\sigma(\check{A},\Sigma))}
\mathscr{U}(p_{\check{A};R(\check{C})})\Psi(\check{A},q\Sigma).
\label{GlawF}
\end{equation}

To conclude:  When the tadpole condition \eqref{tadpole} is
satisfied the line bundle $\Lc$ with connection descends to a line bundle
$\Lb$
with connection over a component (or components) of
$\check{H}^{2\ell+2}(X)$. This line bundle with connection is
completely  determined by a pair
$(\varphi(\check{A},[\check{C}]),k\omega)$ where $\omega$ is the
curvature of the connection (a closed $2$-form with integral
periods) and $\varphi$ is a cocycle
\begin{equation}
\varphi(\check{A};[\check{C}_1]+[\check{C}_2])
=\varphi(\check{A}+R_1;[\check{C}_2])
\,\varphi(\check{A};[\check{C}_1])
\,e^{i\pi k\, \omega(R_1,R_2)}
\end{equation}
which equals $1$ on the flat characters $H^{2\ell}(X;\Rh/\Zh)$. The
cocycle fixes the holonomy of the connection along noncontractible
curves in $\check{H}^{2\ell+2}(X)$. The noncontractible curves are
specified by elements $[\check{C}]$. Note that   the cocycle
$\varphi$ depends both on $\check{A}$ and $[\check{C}]$ as it must
since the curvature is nonzero.  Notice that if we are in the
topologically trivial component, e.g. $a(\check{A})=0$, then we have
a preferred point $\check{A}=0$. Then essentially all nontrivial
information is encoded in the holonomies through the curves
containing the origin \cite{Witten:1999vg}.

\paragraph{Gauss law in local coordinates.}
Each component in the space of objects in
$\check{\mathscr{H}}^{2\ell+2}(X)$ is a contractible space. Thus the
line bundle $\Lc\to \mathscr{C}$
is trivial. To construct a section explicitly we need to choose an
explicit trivialization of this line bundle. To this end we
choose an arbitrary reference differential character
$\check{A}_{\bullet}$ satisfying the tadpole constraint
\eqref{tadpole}. Then an arbitrary field configuration (in the
connected component) can be parameterized by
$\check{A}=\check{A}_{\bullet}+a$ where $a$ is a globally well
defined $(2\ell+1)$-form. Define a canonical nowhere
vanishing section $S$ of unit norm by
\begin{equation*}
S(\check{A}):=
\mathscr{U}(p_{\check{A}_{\bullet};\check{A}-\check{A}_{\bullet}})\, S_{\bullet}
\end{equation*}
where $S_{\bullet}\in \Ch$ and $|S_{\bullet}|=1$. The wave function
$\mathcal{Z}_{p,q}(a,\Sigma)$
is a ratio of two sections $\Psi(\check{A})/S(\check{A})$.

{}From equations \eqref{Gaugetr} and \eqref{GlawF} it follows
that the Gauss law  takes the following form on the wave
function
\begin{equation}
\mathcal{Z}_{p,q}(a+R,\Sigma)=\Omega_0([R])^k\,
e^{-2\pi i k\,\omega(R,\sigma(\check{A}_{\bullet},\Sigma))
-i\pi k\,\omega(R,a)}\,
\mathcal{Z}_{p,q}(a,\Sigma)
\label{GSL2}
\end{equation}
for an arbitrary closed $(2\ell+1)$-form
$R$ with integral periods. The form $\sigma(\check{A},\Sigma)$ is defined in \eqref{sigmadef}.

\paragraph{Dependence on a choice of a base point.}
Recall that the wave function $\mathcal{Z}_{p,q}(a)$ is defined as a ratio
of two sections $\Psi(\check{A})$ and $S(\check{A})$. The section $S$ depends on a choice of the base
point $\check{A}_{\bullet}$, and thus the wave function $\mathcal{Z}_{p,q}$
also depends on this choice.

\begin{wrapfigure}{l}{120pt}
\vspace{-2mm}
\includegraphics[width=110pt]{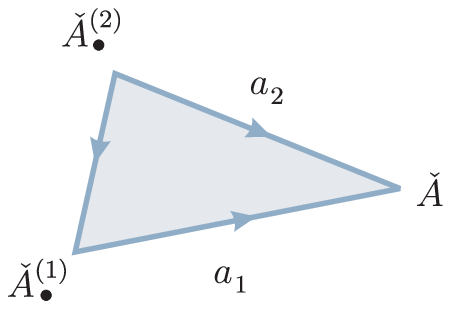}
\vspace{+4mm}
\end{wrapfigure}
Let $\check{A}_{\bullet}^{(1)}$ and $\check{A}^{(2)}_{\bullet}$ be two base points
which solve the tadpole constraint \eqref{tadpole}.
Depending on the choice of a base point we have two different
trivializing sections $S_1$ and $S_2$:
\begin{equation*}
S_i(\check{A}):=\mathscr{U}(p_{\check{A}_{\bullet}^{(i)};\,a_i})\,S_{\bullet}^{(i)}
\end{equation*}
for $i=1,2$ and $a_i:=\check{A}-\check{A}_{\bullet}^{(i)}$.
The coordinates $a_1$ and $a_2$ are related by $a_2=g_{12}(a_1):=a_1+\Delta A_{\bullet}$
where $\Delta A_{\bullet}:= A_{\bullet}^{(1)}-A_{\bullet}^{(2)}$.
In turn the  wave functions
$\mathcal{Z}_1(a_1)$ and
$\mathcal{Z}_{2}(a_2)$ are related by
\begin{equation}
\mathcal{Z}_2(g_{12}(a_1))=\tilde{g}_{12}(a_1)\,\mathcal{Z}_1(a_1)
\label{trchange}
\end{equation}
where the gluing function $\tilde{g}_{12}$ is
\begin{equation}
\tilde{g}_{12}(a_1):=\frac{S_1(\check{A})}{S_2(\check{A})}
=e^{i\pi k\,\omega(a_1,
\Delta A_{\bullet})}\,
(S_{\bullet}^{(2)})^{-1}\mathscr{U}(p_{\check{A}_{\bullet}^{(1)},\Delta A_{\bullet}})
S_{\bullet}^{(1)}.
\label{gcocycle}
\end{equation}

 The dependence on a choice of base point arises because there is
no canonical trivialization, and is related to the fact that the
line bundle is nontrivial over the gauge invariant phase space.
Indeed, one interpretation of these formulas is as follows:
one can think of a choice of   basepoint $\check{A}_{\bullet}$ and
  trivializing section $S(\check{A})$ as a choice of local
coordinate system on the line bundle $\Lc\to \mathscr{C}$ (or $\Lb$
over a component of $\check{H}^{2\ell+2}(X)$). Once we have chosen
$\check{A}_{\bullet}$ we can identify the space of gauge fields with
$\Omega^{2\ell+1}(X)$ (or with
$\Omega^{2\ell+1}(X)/\Omega_{\Zh}^{2\ell+1}(X)$ respectively). In
  local coordinates a section $\Psi(\check{A})$ is described by a
function $\mathcal{Z}_{p,q}(a)$ where
$a:=\check{A}-\check{A}_{\bullet}\in\Omega^{2\ell+1}(X)$. Suppose we
are given two coordinate systems $(\check{A}_{\bullet}^{(1)},S_1)$
and $(\check{A}_{\bullet}^{(2)},S_2)$. The formula \eqref{trchange}
defines the gluing function $\tilde{g}_{12}$ of local sections
$\mathcal{Z}_1$ and $\mathcal{Z}_2$ in coordinate systems $1$ and
$2$. One can easily verify that
$\tilde{g}_{12}\tilde{g}_{23}\tilde{g}_{31}=1$, and thus
\eqref{gcocycle} defines a cocycle. The cocycle $g_{12}(a_1)$ is
globally well defined on $\Lc\to\mathscr{C}$, but it does not
descend to a globally well defined cocycle on
$\Lb\to\check{H}^{2\ell+2}(X)$.

\clearpage
\section{Construction of the partition function}
\label{sec:pfunc} \setcounter{equation}{0} The content of this
section is as follows: To obtain a quantum Hilbert space we need to
choose a polarization on the phase space
$\mathscr{C}=\mathrm{Obj}(\mathscr{H}^{2\ell+2}(X))$. A choice of
Riemannian metric $g_E$ on $X$ defines a complex structure $J=-*_E$
on $T_{\check{A}}\mathscr{C} \cong \Omega^{2\ell+1}(X)$. The quantum
Hilbert space consists of holomorphic sections $\{\Psi\}$ of $\Lc$,
i.e., which satisfy  $D^-\Psi=0$.

Note that there are infinitely many sections of $\mathcal{L}$
which satisfy the Gauss law \eqref{Glaw}, in contrast there are
finitely many \textit{holomorphic} sections which satisfy the Gauss
law \eqref{Glaw}. By choosing a local coordinate system
$(\check{A}_{\bullet},S(\check{A}))$ on $\Lc\to \mathscr{C}$ one can
try to construct a holomorphic solution of the Gauss law explicitly.
The resulting expression will in addition depend on some extra
choices such as a lagrangian decomposition
$\bar{H}^{2\ell+1}(X;\Zh)=\bar{\Gamma}_1\oplus \bar{\Gamma}_2$ of the integral
cohomology modulo torsion. The (local) expression for the partition
function is summarized by Theorem~\ref{thm:1}.

\subsection{Choice of polarization}
Equation \eqref{tadpole} constrains the  connected component in the
space of the gauge fields $\mathscr{C}$. Now by choosing a local coordinate system
$(\check{A}_{\bullet},S)$ we can identify the phase space with the real
vector space $V_{\Rh}=\Omega^{2\ell+1}(X,\Rh)$ by $\check{A}=\check{A}_{\bullet}
+a$, $a\in V_{\Rh}$. \footnote{If we took
the view of constraining and then quantizing then  the phase space
will be a quotient of the space of closed forms.}

The vector space $V_{\Rh}$ has a natural antisymmetric form
\begin{equation}
\omega(\phi_1,\phi_2)=\int_X\phi_1\wedge \phi_2.
\label{omega}
\end{equation}
This $2$-form is closed and nondegenerate and thus it defines
a symplectic structure on the space of gauge fields $\mathscr{C}$.
Moreover a choice of Riemannian metric $g_E$ on $X$ defines the Hodge metric on $V_{\Rh}$
\begin{equation}
g(\phi_1,\phi_2)=\int_X \phi_1\wedge *_E\phi_2.
\end{equation}
Each metric on $V_{\Rh}$ defines for us a compatible complex
structure:
\begin{equation}
g(\phi_1,\phi_2)=\omega(J\cdot\phi_1,\phi_2)\quad\Rightarrow\quad
J\cdot\phi=-*_E\phi.
\label{g}
\end{equation}

Using this complex structure we decompose the space of real forms $V_{\Rh}$ as
\begin{center}
\includegraphics[width=260pt]{sdasd.eps}
\end{center}
Any vector $R^{+}$ of the complex vector space $V^+$ can be
\textit{uniquely} written as
\begin{equation}
R^+=\frac12(R+i*_ER)
\label{Rplus}
\end{equation}
for some real vector $R\in V_{\Rh}$.

This decomposition introduces complex coordinates on the patch
$(\check{A}_{\bullet},S)$. Recall that in \textit{real} local
coordinates we have a covariant derivative $D:=\delta  -i\pi
k\,\omega(\delta a,a)$ which is defined on sections of the line
bundle $\Lc$. Here $\delta$ is the usual differential with respect
to $a$. One can verify that this definition of the covariant
derivative is consistent with the coordinate transformation
\eqref{trchange}. In   \textit{complex} coordinates the covariant
derivative $D$ decomposes as $D=D^++D^-$ where
\begin{equation}
D^+=\delta^+-i\pi k\,\omega(\delta a^+,a^-)
\quad\text{and}\quad
D^-=\delta^--i\pi k\,\omega(\delta a^-,a^+).
\end{equation}
The quantum Hilbert space consists of holomorphic section, i.e.
$D^-\Psi=0$.

In the local coordinates $(\check{A}_{\bullet},S)$ one can identify
holomorphic sections $D^-\mathcal{Z}_{p,q}(a^+,a^-)=0$ with holomorphic functions $\vartheta(a^+)$
via
\begin{equation}
\mathcal{Z}_{p,q}(a^+,a^-;\Sigma)=e^{i\pi k\,\omega(a^-,a^+)}\vartheta(a^+;\Sigma).
\end{equation}
Again one can verify that the corresponding gluing functions
\eqref{gcocycle} for $\vartheta$ depend \textit{holomorphically} on $a^+$.
In this case the Gauss law constraint \eqref{GSL2} takes the following simple
form
\begin{equation}
\vartheta(a^++R^+;\Sigma)=\bigl\{\Omega_0([R])
\,e^{-2\pi i \omega(R,\sigma(\check{A}_{\bullet},\Sigma))}\bigr\}^k\,
\,e^{\frac{\pi k}{2}\,H(R^+,R^+)
+\pi k H(a^+,R^+)}
\,\vartheta(a^+;\Sigma)
\label{theta}
\end{equation}
for all $R\in\Omega^{2\ell+1}_{\Zh}(X)$. Here we have introduced a
hermitian form $H$ on $V^+\times V^+$. It is defined  using the
metric $g$ and symplectic form $\omega$:
\begin{equation}
H(u^+,v^+):=2i\,\omega(u^+,\overline{v^+})
=g(u,v)+i\omega(u,v).
\label{H}
\end{equation}
In our notation $H$ is $\Ch$-linear in the first argument and
$\Ch$-antilinear in the second: $H(u,v)=\overline{H(v,u)}$.

\subsection{Partition function}
\label{subsec:part}

Equation \eqref{theta} looks like a functional equation for a theta
function. The important difference is that the equation for a theta
function is usually defined on a finite dimensional vector space,
while our equation is on the infinite dimensional vector space
$\Omega^{2\ell+1}(X)$. Nevertheless we can use the same technique to
solve it, namely, we will use Fourier analysis.

Note that the function $\vartheta(a^+;\Sigma)$ is not invariant
under translation $a^+\mapsto a^++R^+$. To be able to apply
Fourier analysis we need to have a function which is essentially
invariant (i.e. transforms by a character) under translation of at
least ``half'' of the group $\Omega^{2\ell+1}_{\Zh}(X)$. In the
theory of theta functions this problem is usually solved by
introducing a $\Ch$-\textit{bilinear} form $B$ on $V^+\times V^+$.
Using this $\Ch$-bilinear form we can define a new holomorphic
function
\begin{equation}
\tilde{\vartheta}(a^+;\Sigma)=e^{-\frac{\pi k}{2}\,B(a^+,a^+)}\vartheta(a^+;\Sigma)
\end{equation}
which satisfies the following Gauss law
\begin{equation}
\tilde{\vartheta}(a^++R^+;\Sigma)=\bigl\{\Omega_0([R])
\,e^{-2\pi i \omega(R,\sigma(\check{A}_{\bullet},\Sigma))}\bigr\}^k\,
\,e^{\frac{\pi k}{2}\,(H-B)(R^+,R^+)
+\pi k(H-B)(a^+,R^+)}
\,\tilde{\vartheta}(a^+;\Sigma).
\label{tildetheta}
\end{equation}
The form $(H-B)(\cdot,R^+)$  vanishes on the ``half'' of the group
$\Omega^{2\ell+1}_{\Zh}(X)$. Thus we can solve equation
\eqref{tildetheta} by the Fourier analysis.
So to proceed further we need to define a $\Ch$-bilinear from
on an infinite dimensional space $V^+\times V^+$.
For simplicity we henceforth put $k=1$.

\paragraph{Defining a $\Ch$-bilinear form.}
Recall that we are given an infinite dimensional vector space
$V_{\Rh}=\Omega^{2\ell+1}(X)$, a complex structure on it $J=-*_E$
and a symplectic form $\omega$. In the previous subsection using these structures we defined a
complex vector space $V^+$ together with the hermitian form $H$ on it
which is $\Ch$-linear in the first argument
and $\Ch$-antilinear in the second.

To define a $\Ch$-bilinear form on $V^+\times V^+$
it is sufficient to have a $\Ch$-antilinear involution of $V^+$
\begin{equation*}
\tilde{I}:V^+\to V^+\quad\text{and}\quad
\tilde{I}(z v^+)=\bar{z}\, \tilde{I}(v^+)
\quad\forall z\in \Ch.
\end{equation*}
Indeed, given $\tilde{I}$ we can
define the bilinear form by
\begin{equation}
B(u^+,v^+):=H(u^+,\tilde{I}(v^+)).
\label{BtildeI}
\end{equation}
There is \textit{no} natural choice of such an involution. However a
choice of Lagrangian subspace $V_2\subset V_{\Rh}$
\begin{wrapfigure}{l}{140pt}
\vspace{-5mm}
\includegraphics[width=135pt]{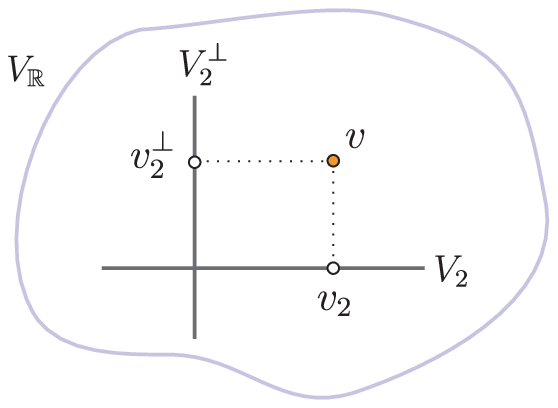}
\end{wrapfigure}
defines one. Note that $V_{\Rh}=V_2\oplus V_2^{\perp}$ with
$V_2^{\perp}:=J(V_2)$ is a Lagrangian decomposition. Indeed, both
$V_2$ and $J(V_2)$ are maximally isotropic subspaces. Thus to prove
the statement it is sufficient to show  that $V_2\cap J(V_2)=\{0\}$.
Since $V_2$ is Lagrangian $V_2\cap J(V_2)$ is a subspace of
$V_{\Rh}$ on which $\omega$ vanishes. But $J$ is a compatible
complex structure, therefore the metric $g$ defined by \eqref{g}
also vanishes. In turn $g$, being the Hodge metric, is nondegenerate
and thus $V_2\cap J(V_2)=\{0\}$.

Every
Lagrangian decomposition defines an involution $I$ of $V_{\Rh}$. Any
vector $v\in V_{\Rh}$ can be uniquely written as
$v=v_2+v_2^{\perp}$. The involution $I$ is defined by changing sign of $v_2^{\perp}$
\begin{equation}
I(v):=v_2-v_2^{\perp}.
\label{defI}
\end{equation}
This involution is compatible with the symplectic structure in
a sense that the $\Rh$-bilinear form $\omega(I\cdot,\cdot)$ is symmetric.
Moreover, it anticommutes with the complex structure, $I\circ J+J\circ I=0$.
The involution $I$ on $V_{\Rh}$ defines a $\Ch$-antilinear involution
$\tilde{I}$ on $V^+$ by\footnote{To prove this
note that for $z=x+iy$ we can write $z v^+ = (x+y J) v^+=((x+y J) v)^+$.
Now using definition \eqref{defI} of the involution $I$ one easily verify
$\tilde{I}(z v^+)=\bar{z}\tilde{I}(v^+)$.
 }
\begin{equation}
\tilde{I}(v^+):=(I(v))^+.
\label{tildeI}
\end{equation}

The form $(H-B)(u^+,v^+)$ can be written in the following
two equivalent ways:
\begin{subequations}
\begin{align}
(H-B)(u^+,v^+)&=2i\, \omega(u,\,\mathcal{F}^-(v))
\quad\text{where}\quad \mathcal{F}^-(v):=v_2^{\perp}-i*_E v_2^{\perp};
\label{H-Bdef2}
\\
&=2g(u_2^{\perp},v_2^{\perp})+2i\,\omega(u_2,v_2^{\perp}).
\label{H-Bdef}
\end{align}
\end{subequations}
Or put differently the form $B$ defined by \eqref{BtildeI} has the following properties
\begin{equation}
B|_{V^+\times V_2^+}=H|_{V^+\times V_2^+}
\quad\text{and}\quad
B|_{V_2^+\times V^+}=H|_{V_2^+\times V^+}-2i\,\omega|_{V_2\times V_{\Rh}}.
\label{Brestr}
\end{equation}
{}From this definition of $B$ it follows that it is
completely determined by the hermitian form $H$
and a choice Lagrangian subspace $V_2$.

Another way to define $B$ proceeds as follows. Note that $V_2^+$
is a real subspace of $V^+$ generating $V^+$ as a complex
vector space. The hermitian form $H$ restricted to $V_2^+\times V_2^+$
defines a symmetric $\Rh$-bilinear form. Its $\Ch$-bilinear extension
to $V^+\times V^+$ yields the form $B$ defined above.

\paragraph{Decomposition of $\Omega^{2\ell+1}_{\Zh}(X)$.}
Having chosen $B$ we can now try to solve equation \eqref{tildetheta}.
{}From \eqref{Brestr} it follows that $(H-B)(\cdot,R^+)$
vanishes for $R\in V_2\cap \Omega^{2\ell+1}_{\Zh}(X)$.
Thus the function $\tilde{\vartheta}(a^+)$ is
essentially invariant (i.e. transforms by a character)
 under
translations by the group $V_2\cap \Omega^{2\ell+1}_{\Zh}(X)$.
Now we need to choose a complementary part of this subgroup
inside $\Omega^{2\ell+1}_{\Zh}(X)$. The complication here is that
$\Omega^{2\ell+1}_{\Zh}(X)$ is \textit{not} a lagrangian subspace.

We define a ``complementary part'' of this subgroup as follows.
The symplectic form $\omega$ on $V_{\Rh}$ defines a symplectic form
on the DeRham cohomology $\Gamma=H^{2\ell+1}_{DR}(X)$. This symplectic
form is integral valued on the image $\bar{\Gamma}:=\bar{H}^{2\ell+1}(X;\Zh)$
 of integral cohomology inside the DeRham cohomology.
In turn, a choice of Lagrangian subspace $V_2\subset
V_{\Rh}$ defines a Lagrangian subspace $\Gamma_2\subset H^{2\ell+1}_{DR}(X)$.
We denote by choose $\bar{\Gamma}_2$ the corresponding lattice inside $\bar{H}^{2\ell+1}(X;\Zh)$.
Now define $\Gamma_1$ to be an arbitrary complementary Lagrangian
subspace of $\Gamma_2$ such that the lattice
$\bar{\Gamma}$ decomposes as $\bar{\Gamma}_1\oplus \bar{\Gamma}_2$. Choose  a subspace
$V_1\subset V_{\Rh}$ to consist of all closed $2\ell+1$-forms
whose DeRham cohomology class lies in $\Gamma_1$:
\begin{equation}
V_1=\{R\in\Omega^{2\ell+1}_{d}(X)\,|\, [R]_{DR}\in\Gamma_1\}.
\label{LdefV1}
\end{equation}
The intersection $V_1\cap \Omega^{2\ell+1}_{\Zh}(X)$ we denote by
$\bar{V}_1$,  i.e. the space  \eqref{LdefV1} with $\Gamma_1$ changed
to $\bar{\Gamma}_1$.

\begin{lemma}
$V_1$ defined by \eqref{LdefV1} is a Lagrangian subspace of $\Omega^{2\ell+1}(X)$.
\label{lem:1}
\end{lemma}
\begin{proof}
Note that by construction $V_1$ is isotropic. We need to prove that
$V_1$ is maximally isotropic. To this end we choose an arbitrary \textit{Riemannian}
metric $h$ on $X$. Then we can rewrite the definition
\eqref{defV1} by changing $\Gamma_1$ to $\Gamma_1^{h}$.
In this form the statement of the lemma obviously follows from
the Hodge decomposition.
\end{proof}

Note that $V_1$ and $V_2$ are \textit{not complementary} lagrangian
subspaces. They have nonzero intersection $V_{12}:=V_1\cap V_2$ where
\begin{equation}
V_{12}=\{\text{exact forms in $V_2$}\}.
\end{equation}

\paragraph{Quadratic function $\Omega_0$.}
To  write down an explicit  solution to the Gauss law \eqref{theta}
we need a simple formula for $\Omega_0$. As we have mentioned above a
choice of $\Omega_0$ with $\mu=0$ is naturally determined by   a
Lagrangian decomposition of
$\bar{H}^{2\ell+1}(X;\Zh)=\bar{\Gamma}_1\oplus \bar{\Gamma}_2$.
Any $R\in\Omega_{\Zh}^{2\ell+1}(X)$ can be written
as $R=R_1+R_2$ where $R_1\in \bar{V}_1$ and $R_2\in V_2\cap\Omega^{2\ell+1}_{\Zh}(X)$.
Since $V_1\cap V_2\ne\{0\}$ this decomposition is not unique
and two different decompositions are related by adding exact forms
in $V_2$. Now define
\begin{equation}
\Omega_{\Gamma_1\oplus \Gamma_2}(R):=e^{i\pi \omega(R_1,R_2)}.
\end{equation}
Since $R_1$ and $R_2$ are closed it follows that $\Omega_{\Gamma_1\oplus \Gamma_2}(R)$
does not depend on a particular choice of decomposition $R=R_1+R_2$.
Moreover $\Omega_{\Gamma_1\oplus\Gamma_2}$ takes values in $\{\pm 1\}$.
\begin{center}
\includegraphics[width=230pt]{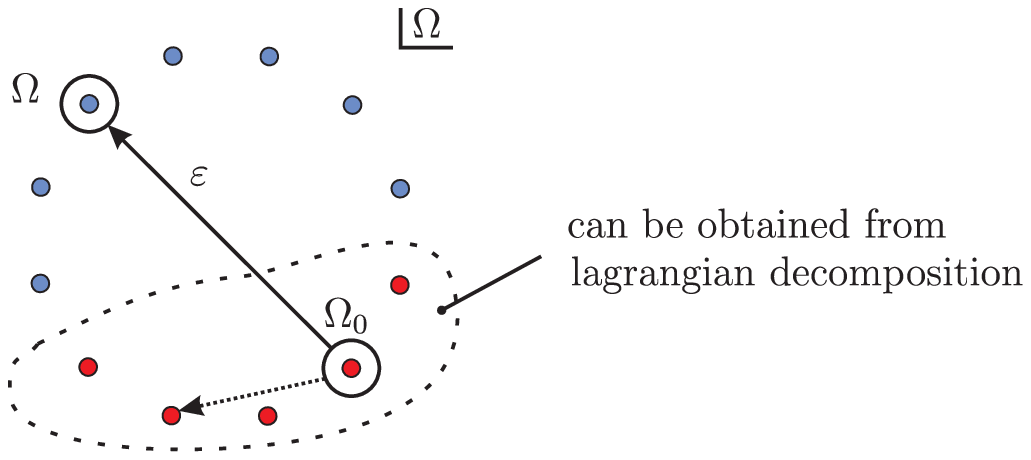}
\end{center}
Given $\Omega_{\Gamma_1\oplus\Gamma_2}$ we can parameterize all solutions with
$\mu=0$ by $[\varepsilon]\in \Omega^{2\ell+1}_d(X)/\Omega^{2\ell+1}_{\Zh}(X)$.
So $\Omega_0$ can be written as
\begin{equation}
\Omega_0(R)=e^{i\pi\omega(R_1,R_2)+2\pi i\omega(\varepsilon,R)}.
\label{Omega0}
\end{equation}
If we want $\Omega_0(R)$ to take values in $\{\pm 1\}$ then
$\varepsilon$ is quantized $[\varepsilon]\in\Omega^{2\ell+1}_{\frac12 \Zh}(X)/
\Omega^{2\ell+1}_{\Zh}(X)$.
In this case a simple calculation shows
that only \textit{even} solutions ($\mathrm{arf}(\Omega)=+1$)
can be obtained by a choice of
Lagrangian decomposition.

We will see that a choice of $\varepsilon$ yields a half integral shift
in the flux quantization condition.
In particular, if $\Omega$ is an odd solution
then the self-dual flux is half-integrally quantized.

\paragraph{Partition function.}
Now one can solve equation \eqref{tildetheta} via   Fourier
analysis. The expression for the partition function can be
summarized by the following theorems:

\begin{thm}
\label{thm:1}
The following Euclidean functional integral
\begin{multline}
\vartheta^{\eta}(a^+)
=\exp\Bigl[-\frac{\pi}{2}(H-B)(\eta^+,\eta^+)
+\frac{\pi}{2}B(a^+,a^+)-\pi(H-B)(a^+,\eta^+)\Bigr]
\\
\times\int_{\bar{V}_1/V_{12}} \mathscr{D} R
\,\exp\Bigl[-\frac{\pi}{2}(H-B)(R^+,R^+)
+\pi(H-B)(a^++\eta^+,R^+)\Bigr]
\label{Epath2}
\end{multline}
(where the integral goes over all closed forms $R\in \bar{V}_1$
modulo exact forms in $V_2$)
$a,\eta\in\Omega^{2\ell+1}(X)$
satisfies the functional equation
\begin{equation}
\vartheta^{\eta}(a^++\lambda^+)=
\Omega_{\Gamma_1\oplus\Gamma_2}(\lambda)\,
e^{2\pi i \,\omega(\eta,\lambda)}\,
e^{\pi H(a^+,\lambda^+)+\frac{\pi}{2}H(\lambda^+,\lambda^+)}\,
\vartheta^{\eta}(a^+)
\label{prop2}
\end{equation}
for all $\lambda\in \Omega^{2\ell+1}_{\Zh}(X)$.
\end{thm}

\begin{rem}
The form $(H-B)$ restricted to $V_1^+\times V_1^+$ is symmetric.
Indeed, the first term in equation \eqref{H-Bdef} is obviously
symmetric. For $u$ and $v$ from a Lagrangian subspace we have
$0=\omega(u,v)= \omega(u_2,v_2^{\perp})+\omega(u_2^{\perp},v_2)$,
which shows that the second term in \eqref{H-Bdef} is also
symmetric. From \eqref{H-Bdef} it also follows that it vanishes on
$V_{12}$ and the $\Re (H-B)|_{V_1^+\times V_1^+}$ is positive
definite on the complement of $V_{12}$ inside $V_1$. In the theory
of theta functions the quadratic  form $(H-B)$ restricted to the
finite dimensional space $\Gamma_1^h:=V_1\cap
\mathscr{H}^{2\ell+1}(X)$
\begin{equation}
\tau(v_1^+) := \frac{i}{2} (H-B)(v_1^+,v_1^+)
\quad\text{for}\quad v_1\in\Gamma_1^h
 \label{cpx.period}
\end{equation}
is known as the complex period matrix.

\end{rem}

\begin{proof}[Proof of the theorem.]
The proof is a straightforward calculation.
First we represent a closed $(2\ell+1)$-form with integral periods $\lambda$
as $\lambda=\lambda_1+\lambda_2$
where $\lambda_1\in \bar{V}_1$ and $\lambda_2\in V_2\cap\Omega^{2\ell+1}_{\Zh}$.
This decomposition is \textit{not} unique. Any two decompositions
are related by adding an exact form from $V_2$. As one will see the result of the calculation
does not depend on a particular choice of such a decomposition.
We shift the integration variable $R\mapsto R+\lambda_1$:
\begin{multline*}
\vartheta^{\eta}(a^++\lambda^+)=
\exp\Bigl[-\frac{\pi}{2}(H-B)(\eta^+,\eta^+)
+\frac{\pi}{2}B(a^++\lambda^+,a^++\lambda^+)-\pi(H-B)(a^++\lambda^+,\eta^+)\Bigr]
\\
\times
\int_{\bar{V}_1/V_{12}}\Ds R
\,\exp\Bigl[-\frac{\pi}{2}(H-B)(R^+,R^+)
-\frac{\pi}{2}(H-B)(\lambda_1^+,\lambda_1^+)
-\pi(H-B)(\lambda^+,R^+)
\\
+\pi(H-B)(a^++\eta^+,R^++\lambda_1^+)
+\pi(H-B)(\lambda^+,R^++\lambda_1^+)
\Bigr].
\end{multline*}
To obtain this expression we used the fact that $(H-B)$ restricted
to a lagrangian subspace $V_1^+$ is symmetric. One sees that the
terms containing $R^+$ and $\lambda^+$ cancel. So we can rewrite
this expression as
\begin{multline*}
\vartheta^{\eta}(a^++\lambda^+)=
\vartheta^{\eta}(a^+)
\exp\Bigl[\frac{\pi}{2}H(\lambda^+,\lambda^+)+\pi H(a^+,\lambda^+)
\Bigr]
\\
\times\exp\Bigl[\frac{\pi}{2}(H-B)(\lambda_2^+,\lambda^+)
-\pi(H-B)(\lambda^+,\eta^+)+\pi(H-B)(\eta^+,\lambda^+)
\Bigr].
\end{multline*}
To obtain this expression we used the fact that $(H-B)|_{V^+\times
V_2^+}=0$. Now since $B$ is symmetric it cancels in the last two
terms in the second line. Using the formula
$H(u^+,v^+)=g(u,v)+i\omega(u,v)$ and \eqref{Brestr} one finds that
\eqref{Epath2} satisfies the  functional equation \eqref{prop2}.
\end{proof}

\begin{cor}
\label{cor61}
The partition function $\mathcal{Z}_{1,1}(a^+,a^-;\Sigma)$ is
\begin{equation}
\mathcal{Z}^{\varepsilon}(a^+,a^-;\Sigma)
=e^{i\pi \omega(a^-,a^+)}\vartheta^{\varepsilon+\sigma(\check{A}_{\bullet},\Sigma)}(a^+)
\label{Zfinal}
\end{equation}
where $\varepsilon\in\Omega^{2\ell+1}_d(X)$ is defined in
\eqref{Omega0}.
\end{cor}

\begin{rem}
Note that $\sigma(\check{A}_{\bullet},\Sigma)$ is a singular
differential form, so the term in \eqref{Epath2} which is quadratic
in $\eta$ requires regularization.
\end{rem}

\begin{thm}
\label{thm62} The functional integral \eqref{Epath2} for
$\eta,\,a\in\mathscr{H}^{2\ell+1}(X)$ is equal to $\Nc_g\,
\vartheta\left[\begin{smallmatrix} \eta_1
\\
\eta_2
\end{smallmatrix}\right](a^+)
$ where $\mathcal{N}_g$ is an important metric dependent factor coming from
the integration over the exact forms in \eqref{Epath2}, and
$\vartheta\left[\begin{smallmatrix}
\eta_1
\\
\eta_2
\end{smallmatrix}\right](a^+)
$ is the canonical theta function on the finite dimensional
torus $\mathscr{H}^{2\ell+1}/\mathscr{H}^{2\ell+1}_{\Zh}$
\begin{multline}
\vartheta\left[\begin{smallmatrix}
\eta_1
\\
\eta_2
\end{smallmatrix}\right](a^+)
=\exp\Bigl[-\frac{\pi}{2}(H-B)(\eta^+_2,\eta^+_1)
+\frac{\pi}{2}B(a^+,a^+)-\pi(H-B)(a^+,\eta^+)\Bigr]
\\
\times \sum_{R\in\bar{\Gamma}_1^{h}-\eta_1}
\,\exp\Bigl[-\frac{\pi}{2}(H-B)(R^+,R^+)
+\pi(H-B)(a^++\eta^+_2,R^+)\Bigr].
\label{Epath3}
\end{multline}
Here $\eta=\eta_1+\eta_2$ according to the Lagrangian decomposition
of the space of harmonic forms
$\Gamma^h=\Gamma_1^h+\Gamma_2^h$.
\end{thm}

\paragraph{Quantum equation of motion.}
The infinitesimal version of the Gauss law \eqref{GSL2} for $R=dc$ yields
the differential equation on $\mathcal{Z}$
\begin{equation}
d(D_a\mathcal{Z}_{1,1})(a;\Sigma)=-2\pi i\Bigl[F(\check{A})-\delta(\Sigma)\Bigr]
\mathcal{Z}_{1,1}(a;\Sigma).
\label{qEOM}
\end{equation}
Now we can apply this equation to the partition function \eqref{Zfinal}
restricted to the real slice in the space of complexified gauge fields: $(a^+)^*=a^-$.
Taking into account that $\mathcal{Z}$ is a holomorphic section,
$D^-\mathcal{Z}=0$ we obtain the following

\begin{thm}[Quantum equation of motion]
The infinitesimal Gauss law yields the quantum equation of motion
\begin{equation}
\bigl\langle d\mathcal{F}^-(R-\varepsilon-\sigma)\bigr\rangle_{\check{A},\Sigma}
=\delta(\Sigma)-F(\check{A})
\label{qEOM2}
\end{equation}
where $\mathcal{F}^-(a):=a_2^{\perp}-i*_Ea_2^{\perp}$ for any
$a\in\Omega^{2\ell+1}(X)$, $\sigma=\sigma(\check{A};\Sigma)$ is
defined in \eqref{sigmadef}. $\langle
\Oc(R)\rangle_{\check{A},\Sigma}$ is the normalized correlation
function defined as the ratio of the Euclidean functional integral
\eqref{Epath2} with the insertion of $\Oc(R)$ and the same integral
without the insertion.
\end{thm}

\begin{proof}
{}From \eqref{H-Bdef2} it follows that
$(H-B)(a^+,b^+)=2i\,\omega(a,\,\mathcal{F}^-(b))$. A straightforward
calculation yields
\begin{equation*}
\frac{1}{2\pi i\, \mathcal{Z}^{\varepsilon}}\,D^+\mathcal{Z}^{\varepsilon}(a^+,a^-;\Sigma)
=\omega(\delta a,\, \bigl\langle \mathcal{F}^-(R-\varepsilon
-\sigma)\bigr\rangle_{\check{A},\Sigma})+\omega(\delta a^+,(a^+)^*-a^-).
\end{equation*}
Now if we restrict it to the real slice $(a^+)^*=a^-$ the last term disappears.
To obtain \eqref{qEOM2} one needs to substitute $\delta a=d \alpha$ into
the equation above, integrate by parts and compare with \eqref{qEOM}.
\end{proof}

\begin{rem}
Suppose that we have chosen another complementary Lagrangian
subspace $\Gamma_1^{\prime h}$ so that $\bar{\Gamma}_1^{\prime h}\oplus\bar{\Gamma}_2^h$
is a decomposition for the lattice $\bar{\Gamma}^h$.
The Lagrangian decompositions
$\Gamma_1'\oplus\Gamma_2$ and $\Gamma_1\oplus\Gamma_2$ are
related by a linear transformation $f:\Gamma_1^h\to \Gamma_2^h$ satisfying
\begin{equation}
\omega(f(v_1),u_1)+\omega(v_1,f(u_1))=0.
\end{equation}
The function $\omega(\bar{f}(\lambda_1),\lambda_1)\mod 2$ is a
linear function on the lattice $\bar{\Gamma}_1^h$. Thus there exists
a vector $w_f$ representing the equivalence class $[w_f]\in
\bar{\Gamma}_2^h/2\bar \Gamma_2^h$ such that
\begin{equation}
\omega(w_f,\lambda_1)=\omega(f(\lambda_1),\lambda_1)\mod 2.
\end{equation}
The theta functions \eqref{Epath3} corresponding to lagrangian decompositions
are related by
\begin{equation}
\vartheta_{\Gamma_1'\oplus\Gamma_2}\left[\begin{smallmatrix}
\varepsilon_1
\\
\varepsilon_2
\end{smallmatrix}
\right](a^+)=e^{i\pi \omega(w_f,\varepsilon_1)}
\vartheta_{\Gamma_1\oplus\Gamma_2}\left[\begin{smallmatrix}
\varepsilon_1
\\
\varepsilon_2+\frac12w_f
\end{smallmatrix}
\right](a^+).
\label{L1dept}
\end{equation}
\end{rem}

\paragraph{The partition function in a simple case.}
In this paragraph we describe the properties of the partition function
of a pure self-dual field, meaning that it couples
neither to external gauge field nor to sources.
To motivate the construction consider first
the chiral scalar in two dimensions. Recall that
the chiral scalar is isomorphic to a chiral Weyl fermion.
It is a section of a spin bundle $S_{\sigma}$.
To define the spin bundle $S_{\sigma}$ globally one needs to choose a spin structure, $\sigma$.
The spin structure is usually not unique. The space of spin structures
is an affine space with translation group $H^1(X;\Zh/2)$.
One can think of this group as the space of isomorphism classes of topologically trivial
line bundles $\Lc_{\alpha}$ with  flat connection
specified by $\alpha\in H^1(X;\Zh/2)$. The group
acts by $S_{\sigma}\mapsto S_{\sigma+\alpha}:=S_{\sigma}\otimes \Lc_{\alpha}$.

Now we choose a spin structure and couple the fermion to the external
$U(1)$ gauge field $A$. This means that we are given a trivial line bundle $\Lc$ with
connection, and the fermion is a section of $S_{\sigma}\otimes\Lc$.
We want now to turn off the gauge field. The surprising fact is that
there are several ways to do it: one can either put the gauge field
to zero and get back the fermion in $S_{\sigma}$,
or one can put $A$ to be \textit{a flat} connection specified
by $\alpha\in H^1(X;\Zh/2)$, in this case one gets a fermion living in
$S_{\sigma+\alpha}$.

Let us go back to our initial problem. We are given a partition
function of the self-dual field coupled to an external $U(1)$ gauge
field $[\check{A}]$. We would like now to turn it off. There are
several ways to do it corresponding to different choices of QRIF
with values in $\{\pm 1\}$.

So we assume that the characteristic class $\mu$ of the QRIF
vanishes and there are no Wilson surfaces. In this case the tadpole
constraint \eqref{tadpole} says that $\check{A}$ is topologically
trivial, $[\check{A}]=[A]$. For our purpose it is enough to consider
only flat gauge fields, so  $dA=0$. In addition we put
$a=A-A_{\bullet}$ to zero. In this case the Gauss law \eqref{GSL2}
takes the form
\begin{equation*}
\mathcal{Z}_{1,1}(R)=\Omega_0([R])\,
e^{-2\pi i\,\omega(R,A_{\bullet})}\,
\mathcal{Z}_{1,1}(0)
\end{equation*}
for any $R\in\Omega^{2\ell+1}_{\Zh}(X)$.
The only parameter we are left to play with is the
holonomy of the base point: $[A_{\bullet}]\in
\Omega^{2\ell+1}_d(X)/\Omega_{\Zh}^{2\ell+1}(X)$.
To fix it we require that the quadratic refinement
\begin{equation*}
\Omega(R):=\Omega_0(R)\,e^{-2\pi i\,\omega(R,A_{\bullet})}
\end{equation*}
take values in $\{\pm 1\}$. The space of such QRIF's can be
parameterized by $[\varepsilon]\in
\Omega^{2\ell+1}_{\frac12\Zh}/\Omega^{2\ell+1}_{\Zh}$:
\begin{equation}
\Omega_{\varepsilon}(R):=\Omega_{\Gamma_1\oplus\Gamma_2}(R)\,e^{2\pi i \omega(\varepsilon,R)}.
\end{equation}
{}From Corollary~\ref{cor61} and Theorem~\ref{thm:1} we obtain the following

\begin{thm}
The set of partition functions for a pure self-dual field forms a
torsor for $\Omega^{2\ell+1}_{\frac12\Zh}/\Omega^{2\ell+1}_{\Zh}$.
The basis can be chosen as
\begin{equation}
\mathcal{Z}\left[\begin{smallmatrix}
\varepsilon_1
\\
\varepsilon_2
\end{smallmatrix}\right]
=e^{-i\pi\omega(\varepsilon_1,\varepsilon_2)}
\,\int_{\bar{V}_1(\varepsilon_1)/V_{12}} \mathscr{D} R
\,\exp\Bigl[-\frac{\pi}{2}(H-B)(R^+,R^+)
+2\pi i\,\omega(\varepsilon_2,R)\Bigr]
\label{EpathSD}
\end{equation}
where $[\varepsilon_1]\in\frac12\bar{\Gamma}_1/\bar{\Gamma}_1$,
$[\varepsilon_2]\in\frac12\bar{\Gamma}_2/\bar{\Gamma}_2$,
we integrate over all forms in $\bar{V}_{12}(\varepsilon_1)$
modulo exact forms in $V_2$ and
\begin{equation}
\bar{V}_{12}(\varepsilon_1)=\{R\in\Omega_{\Zh}^{2\ell+1}(X)\,|\,
[R+\varepsilon_1]_{DR}\in\bar{\Gamma}_1\}.
\end{equation}
\end{thm}

\begin{rem}
{}From \eqref{EpathSD} one obtains that $\varepsilon_1$ has interpretation
of the half-integral shift in the flux quantization condition while
$\varepsilon_2$ is the topological theta angle.
\end{rem}

\subsection{Normalization}
\label{subsec:norm}

In this section we discuss some properties of the metric dependent
normalization factor $\Nc_g$ which appeared in Theorem~\ref{thm62}.
We present a calculation which fixes $\|\Nc_g\|^2$.

The partition function $\mathcal{Z}(a;\Sigma)$ restricted
to $P_{q\Sigma}\cong \Omega^{2\ell+1}_d/\Omega^{2\ell+1}_{\Zh}$ defines
an element of the finite dimensional Hilbert space $\Hc^{qu}$.
One can normalize the wave function, $\|\mathcal{Z}\|^2=1$, with respect to the inner product in
$\Hc^{qu}$ and in this way fix the norm square of $\Nc_g$.

It is clear that $\Nc_g$ does not depend on the source $\Sigma$. So
to simplify the calculation we put it zero, and assume that the
characteristic class of the QRIF $\mu=0$. In this case we can choose
the base point $\check{A}_{\bullet}=0$. This means that
$\sigma(\check{A}_{\bullet})=0$ and
$a=\check{A}-\check{A}_{\bullet}$ is a closed form. In Theorem~\ref{thm62}
we introduced a normalization factor $\Nc_g$. This must be regarded
as a section of a Hermitian line bundle $\mathscr{L}$  over the space
of metrics. The norm on the Hilbert space $\Hc^{qu}$ is just the
$L^2$-norm on $\mathscr{L}\otimes \Lb$:
\begin{equation}
\|\mathcal{Z}\|^2_{L^2}:=\int_{\Omega^{2\ell+1}_d/\Omega^{2\ell+1}_{\Zh}}\Ds
a\, \|\mathcal{Z}(a)\|^2 \label{norm}
\end{equation}
where the second set of $\| \cdot \|^2$ denotes the norm on
$\mathscr{L}$.

 {}From Theorem~\ref{thm62} and Corollary~\ref{cor61} we
learn that the partition function restricted to the real slice
$a^-=(a^+)^*$ can be written as
\begin{equation}
\mathcal{Z}\left[\begin{smallmatrix}
\varepsilon_1
\\
\varepsilon_2
\end{smallmatrix}
\right](a)=\mathcal{N}_g\,e^{-i\pi\omega(\varepsilon_2,\varepsilon_1)}
\sum_{R\in \bar{\Gamma}_1^h-\varepsilon_1}
e^{-\frac{\pi}{2}\,(H-B)(R^+-a_1^+,R^+-a_1^+) +2\pi i
\omega(a_2+\varepsilon_2,R) +i\pi\omega(a_1,a_2)}
\end{equation}
where $a$ is \textit{a harmonic} form, $a=a_1+a_2$ according to the Lagrangian
decomposition $\Gamma^h=\Gamma_1^h\oplus\Gamma_2^h$.

To calculate the norm \eqref{norm} we need to fix a gauge in this
functional integral. This can be done by using equation
\eqref{integral}. By evaluating the Gaussian integral and solving
the equation $\|\mathcal{Z}\|^2_{L^2}=1$ for ${\Nc}_g$ one finds
\begin{equation}
\|\mathcal{ N}_g\|^{2} = \left[\frac{1}{\det(\Im
\tau)}\prod_{p=0}^{2\ell}
\bigl[V_p^{-2}\,\det{}'(L^2d^{\dag}d|_{\Omega^p(X)\cap \mathrm{im}\,
d^{\dag}})\bigr]^{(-1)^{p}}\right]^{-1/2} \label{N}
\end{equation}
where $\tau$ is the complex period matrix defined by
\eqref{cpx.period} (and in appendix~\ref{app:sum}), $V_p$ is the
volume of the torus of harmonic $p$-forms
$\mathscr{H}^p/\mathscr{H}^p_{\Zh}$ defined by \eqref{C:Vp}.

{}From equation \eqref{N} it follows that $\Nc_g$ is some kind of
square root of the right hand side of \eqref{N}. We now conjecture
that there is a very natural squareroot provided we view $\Nc_g$ as
a section of some determinant line bundle. We expect that we should
set
\begin{equation}
\det(\Im \tau)^{-1}\,
\prod_{p=0}^{2\ell}
\bigl[V_p^{-2}\,\det{}'(L^2d^{\dag}d|_{\Omega^p(X)\cap \mathrm{im}\, d^{\dag}})\bigr]^{(-1)^{p}}
=\|\det \overline{\mathcal{D}}\|^2_Q
\label{Quillen}
\end{equation}
where the right hand side is the Quillen norm of a section $\det
\overline{\mathcal{D}}$ of some determinant line bundle
$\texttt{DET}(\overline{\mathcal{D}})$ over the space of metrics on
$X$. In the case of the chiral scalar $\det \overline{\mathcal{D}} =
\det \bar \pd$. Here one can check that indeed on a two dimensional
torus with metric $ds^2=\frac{1}{\tau_2}|d\sigma^1-\tau
d\sigma^2|^2$ the integral over exact forms in \eqref{Epath2} equals
$1/\eta(\tau)$. This is consistent with \eqref{Quillen} for
$\ell=0$. So we conjecture that this will continue to be true for
$\ell>0$.

\section{Action and equations of motion}
\label{sec:eom}
\setcounter{equation}{0}
The action for the self-dual field is essentially
the complex period matrix \eqref{cpx.period} extended
from the cohomology to the space of closed forms.
The purpose of this section is
to describe this extension.

\subsection{Classical action}
First we need to extend the definition of the complex period matrix
\eqref{cpx.period} defined on the cohomology to the infinite
dimensional  symplectic  vector space $V_{\Rh}=\Omega^{2\ell+1}$.

\paragraph{$(X,g_E)=$ Riemannian manifold.}
Following the  discussion in section~\ref{subsec:part} we choose an
orthogonal coordinate system on $V_{\Rh}$ to be $V_2\oplus
V_2^{\perp}$ where $V_2$ is a Lagrangian subspace and
$V_2^{\perp}=*_EV_2$ is its orthogonal complement with respect to
the Hodge metric. From the positivity of $g_E$ it follows that
$V_2\cap V_2^{\perp}=\{0\}$. Recall that the Hodge complex structure
is compatible with the symplectic
\begin{wrapfigure}{l}{150pt}
\includegraphics[width=145pt]{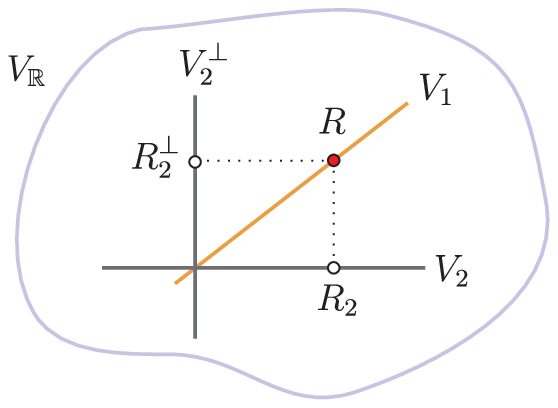}
\end{wrapfigure}
structure, thus
$V_{\Rh}=V_2\oplus V_2^{\perp}$ is a
Lagrangian decomposition.
So any form $v\in V_{\Rh}$ can be uniquely
written in the form
$v=v_2+v_2^{\perp}$ for some $v_2\in V_2$ and $v_2^{\perp}\in V_2^{\perp}$.

Let $V_1$ be another Lagrangian subspace. A choice of Lagrangian
decomposition $\Gamma=\Gamma_1\oplus \Gamma_2$ of the cohomology
$\Gamma=H^{2\ell+1}_{DR}(X)$ defines a \textit{canonical} choice of
$V_1$. However we postpone this discussion till the next paragraph.
Now any element $R$ from the Lagrangian subspace $V_1$ can be
written as
\begin{equation}
R=R_2+R_2^{\perp}
\label{RRperp_dec}
\end{equation}
where $R_2$ and $R_2^{\perp}$ are not independent but related by some linear function
(see the figure). From \eqref{H-Bdef} it follows
that the Euclidean action is
\begin{equation}
S_E(R^+):=\pi\int_X \bigl(R_2^{\perp}\wedge *_E R_2^{\perp}-iR_2\wedge R_2^{\perp}\bigr).
\label{Eaction}
\end{equation}

\paragraph{$(M,g)=$ Lorentzian manifold.}
The action in the Lorentzian signature can obtained
from \eqref{Eaction} by Wick rotation:
\begin{equation}
S_L(R):=\pi\int_M \bigl(R_2^{\perp}\wedge *R_2^{\perp} +R_2\wedge R_2^{\perp}\bigr).
\label{Maction}
\end{equation}
This action depends on a choice of Lagrangian subspace $V_2$. For a
Riemannian manifold a choice of $V_2$ automatically defines the
Lagrangian decomposition $V_{\Rh}=V_2\oplus  *_EV_2$. For a
Lorentzian manifold this is not true, and we need \textit{to
constrain} the choice of $V_2$ by the requirement
\begin{equation}
V_2\cap *V_2 =\{0\}.
\label{V2constr}
\end{equation}

In principle, $V_2$ can be an arbitrary Lagrangian subspace
satisfying the constraint \eqref{V2constr}. Recall that on any Lorentzian
manifold $M$ there exists a nowhere vanishing
timelike vector field $\xi$. It can be
used to define a Lagrangian subspace
$V_2\subset \Omega^{2\ell+1}(M)$:
\begin{equation*}
V_2(\xi):=\{\omega\in \Omega^{2\ell+1}(M)\,|\,i_{\xi}\omega=0\}.
\end{equation*}

\paragraph{A choice of Lagrangian subspace $V_1$.}
The symplectic form $\omega$ on $V_{\Rh}$ defines a symplectic form
on the DeRham cohomology $\Gamma=H^{2\ell+1}_{DR}(M)$.
In turn, a choice of Lagrangian subspace $V_2\subset
V_{\Rh}$ defines a Lagrangian subspace $\Gamma_2\subset \Gamma$. We
choose $\Gamma_1$ to be an arbitrary complementary Lagrangian
subspace, so $\Gamma=\Gamma_1\oplus \Gamma_2$. Choose  a subspace
$V_1\subset V_{\Rh}$ to consists of all closed $2\ell+1$-forms
whose DeRham cohomology class lies in $\Gamma_1$:
\begin{equation}
V_1=\{R\in\Omega^{2\ell+1}_d(M)\,|\, [R]_{DR}\in\Gamma_1\}.
\label{defV1}
\end{equation}
$V_1$ defined by \eqref{defV1} is a Lagrangian subspace of $\Omega^{2\ell+1}(M)$
(see Lemma~\ref{lem:1} for proof).

\paragraph{Equations of motion.} The variational problem for the action \eqref{Maction}
is summarized by the following theorem:
\begin{thm}
\label{thm:71}
Let $V_1\subset V_{\Rh}$ be a Lagrangian subspace defined by
\eqref{defV1}, and let $R\in V_1$ be a closed form. Then the action
\begin{equation}
S_L(R)=\pi\int_M \bigl(R_2^{\perp}\wedge *R_2^{\perp} + R_2\wedge R_2^{\perp}\bigr)
\label{SMaction}
\end{equation}
has the following properties:
\begin{subequations}
\begin{enumerate}
\item[a.] Variation with respect to $R\mapsto R+d\delta c$
where $\delta c\in\Omega^{2\ell}_{cpt}(M)$ is
\begin{equation}
\delta S_L(R)=-2\pi\int_M \delta c
\wedge d (*R_2^{\perp}- R_2).
\label{nSDflux}
\end{equation}

\item[b.] Stationary points of the action are the solutions of equations:
\begin{align}
\text{Bianchi identity:}&\quad d(R_2+R_2^{\perp})=0
\label{bianchi}
\\
\text{equation of motion:}&\quad d (*R_2^{\perp}- R_2)=0.
\label{eomthm}
\end{align}

\item[c.] From \eqref{bianchi} and \eqref{eomthm} it follows
that the following anti-self dual form is closed:
\begin{equation}
\mathcal{F}^{+}(R):=R_2^{\perp}+ *R_2^{\perp},
\qquad d\mathcal{F}^{+}(R)=0.
\label{decF-}
\end{equation}

\item[d.] The variation \eqref{nSDflux} can be written as
\begin{equation}
\delta S_L(R)=-2\pi\int_M \delta c
\wedge d \mathcal{F}^{+}(R).
\label{SDflux}
\end{equation}

\end{enumerate}
\end{subequations}
\end{thm}

\begin{proof}
The variation of the action is
\begin{equation*}
\delta S_L=\pi\int_{M} \bigl(
2\delta R_2^{\perp}\wedge *R_2^{\perp} + \delta R_2\wedge R_2^{\perp}
+R_2\wedge \delta R_2^{\perp}\bigr).
\end{equation*}
The variations $\delta R_2$ and $\delta R_2^{\perp}$ are not independent
but come from the variation $\delta R$. $R$ is an element
of Lagrangian subspace and thus so must be $R+\delta R$:
\begin{equation}
0=\omega(\delta R, R)=\int_{M}\bigl(\delta R_2\wedge R_2^{\perp}
+\delta R_2^{\perp}\wedge R_2\bigr).
\label{trick}
\end{equation}
Using this constraint one easily finds that the variation of the action is
\begin{equation*}
\delta S_L=2\pi\int_M \delta R_2^{\perp}
\wedge (\underbrace{*R_2^{\perp} -R_2}_{\in V_2}).
\end{equation*}
Notice that the expression in the brackets is in the Lagrangian
subspace $V_2$. Thus we can substitute $\delta R$ instead of $\delta
R_2^{\perp}$:
\begin{equation*}
\delta S_L=2\pi\int_M \delta R
\wedge (*R_2^{\perp} -R_2).
\end{equation*}
Recall that we must vary $R$ within a fixed cohomology class, so $\delta R=d\,\delta c$
where $\delta c\in \Omega^{2\ell}_{cpt}(M)$:
\begin{equation}
\delta S_L=-2\pi\int_M \delta c
\wedge d(*R_2^{\perp} -R_2)
\stackrel{dR=0}{=}
-2\pi\int_M \delta c
\wedge d(\underbrace{R_2^{\perp}+*R_2^{\perp}}_{=:\mathcal{F}^+})
\label{variation}
=-2\pi\int_M \delta c
\wedge d\mathcal{F}^+.
\end{equation}
Note that $\mathcal{F}^+(R)$ is \textit{automatically} self dual.
Since $\delta c$ is an arbitrary $2\ell$-form  it immediately
follows that $d\mathcal{F}^+=0$.
\end{proof}

\begin{rem}
The action for anti self-dual field can be obtained from
\eqref{SMaction} by changing the sign of the second term.
\end{rem}

\begin{thm}
An arbitrary \textbf{closed} self dual form $\mathcal{F}^+$ can
be written in form $\mathcal{F}^+(R)$ for some $R\in V_1$.
\end{thm}
\begin{proof}
A closed self-dual form defines the self-dual
DeRham cohomology class $[\mathcal{F}^+]_{DR}\in H^{2\ell+1}_{DR}(M)$.
If this DeRham cohomology class is zero then $\mathcal{F}^+$ is
an exact form, and we can take $R=\mathcal{F}^+\in V_1$.

If $[\mathcal{F}^+]_{DR}\ne 0$ then
we choose a self-dual representative $\alpha^+$ of this
cohomology class. Then $\mathcal{F}^+-\alpha^+$ is the self-dual
\textit{exact} form. From the constraint $V_2\cap *V_2=\{0\}$
it follows that $\Gamma_2\cap [*\Gamma_2]=[0]$ in the DeRham cohomology.
Thus the class $\alpha^+$ can be written as $\alpha_2+*\alpha_2$
for some representative $\alpha_2$ of the cohomology class
$[\alpha_2]\in \Gamma_2$. From the fact that
$\Gamma_1$ and $\Gamma_2$ are  complementary Lagrangian subspaces
it follows that the class $[*\alpha_2]$ can be obtained
by the orthogonal projection from some class $[\alpha_1]\in \Gamma_1$.
So choose an arbitrary representative $\alpha_1$ of the class $[\alpha_1]$
and consider $R=\alpha_1+(\mathcal{F}^+-\alpha^+)$. By the construction $R$
is in $V_1$ and $\mathcal{F}^+(R)=\mathcal{F}^+$.
\end{proof}

\paragraph{Gauge symmetries.} The action \eqref{SMaction} in addition
to the standard gauge symmetry $C\mapsto C+\omega$ where $\omega$ is
closed $2\ell$-form with integral periods has an extra gauge symmetry.
Indeed, the functional  $S_L(R)$ vanishes on the Lagrangian subspace $V_2$.
The Lagrangian subspaces $V_1$ and $V_2$ have non-zero intersection which
we denote by
\begin{equation}
V_{12}:=V_1\cap V_2=\{\text{exact forms in }V_2\}.
\end{equation}
For any $R\in V_1$ and $v_{12}\in V_{12}^{cpt}$ (compactly
supported forms in $V_{12}$) it follows that
\begin{equation}
S_L(R+v_{12})=S_L(R)+S_L(v_{12})=S_L(R).
\end{equation}
This properties can be summarized by the following theorem:
\begin{thm} The action \eqref{SMaction} has two types of gauge symmetries:
\begin{enumerate}
\item[a.] It manifestly invariant with respect $C\mapsto C+\omega$
where $\omega\in \Omega^{2\ell}_{\Zh}(M)$.
\item[b.] It is invariant under a shift $R\mapsto R+v_{12}$ where $v_{12}\in (V_1\cap V_2)_{cpt}$:
\begin{equation}
S_L(R+v_{12})=S_L(R).
\label{extragauge}
\end{equation}
The anti self-dual field \eqref{decF-} does not depend on $v_{12}$:
\begin{equation}
\mathcal{F}^+(R+v_{12})=\mathcal{F}^+(R).
\end{equation}
\end{enumerate}
\end{thm}

{}From this theorem it follows that
the gauge symmetry $R\mapsto R+v_{12}$ does not affect classical equations.
However this extra gauge symmetry has to be taken into account
in the quantum theory.

\paragraph{Coupling to the sources.}
One can use Theorem~\ref{thm:1} to write a coupling
of the self-dual field to a brane.

\begin{thm}
Let $\Sigma$ be a topologically trivial $2\ell$-cycle on $M$
(it might have several connected components). The differential
character $[\check{\delta}(\Sigma)]$ is topologically trivial
and is represented by a (singular) $(2\ell+1)$-form
$\sigma\in[\sigma]\in\Omega^{2\ell+1}(M)/\Omega^{2\ell+1}_{\Zh}(M)$:
$d\sigma = \delta(\Sigma)$. The action for the self-dual field
coupled to the brane is
\begin{equation}
S_L(R):=\pi\int_M (R_2^{\perp}\wedge * R_2^{\perp}
+R_2\wedge R_2^{\perp})
+2\pi \int_M R_2^{\perp}\wedge (*\sigma_2^{\perp}-\sigma_2).
\label{SDactSigma}
\end{equation}
The variation of the action with respect to $R\mapsto R+d\delta c$
for $\delta c\in\Omega^{2\ell}_{cpt}(M)$ is
\begin{equation}
\delta S_L(R)=2\pi \int_M \delta c\wedge \bigl[
\delta(\Sigma)-d\mathcal{F}^+(R+\sigma)
\bigr].
\label{SDSvar}
\end{equation}
\end{thm}

\begin{rem}
The equations of motion can be written as
\begin{equation}
dR=0,\quad d\mathcal{F}^+(R+\sigma)=\delta(\Sigma)\quad\text{for}
\quad R\in V_1\quad\text{and}\quad
[\sigma]=[\check{\delta}(\Sigma)].
\end{equation}
\end{rem}

\begin{rem}
The condition $[\sigma]=[\check{\delta}(\Sigma)]$ fixes only
an equivalence class $[\sigma]\in\Omega^{2\ell+1}(M)/\Omega^{2\ell+1}_{\Zh}(M)$
but not $\sigma$ itself.  The action \eqref{SDactSigma} depends on a choice
of a representative $\sigma$ of $[\sigma]$. However the partition function
\eqref{Epath2} does not depend on such a choice.
\end{rem}

\begin{proof}
The action \eqref{SDactSigma} can be directly obtained from
Theorem~\ref{thm:1} and Corollary~\ref{cor61} by setting $\check{A}=0$.
Using the result of Theorem~\ref{thm:71} one finds
\begin{equation*}
\delta S_L(R)=-2\pi \int_M \delta c\wedge d\mathcal{F}^+(R)
+2\pi \int_M \delta R_2^{\perp}\wedge (*\sigma_2^{\perp}-\sigma_2).
\end{equation*}
The term in the brackets lies in the Lagrangian subspace $V_2$ thus
we can change $\delta R_2^{\perp}$ to $\delta R=d \delta c$.
Integrating by parts and using that $d\sigma=\delta(\Sigma)$ and
 $\mathcal{F}^+(R):=R_2^{\perp}+*R_2^{\perp}$
one arrives at equation \eqref{SDSvar}.
\end{proof}

\subsection{Examples}
\label{sec:examples}
In this section we consider two examples: a chiral scalar field on $\Rh^{1,1}$
and a self-dual field on a production manifold $M=\Rh\times N$
where $N$ is compact $4\ell+1$-manifold.

\subsubsection{Chiral scalar on $\Rh^{2}$.}
Consider a Lorentzian manifold $M=\Rh^{2}$ equipped
with the metric $ds^2=e^{2\varphi(x,t)}(-dt^2+dx^2)$.
Choose the Lagrangian subspace $V_2$ as
\begin{equation}
V_2=\{dt\,\omega_t(x,t)\}\quad\Rightarrow\quad
V_2^{\perp}=\{dx\,\omega_x(x,t)\}.
\end{equation}
The Lagrangian subspace $V_1$ is just the space of
all exact $1$-forms $\Omega^{1}_{\text{exact}}(M)$. So $R\in V_1$ decomposes as
\begin{equation}
R=\underbrace{dx\,R_x}_{R_2^{\perp}}+\underbrace{dt\,R_t}_{R_2}.
\end{equation}
The action \eqref{SMaction} takes the form
\begin{equation*}
S_L(R)=\pi\int_{\Rh^2}dtdx\,R_x(R_x+R_t).
\end{equation*}
Now for $R=d\phi$ the action becomes
\begin{equation}
S_L(\phi)=\pi\int_{\Rh^2}dtdx\,
\Bigl[(\pd_x\phi)^2+(\pd_x\phi)(\pd_t\phi)\Bigr].
\end{equation}
This  action for the chiral scalar has been proposed before
\cite{Floreanini:1987as}. The equation of motion is
\begin{equation}
(\pd_x+\pd_t)\pd_x\phi=0.
\end{equation}
Thus the general solution is $\phi(x,t)=f(t)+\phi_L(x-t)$. It seems
that we get an extra degree of freedom represented by an arbitrary
function of time $f(t)$. However the anti self-dual field
$\mathcal{F}^+$ depends only on $\phi_L(x-t)$. Indeed, substituting
the solution to equation \eqref{decF-} one finds
\begin{equation}
\mathcal{F}^+(\phi)=(dx-dt)\phi_L'(x-t)
\end{equation}
where $\phi_L^{\prime}$ denotes the derivative of $\phi_L$ with respect to
the argument.

The extra degree of freedom $f(t)$ is the gauge
degree of freedom \eqref{extragauge}, and it can be removed by the gauge
transformation $R\mapsto R-df(t)$ where $-df(t)\in V_1\cap V_2$.

\clearpage
\subsubsection{Action for self-dual field on a product manifold}
\label{subsec:ex}
\begin{wrapfigure}{l}{190pt}
\includegraphics[width=180pt]{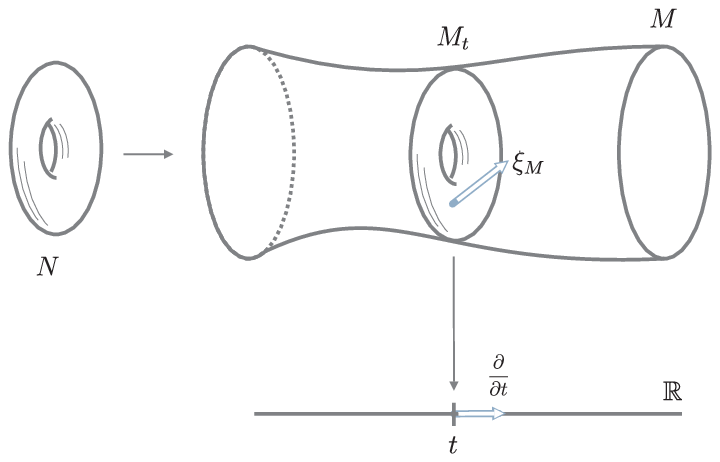}
\end{wrapfigure}
Not every $(4\ell+2)$-manifold $M$ admits a smooth Lorentzian
metric\footnote{Recall that oriented manifold $M$ admits a
Lorentzian metric if and only if there exists a nowhere vanishing
vector field.}. However if $M$ is a product manifold say $\Rh\times
N$
 then there exists a
nowhere vanishing vector field, and thus $M$ admits a Lorentzian metric.
It is convenient to think of $M=\Rh\times N$ as a topologically trivial fibration
$M$ over $\Rh$ with fiber $N$.

On $\Rh$ there is a canonical nowhere vanishing vector field $\pd/\pd t$.
To lift it to $M$ we need to choose a connection on $M$.
A connection $\Theta$ is a globally well defined $1$-form on $M$
with values in the Lie algebra of $\mathrm{Diff}^+(N)$, i.e. $\Theta \in \Omega^1(M;\mathrm{Vect}(N))$.
A choice of Riemannian metric $g_N$ on $N$ and Lorentzian metric $ds^2_{\Rh}=-\rho^2dt^2$
on $\Rh$ defines a Lorentzian metric on $M$
\begin{subequations}
\begin{equation}
ds^2_M=-\rho^2\,dt^2+g_N(\Theta,\Theta).
\label{exmet}
\end{equation}
In local coordinates $(t,x^i)$ the connection $\Theta$ can be written as
\begin{equation}
\Theta=(dx^i-\xi^i dt)\otimes \frac{\pd}{\pd x^i},
\end{equation}
and the metric takes the familiar form
\begin{equation}
ds_M^2=-\rho^2 dt^2+(g_N)_{ij}(dx^i-\xi^i dt)(dx^j-\xi^j dt).
\label{lsmetric}
\end{equation}
\end{subequations}
In general relativity $\rho$ is usually called lapse and $\xi^j$ the
shift.

The vector field $\pd/\pd t$ on $\Rh$ lifts to a vector field $\xi_M$ on $M$:
\begin{equation}
i_{\xi_M}\Theta=0\quad\Rightarrow\quad \xi_M:=\frac{\pd}{\pd t}+\xi^i\frac{\pd}{\pd x^i}.
\end{equation}
The connection $\Theta$ defines a decomposition of tangent bundle $T_xM$
on horizontal and vertical vector fields. This decomposition is the orthogonal
decomposition in the metric \eqref{exmet}.

\paragraph{Decomposition of forms.}
An orthogonal projector onto the space of horizontal vectors is
defined by
\begin{equation*}
P(\eta):=\xi_M\,\frac{g(\xi_M,\eta)}{g(\xi_M,\xi_M)}=\xi_M\, dt(\eta).
\end{equation*}
The dual projector $P^*:=dt\otimes i_{\xi_M}$ defines a decomposition
of the differential forms onto vertical and horizontal:
\begin{equation}
\Omega^{2\ell+1}(M)=\underbrace{P^*\Omega^{2\ell+1}(M)}_{horizontal}\oplus
\underbrace{(1-P^*)\Omega^{2\ell+1}(M)}_{vertical}.
\label{HV}
\end{equation}

Since $M$ is a product manifold there is another decomposition of
\begin{equation}
\Omega^{2\ell+1}(M)=\Omega^{2\ell+1}(N)\otimes \Omega^0(\Rh)\oplus
\Omega^{2\ell}(N)\otimes \Omega^1(\Rh).
\label{odec}
\end{equation}
These two decomposition are related in the following way
\begin{equation}
R=\bar{R}+dt\wedge \bar{R}_0=
\underbrace{(\bar{R}-dt\wedge i_{\xi}\bar{R})}_{vertical}
+\underbrace{dt\wedge(\bar{R}_0+i_{\xi}\bar{R})}_{horizontal}.
\label{hvdec}
\end{equation}

The Hodge $*_M$ exchanges the vertical and horizontal forms
\begin{equation}
*_M R=\underbrace{-\rho\,dt\wedge *_N\bar{R}}_{horizontal}
+\underbrace{\bigl\{-\rho^{-1}*_N(\bar{R}_0+i_{\xi}\bar{R})
+dt\wedge i_{\xi}[\rho^{-1}*_N(\bar{R}_0+i_{\xi}\bar{R})]\bigr\}}_{vertical}
\label{hodge*}
\end{equation}

\paragraph{Lagrangian subspaces.}
We choose a Lagrangian subspace $V_2=\Omega^{2\ell}(N)\otimes \Omega^1(\Rh)$.
Notice that it coincides with the space of horizontal $2\ell+1$-forms,
thus $*_M V_2 = \{\text{vertical forms}\}$.

The DeRham cohomology decomposes as
\begin{equation}
H^{2\ell+1}_{DR}(M)\cong H^{2\ell+1}_{DR}(N)\oplus H^{2\ell}_{DR}(N).
\end{equation}
A choice of Lagrangian subspace $V_2$ yields a choice for $\Gamma_2$:
$\Gamma_2\cong H^{2\ell}(N)$. Choose $\Gamma_1\cong H^{2\ell+1}(N)$.
Now the Lagrangian subspace $V_1$ is defined by
\begin{equation}
V_1=\{ R\in\Omega^{2\ell+1}_d(M)\,|\,[R]_{DR}\in H^{2\ell+1}_{DR}(N)\}.
\end{equation}

\paragraph{Action.}
Given a form $R\in V_1$ we can write it as in \eqref{hvdec}
\begin{equation}
R=\underbrace{(\bar{R}-dt\wedge i_{\xi}\bar{R})}_{R_2^{\perp}}
+\underbrace{dt\wedge(\bar{R}_0+i_{\xi}\bar{R})}_{R_2}.
\end{equation}
Now substituting this decomposition into the action \eqref{Maction}
and using \eqref{hodge*} one finds
\begin{equation}
S_L(R)=\pi\int_{\Rh}dt
\int_N \bigl[
\rho\,\bar{R}\wedge *_N\bar{R}+(\bar{R}_0+i_{\xi}\bar{R})\wedge \bar{R}
\bigr].
\label{MAaction}
\end{equation}

\paragraph{Gauge symmetry.}
The action \eqref{MAaction} has an extra gauge symmetry \eqref{extragauge}:
\begin{equation}
V_{12}=V_1\cap V_2=\{dt\wedge (\dot{\bar{c}}-d_N\bar{c}_0)\,|\,
\bar{c}\in \Omega^{2\ell}_{d_N}(N)\otimes \Omega^0(\Rh)
\;\;\text{and}\;\;
\bar{c}_0\in\Omega^{2\ell-1}(N)\otimes\Omega^1(\Rh)\}.
\label{exV12}
\end{equation}

\paragraph{Equations of motion.}
The equation of motion \eqref{eomthm} take the form
\begin{equation}
d_N[\rho *_N\bar{R}+\bar{R}_0+i_{\xi}\bar{R}]=0,
\text{ or using Bianchi identity } \dot{\bar{R}}=d_N (\rho *_N\bar{R}-i_{\xi}\bar{R}).
\label{exeom}
\end{equation}
Let $\bar{\alpha}$ be a time-independent
$2\ell+1$-form on $N$ representing the DeRham cohomology class of $[R]_{DR}\in \Gamma_1$.
$R$ can be written as
\begin{equation*}
R=\alpha + d_N\bar{c}+dt\wedge (\dot{\bar{c}}-d_N\bar{c}_0).
\end{equation*}
Substituting this to \eqref{exeom} one finds
\begin{equation}
d_N\bigl[\dot{\bar{c}}-\rho *_N(\bar{\alpha}+d_N\bar{c})
+i_{\xi}(\bar{\alpha}+d_N\bar{c})\bigr]=0.
\label{eom1}
\end{equation}
Notice that arbitrary ($t$-dependent) closed $2\ell$-form
solves this equation. This set of solution corresponds to
the gauge degrees of freedom containing in \eqref{exV12}.
One should not worry about this set of solutions because they
are projected out from the expression for the self-dual field:
\begin{equation}
\mathcal{F}^+=\left.\bar{R}-dt\wedge \bigl[\rho *_N\bar{R}-i_{\xi}\bar{R}\bigr]
\right|_{\bar{R}=\bar{\alpha}+d_N\bar{c}}
.
\end{equation}

\subsection{Comparison with Henneaux-Teitelboim action}
\label{sec:other}

In this section we compare with the previous work of Henneaux and
Teitelboim. We begin by reviewing briefly  their work \cite{Henneaux:1988gg}.

\paragraph{Henneaux-Teitelboim action.}

Let $M=\Rh\times N$ be a product manifold equipped with the
Lorentzian metric \eqref{lsmetric}. Consider a \textit{closed}
$2\ell+1$-form $F$ on $M$ (locally $F=dC$). It can be decomposed as
in \eqref{hvdec}
\begin{equation}
F=dt\wedge \bar{F}_0+\bar{F},\quad
d_N\bar{F}=0\quad\text{and}\quad
\dot{\bar{F}}=d_N\bar{F}_0.
\end{equation}
The equation of motion for $F$ is the self-duality constraint.
The self-duality constraint $*_M F=F$ is equivalent to
\begin{equation}
\rho*_N\bar{F}+\bar{F}_0+i_{\xi}\bar{F}=0.
\label{HT:SD}
\end{equation}
The
action of Henneaux and Teitelboim \cite{Henneaux:1988gg}
can be summarized by the following two theorems:

\begin{thm} The action
\begin{equation}
S_{HT}(F):=\int_{\Rh} dt\int_N \bigl[
\rho\bar{F}\wedge *_N\bar{F}+(\bar{F}_0+i_{\xi}\bar{F})\wedge \bar{F}
\bigr]
\label{HT:action}
\end{equation}
for the closed $2\ell+1$-form $F=dt\wedge \bar{F}_0+\bar{F}$ has the following properties:
\begin{enumerate}
\item it is manifestly invariant under $C\mapsto C+\omega$ where $\omega\in \Omega^{2\ell+1}_{\Zh}(M)$.

\item the variation of the action under $F\mapsto F+d_N\delta\bar{c}
+dt\wedge (\delta\dot{\bar{c}}-d_N\bar\delta \bar{c}_0)$  is
\begin{equation}
\delta S_{HT}=-2\int_{\Rh} dt\int_N\delta\bar{c}\wedge
d_N\bigl[\rho*_N\bar{F}+\bar{F}_0+i_{\xi}\bar{F}\bigr].
\label{HT:eom}
\end{equation}
Here $\delta\bar{c}$ and $\delta \bar{c}_0$ are $t$-dependent $2\ell$-
and $(2\ell-1)$-forms on $N$ with compact support.

\item it has an additional gauge symmetry $\bar{F}_0\mapsto \bar{F}_0+
\dot{\bar{\lambda}}-d_N\bar{\lambda}_0$ where $\bar{\lambda}$ is a
closed $t$-dependent $2\ell$-from on $N$ with compact support, and
$\bar{\lambda}_0$ is an arbitrary $t$-dependent $(2\ell-1)$-from on
$N$.
\end{enumerate}
\end{thm}

\begin{thm} A family
of closed forms
\begin{equation}
F_{\bar{\lambda},\,\bar{\lambda}_0}=\bar{F}+dt\wedge (\bar{F}_0+\dot{\bar{\lambda}}-d_N\bar{\lambda}_0)
\end{equation}
where $\bar{\lambda}\in \Omega^{2\ell}_{d_N}(N)\otimes \Omega^0(\Rh)$
and $\bar{\lambda}_0\in\Omega^{2\ell-1}(N)$
contains a self-dual form $\mathcal{F}^+=F_{\bar{\lambda}^*,\,\bar{\lambda}_0^*}$
if and only if
\begin{equation}
d_N\bigl[\rho*_N\bar{F}+\bar{F}_0+i_{\xi}\bar{F}\bigr]=0,
\label{eqthm}
\end{equation}
i.e. iff the family $F_{\bar{\lambda},\,\bar{\lambda}_0}$ is an extremum of the action \eqref{HT:action}.
\end{thm}

The idea of the proof is as follows \cite{Bekaert:1998yp}.
If $F$ is closed and self-dual then
because of equation \eqref{HT:SD} it satisfies \eqref{eqthm}.
The converse goes as follows: If \eqref{eqthm} is satisfied then
\begin{equation*}
\rho*_N\bar{F}+\bar{F}_0+i_{\xi}\bar{F}=\omega_{2\ell}(t)
\end{equation*}
where $\omega_{2\ell}$ is a closed $t$-dependent $2\ell$-form on $N$.
Using the Riemannian metric $g_N$ on $N$ we can decompose $\omega_{2\ell}$
on harmonic and exact parts
$\omega_{2\ell}=\omega_{2\ell}^h(t)+d_N\alpha_{2\ell-1}(t)$.
Now choose
\begin{equation*}
\bar{\lambda}^*(t):=\int^tdt'\,\omega_{2\ell}^h(t')
\quad\text{and}\quad
\bar{\lambda}_0^*(t):=-\alpha_{2\ell-1}(t).
\end{equation*}
The form $F_{\bar{\lambda}^*,\,\bar{\lambda}_0^*}$ is self-dual.

\paragraph{Comparison of the two actions.}
By comparing expressions \eqref{HT:action} and \eqref{MAaction} one
finds that modulo a  change of notation they are identical. However
there is an important difference: in the Henneaux-Teitelboim
approach one tries to get a self-duality constraint from the
variation of the action. By contrast,  in our approach we get a
condition that a certain self-dual form is closed.

\section{Dependence on metric}
\label{sec:metric}
\setcounter{equation}{0}
Let us recall the general strategy in geometric quantization of
Chern-Simons theories.  The partition function is a covariantly
constant section of the vector bundle
$\mathcal{H}^{qu}_J\to\mathcal{T}$:
\begin{equation}
(\nabla_J \mathcal{Z})(A^+,A^-;J)=0.
\label{nablaJ}
\end{equation}
A change of polarization can be interpreted as a change of
 creation/annihilation operators corresponding to a  Bogoliubov transformation.
Thus, a change of complex structure can be compensated by a
Bogoliubov transformation. The Bogoliubov transformation is
implemented by a quadratic exponential, and infinitesimally it is
represented by second order differential operator:
\begin{equation*}
\nabla^{1,0}_J=\delta_J^{1,0}-(\dots)D^+D^+
\quad\text{and}\quad
\nabla^{0,1}_J=\delta^{0,1}_J.
\end{equation*}
In our case the complex structure is determined by the Hodge metric:
$J=-*_E$. Thus eq.~\eqref{Sca}  implies that the metric $g_X$ couples only
to $\mathcal{F}^+$, the self-dual part of $\check{C}$.

\paragraph{A trick with metric variation.}
In this section we study dependance of the action on the  choice of
metric $g$. In particular we need to calculate the variation
$\delta_g(*\omega_k)$. In local coordinates the Hodge $*$ is defined
by \footnote{We use convention in which
\begin{equation*}
\alpha_k\wedge *\beta_k=\frac{1}{k!}\,\alpha_{\mu_1\dots \mu_k}
\beta_{\nu_1\dots \nu_k}\,g^{\mu_1\nu_1}\dots g^{\mu_k\nu_k}
\,\vol(g).
\end{equation*}
}
\begin{equation}
*\omega_k=\frac{1}{k!(d-k)!}\,|\det g|^{1/2}
\,\omega_{\mu_1\dots\mu_k}g^{\mu_1\nu_1}
\dots g^{\mu_k\nu_k}\varepsilon_{\nu_1\dots \nu_k
\alpha_1\dots \alpha_{d-k}}\,
dx^{\alpha_1}\dots dx^{\alpha_{d-k}}.
\end{equation}
Now it is easy to see that the variation of this expression with
respect to the metric consists of two parts which can be elegantly
written as (for a  metric of any signature)
\begin{subequations}
\begin{equation}
\delta_g(*)\omega_k=-\frac12\,\tr(\delta g^{-1}g)*\omega_k
+*(\xi_g\omega_k)
\label{metric-variat}
\end{equation}
where
\begin{equation}
\xi_g:=(\delta g^{-1} g)^{\mu}{}_{\nu}
\,dx^{\nu}\wedge i(\tfrac{\pd}{\pd x^{\mu}})
\label{xi}
\end{equation}
\end{subequations}
The formula \eqref{metric-variat} is the main computational
tool in this section.

\subsection{Stress-energy tensor}
Let us recall a few facts about the stress-energy tensor of a
$(2\ell+1)$-form on a $(4\ell+2)$-dimensional Lorentzian manifold
$(M,g)$. The stress-energy tensor for an unconstrained
$(2\ell+1)$-form $F$ is
\begin{equation*}
T_{\mu\nu}(F):=\frac{\pi}{2(2\ell+1)!}\Bigl[
(2\ell+1)F_{\mu\alpha_1\dots\alpha_{2\ell}}
F_{\nu}{}^{\alpha_1\dots \alpha_{2\ell}}
-\frac12\,g_{\mu\nu}F_{\alpha_1\dots\alpha_{2\ell+1}}
F^{\alpha_1\dots\alpha_{2\ell+1}}
\Bigr].
\end{equation*}
Working with such expressions in local coordinates proves to be
cumbersome, and it is better to proceed in the following
coordinate-independent way. This stress-energy tensor was obtained
from the following action
\begin{equation*}
S=\frac{\pi}{2}\int_X F\wedge *F,
\end{equation*}
and, using \eqref{metric-variat}  one can easily verify that
\begin{equation}
\delta g^{\mu\nu}T_{\mu\nu}(F)\vol(g)
:=
\frac{\pi}{2}\bigl[\xi_gF\wedge * F-\frac12\tr(\delta g^{-1} g) F\wedge *F
\bigr]
.
\label{Tmunu}
\end{equation}
We will use this coordinate free expression as a working definition
for the stress-energy tensor.

The form
$F$ can be written as a sum $F^++F^-$ of self-dual and anti
self-dual forms: $F^{\pm}:=\frac12(F\pm *F)$. $T_{\mu\nu}(F^{\pm})$ have the
following expression
in terms of $F$ (see appendix~\ref{app:metric} for derivation):
\begin{equation}
\delta g^{\mu\nu}T_{\mu\nu}(F^{\pm})\vol(g) =\frac{\pi}{2} F^{\pm}\wedge
\xi_g F^{\pm} =  \frac{\pi}{4}\Bigl[\xi_gF\wedge(*F\pm
F)-\frac12\tr(\delta g^{-1}g)\,F\wedge *F\Bigr]. \label{stresspm}
\end{equation}
{}From here it follows that the stress-energy tensor for $2\ell+1$-form factorizes as
\cite{Alvarez-Gaume:1983ig}
\begin{equation}
T_{\mu\nu}(F^++F^-)=T_{\mu\nu}(F^+)
+T_{\mu\nu}(F^-).
\end{equation}
This is the reason for the existence of a chiral splitting of the normalization
function $\|\mathcal{N}_g\|^2$ in \eqref{N}.

\subsection{Stress-energy tensor for the anti self-dual field}
In section~\ref{sec:eom} we derived the following action for the
self-dual field on a Lorentzian manifold $(M,g)$:
\begin{equation}
S_L(R)=\pi\int_M \bigl(R_2^{\perp}\wedge *R_2^{\perp}
+ R_2\wedge R_2^{\perp}\bigr)
\label{Mact}
\end{equation}
where $R$ is a closed form belonging to the Lagrangian subspace $V_1\subset \Omega^{2\ell+1}(M)$
 and $R=R_2+R_2^{\perp}$
according to the Lagrangian decomposition $\Omega^{2\ell+1}(M)=
V_2\oplus I_L(V_2)$, $I_L:=*_g$.

In this section we prove the following theorem
\begin{thm}
The variation of the action \eqref{Mact} with respect to the metric
is
\begin{equation}
\delta_g S_L(R)=\frac{\pi}{2}\int_M \mathcal{F}^+
\wedge \xi_g(\mathcal{F}^+)
=\int_M\delta g^{\mu\nu}T_{\mu\nu}(\mathcal{F}^{+})\vol(g)
\label{dmetricS}
\end{equation}
where $\mathcal{F}^+:=R_2^{\perp}+*R_2^{\perp}$, the  operator
$\xi_g$ is defined in \eqref{xi}, and the stress-energy tensor
$T_{\mu\nu}(\mathcal{F}^+)$ is defined by \eqref{stresspm} for
$F=2R_2^{\perp}$. The derivation of \eqref{dmetricS} relies only on
the fact that $V_2$ is a Lagrangian subspace, and  that the subspace
$V_1$ does not depend on a choice of metric.
\end{thm}

\noindent\textit{Proof.}
First notice that the  coordinates $R_2$ and $R_2^{\perp}$ change when we
vary the metric (see figure
\begin{wrapfigure}{l}{170pt}
\vspace{-5mm}
\includegraphics[width=150pt]{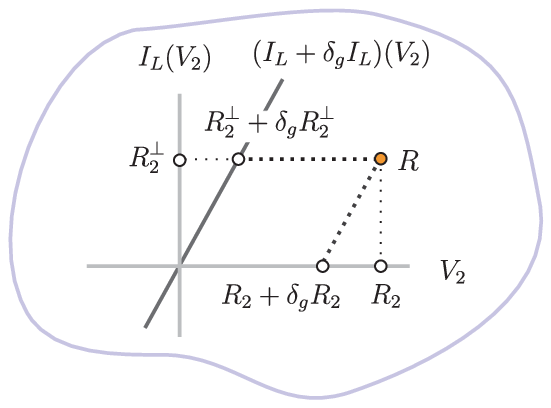}
\vspace{12mm}
\end{wrapfigure}
on the left):
\begin{equation}
0=\delta_g R=\delta_g R_2+\delta_g R_2^{\perp}.
\label{dg}
\end{equation}
Moreover from $*R_2^{\perp}\in V_2$ it follows
that $\delta_g(*R_2^{\perp})\in V_2$ and thus
\begin{equation}
\delta_g R_2^{\perp}+*\delta_g(*)R_2^{\perp}\in V_2^{\perp}.
\label{dgperp}
\end{equation}
The variation of the action is
\begin{equation*}
\delta_g S_L(R)=\pi\int_M \Bigl[
2\delta R_2^{\perp}\wedge *R_2^{\perp}
+R_2^{\perp}\wedge \delta_g(*)R_2^{\perp}
-\delta_gR_2^{\perp}\wedge R_2-R_2^{\perp}\wedge \delta_g R_2\Bigr].
\end{equation*}
The first and the third term in this expression vanish because according
to \eqref{dg} $\delta R_2^{\perp}\in V_2$
and $V_2$ is a Lagrangian subspace. Now using equations \eqref{dg}, \eqref{dgperp},
\eqref{metric-variat}
and Lagrangian condition for $V_2$ one finds:
\begin{align}
\delta_gS_L(R)&=\pi\int_M \delta_g(*)R_2^{\perp}\wedge (*R_2^{\perp}+R_2^{\perp})
\notag
\\
&\stackrel{\ref{metric-variat}}{=}\pi\int_M\Bigl[\xi_gR_2^{\perp}\wedge (*R_2^{\perp}+R_2^{\perp})
-\frac{1}{2}\tr(\delta g^{-1}g)\,R_2^{\perp}\wedge *R_2^{\perp}\Bigr].
\end{align}
After identifying  $F=2R_2^{\perp}$ this expression coincides with
the expression for the stress-energy tensor \eqref{stresspm} of the
anti-self dual field $\mathcal{F}^+$.
\begin{flushright}
\vspace{-6mm}
$\Box$
\end{flushright}

\subsection{Diffeomorphism invariance of the action}
The group of diffeomorphisms $\mathrm{Diff}^+(M)$ of an oriented
manifold $M$ has a normal subgroup $\mathrm{Diff}^+_0(M)$ consisting
of the diffeomorphisms which can be smoothly deformed to the
identity. The discrete group
$\mathrm{MCG}(M)=\mathrm{Diff}^+(M)/\mathrm{Diff}^+_0(M)$ is known
as a mapping class group of $M$. Note that any diffeomorphism $f\in
\mathrm{Diff}_0^+(M)$ from the connected  component of the identity
preserves the  Lagrangian subspace $V_1$: $f^*(V_1)=V_1$.

\begin{lemma}
The action \eqref{MAaction} is invariant under action of
$\mathrm{Diff}_0^+(M)$. The classical stress-energy tensor
$T_{\mu\nu}(\mathcal{F}^+)$ is conserved for solutions of   the
equations of motion.
\end{lemma}

\begin{proof} Let $\eta$ be a vector field on $M$ with
a compact support, then the variation of the action is
\begin{equation*}
\delta_{\eta}S(R)=\pi\int_M\bigl(
2\delta_{\eta}R_2^{\perp}\wedge *R_2^{\perp}
+R_2^{\perp}\wedge \delta_{\eta}(*)R_2^{\perp}
+\delta_{\eta}R_2\wedge R_2^{\perp}
+R_2\wedge \delta_{\eta}R_2^{\perp}
\bigr).
\end{equation*}
The variation $\delta_{\eta}R_2$ consists of two
terms: the first comes from the variation of
$R\mapsto R+L_{\eta}R$ and the second comes
from the variation of the metric $g_{\mu\nu}\mapsto
g_{\mu\nu}+(L_{\eta}g)_{\mu\nu}$.
The first variation is just the variation \eqref{variation}
with $\delta c=i_{\eta}R$. The second variation is equal to
\eqref{dmetricS} with $\delta g^{\mu\nu}=\nabla^{(\mu}\eta^{\nu)}$
where $\nabla_{\mu}$ denotes the Levi-Civita covariant derivative.
Thus we obtain
\begin{equation}
\delta_{\eta}S(R)=2\pi\int_{M}i_{\eta}R\wedge d\mathcal{F}^+(R)
+\int_M\vol(g) \nabla^{\mu}\eta^{\nu}T_{\mu\nu}(\mathcal{F}^+).
\end{equation}
{}From these equation it follows that, using the equations of motion
$\nabla^{\mu}T_{\mu\nu}(\mathcal{F}^+)=0$.
\end{proof}
Of course, quantum mechanically there is a gravitational anomaly.
Understanding this fully is part of the understanding of the factor
$\mathcal{N}_g$.

\section*{Acknowledgments}
We would like to thank E.~Diaconescu,  D.~Freed,  M.~Freedman, M. ~
Hopkins, I.~Singer, L.~Pando Zayas, and E. ~ Witten for discussions
and correspondence. D.B. thanks I.~Arefeva, X.~Bekaert, S.~Gukov,
C.~Hull, I.~Kalkkinen, D.~Panov, A.~Tseytlin, B.~Vancil, I.~Volovich
and  D.~Waldram for discussions. We would like to acknowledge the
hospitality of Kavli Institute for Theoretical Physics where a part
of this work have been done. G.M. Thanks the Institute for Advanced
Study and the Monell foundation for hospitality. D.B. thanks New
High Energy Theory Center at Rutgers University, Department fur
Physik, Ludwig-Maximilians-Universitat and Department of Theoretical
Physics at Steklov Mathematical Institute for hospitality. This work
was supported  in part by DOE grant DE-FG02-96ER40949, and by the
National Science Foundation under Grant No. PHY99-07949. D.B. was
supported by PPARC and in part by RFBR grant 05-01-00758.

\clearpage
\appendix
\renewcommand {\theequation}{\thesection.\arabic{equation}}

\section{Differential cocycles and cohomology}
\label{sec:appA}
\setcounter{equation}{0}

{}From \cite{Hopkins:2002rd} page 9.

\paragraph{Differential cocycle.}
For a manifold $X$ the category of \textit{differential
$n$-cocycles}, $\check{\Hc}^n(X)$ is the category whose
\textbf{objects} are triples
\begin{equation*}
(c,h,\omega)\subset C^n(X;\Zh)\times C^{n-1}(X;\Rh)\times
\Omega^n(X),
\end{equation*}
satisfying (the cocycle condition)
\begin{equation}
d(c,h,\omega):=(\delta c, \omega-c-\delta h,d\omega)=0.
\end{equation}
The set of these cocycles is denoted $\check{Z}^n(X)$.  Note that
$d^2=0$. Here $C^n(X;\Zh)$ and $C^{n-1}(X;\Rh)$ are $n$-cochains and
$(n-1)$-cochains with integral and real coefficients respectively;
$\Omega^n(X)$ denotes the space of $n$-forms on $X$. A
\textbf{morphism} from $(c_1,h_1,\omega_1)$ to $(c_2,h_2,\omega_2)$
is defined by an equivalence class of pairs
\begin{equation*}
(b,k,0)\in C^{n-1}(X;\Zh)\times C^{n-2}(X;\Rh)
\end{equation*}
with the action
\begin{equation}
(c_1,h_1,\omega_1)=(c_2,h_2,\omega_2)+d(b,k,0).
\end{equation}
The equivalence relation on $(b,k)$ is generated by
\begin{equation}
(b,k,0)\sim (b,k,0)-d(a,k',0).
\end{equation}

   The set of isomorphism classes of objects in the
category $\check{\Hc}^n(X)$ is the  Cheeger-Simons cohomology group
$\check{H}^n(X)$.

\paragraph{Integral Wu-structures}
Let $p:E\to S$ be a smooth map, and fix a cocycle $\nu\in
Z^{2k}(E;\Zh/2)$ representing the Wu-class $\nu_{2k}$ of the relative
normal bundle.  A {\em differential integral Wu-structure of degree
$2k$} on $E/S$ is a differential cocycle
\[
\lambda=(c,h,\omega)\in\check{Z}^{2k}(E)
\]
with the property that $c\equiv\nu\mod 2$.

\paragraph{Category of shifted differential cocycles.}
Let $M$ be a manifold,
\[
\nu\in Z^{2k}(M;\Rh/\Zh)
\]
a smooth cocycle, and $2k\ge0$ an integer.  The category of {\em
$\nu$-twisted differential $2k$-cocycles}, $\chcat^{2k}_{\nu}(M)$, is the
category whose objects are triples
\[
(c,h,\omega)\subset C^{2k}(M;\Rh)\times C^{2k-1}(M;\Rh)\times \Omega^{2k}(M),
\]
satisfying
\begin{equation}
c\equiv\nu\mod\Zh
\quad\text{and}\quad
d(c,h,\omega)=0.
\end{equation}
  A morphism from
$(c_1,h_1,\omega_1)$ to $(c_2,h_2,\omega_2)$ is an equivalence class
of pairs
\[
(b,k)\in C^{2k-1}(M;\Zh)\times C^{2k-2}(M;\Rh)
\]
satisfying $(c_2,h_2,\omega_2)=(c_1,h_1,\omega_1)+d(b,k,0)$.
The equivalence relation is generated by
\[
(b,k)\sim(b-\delta a, k+\delta k' + a).
\]

\section{Theta function}
\label{sec:appB} \setcounter{equation}{0} This appendix summarizes
several chapters from \cite{Birkenhake}.

\paragraph{Line bundles over a complex torus.} Let $V_{\Rh}$
denotes a real vector space of dimension $2g$ and $\Lambda$ a
lattice inside $V_{\Rh}$. The lattice $\Lambda$ is a discrete
subgroup of $V_{\Rh}$ of rank $2g$. It acts on $V_{\Rh}$ by
addition. Choose a complex structure $J$ and define a complex vector
space $V^+$ by $V_{\Rh}\otimes \Ch=V^+\oplus V^-$. This
decomposition defines the embedding $\Lambda^+$ of the lattice $\Lambda$ into
$V^+$. The quotient $T=V^+/\Lambda^+$ is a complex torus.

Denote the group of holomorphic line bundles on $T$ by $\mathrm{Pic}(T)$.
Each element of the Picard group defines a holomorphic line bundle.
The Picard group is described by the following exact sequence
\begin{equation}
1\to\mathrm{Pic}^0(T)\to\mathrm{Pic}(T)\to\mathrm{NS}(T)\to 0.
\end{equation}
Here $\mathrm{NS}(T)$ is called Neron-Severi group and
$\mathrm{Pic}^0(T)$ is the connected component of $0$. For
$T=V^+/\Lambda^+$ one can think of it either as $\Rh$-linear real
alternating forms on $V_{\Rh}$ satisfying
$E(\Lambda,\Lambda)\subseteq \Zh$ and $E(Jx,Jy)=E(x,y)$, or as the
set of hermitian forms on $V^+$ with $\Im H(\Lambda^+,\Lambda^+)\subseteq
\Zh$. There is a one-to-one correspondence between hermitian forms
and alternating $\Rh$-bilinear forms on $V_{\Rh}$ satisfying
$E(Jx,Jy)=E(x,y)$ for $x,y\in V_{\Rh}$. Indeed, any element $u$ of
$V^+$ can be \textit{uniquely} written as $u=x^+:=\frac12(1-iJ)x$
for some $x\in V_{\Rh}$. Define
\begin{equation*}
H(x^+,y^+):=E(Jx,y)+iE(x,y).
\end{equation*}
It is easy to verify that $H$ defines a hermitian form on the complex vector space $V^+$.
In our conventions $H$ is $\Ch$ linear in the first argument
and satisfies $H(u,v)=\overline{H(v,u)}$ for $u,v\in V^+$.

One can show that there are isomorphisms
\begin{equation}
\mathrm{Pic}^0(T)\cong \mathrm{Hom}(\Lambda,U(1))
\quad\text{and}\quad
\mathrm{Pic}(T)\cong \mathcal{P}(\Lambda)
\end{equation}
where $\mathcal{P}(\Lambda)$ is the set of pairs $(H,\chi)$ where
$H\in\mathrm{NS}(T)$ and $\chi$ is a semicharacter for $H$. A
\textit{semicharacter} for a hermitian form $H$ is a map
$\chi:\Lambda\to U(1)$ satisfying
\begin{equation}
\chi(\lambda+\mu)=\chi(\lambda)\chi(\mu)\,e^{-i\pi\Im
H(\lambda,\mu)}
\end{equation}
for all $\lambda,\,\mu\in\Lambda$.
Now using the realization of the Picard group in terms of $\mathcal{P}(\Lambda)$
we can describe holomorphic line bundles over $T$ by
the factor of automorphy.
Having a pair $(H,\chi)\in\mathcal{P}(\Lambda)$ we define a
\textit{canonical factor of automorphy}
$a_{(H,\chi)}:\Lambda\times V\to \Ch^*$ by
\begin{equation}
a_{(H,\chi)}(\lambda^+,v^+):=\chi(\lambda)\,e^{\pi H(v^+,\lambda^+)
+\frac{\pi}{2}\,H(\lambda^+,\lambda^+)}
\end{equation}
This map satisfies the cocycle relation
\begin{equation*}
a_{(H,\chi)}(\lambda^++\mu^+,v^+)=a_{(H,\chi)}(\lambda^+,v^++\mu^+)\,a_{(H,\chi)}(\mu^+,v^+).
\end{equation*}
The cocycle defines a line bundle $L(H,\chi)$ by
\begin{equation}
L(H,\chi)\cong ( V\times \Ch)/\Lambda
\end{equation}
where $\Lambda$ acts on $V\times \Ch$ by $\lambda\circ(v^+,t)=
(v^++\lambda^+,\,a_{(H,\chi)}(\lambda^+,v^+)\,t)$. This means that
sections of the line bundle $L(H,\chi)$ over $T$ are those
section of the trivial line bundle over $V$ which satisfy
equation
\begin{equation*}
\vartheta(v^++\lambda^+)=a_{(H,\chi)}(\lambda^+,v^+)\,\vartheta(v^+)
\end{equation*}
One can prove that for every line bundle $L$ over $T$ there is a unique
pair $(H,\chi)$ such that $L\simeq L(H,\chi)$.

\paragraph{Characteristics.}
Recall that $H$ is a hermitian form on $V$ whose
alternating form $E=\Im H$ is integral valued on the lattice $\Lambda$.
The alternating form $E$ defines a symplectic structure on the vector space $V$.
To construct a section of the line bundle $L(H,\chi)$ we have
to choose a Lagrangian decomposition of $V$:
\begin{equation}
V=V_1\oplus V_2.
\end{equation}
The Lagrangian decomposition of $V$ must be such that $(V_1\cap \Lambda)\oplus (V_2\cap \Lambda)$ is a Lagrangian
decomposition of $\Lambda$.

Such a decomposition leads to an explicit description of all line
bundles $L$ in $\mathrm{Pic}^H(T)$. For this we define a map
$\chi_0:V\to\Ch$ by
\begin{equation}
\chi_0(v)=e^{i\pi E(v_1,v_2)}
\end{equation}
where $v=v_1+v_2$ with $v_1\in V_1$ and $v_2\in V_2$.
The map $\chi_0$ satisfies equation
\begin{equation}
\chi_0(u+v)=\chi_0(u)\chi_0(v)\,e^{i\pi E(u,v)}\,e^{-2\pi i E(u_2,v_1)}.
\end{equation}
Thus $\chi_0|\Lambda$ is a semicharacter for $H$. Let
$L_0=L(H,\chi_0)$ denote the corresponding line bundle. Thus the
decomposition $V=V_1\oplus V_2$ distinguishes a line bundle $L_0$ in
$\mathrm{Pic}^H(T)$. For every line bundle $L=L(H,\chi)$ there is a
point $c\in V$ uniquely determined up to translation by an element
of $\Lambda(L)$, such that $L\simeq t_{\bar{c}}^*L_0$ or
equivalently $\chi=\chi_0\,e^{2\pi i E(c,\cdot)}$. Here $\Lambda(L)$
is a discrete subset of $V$ defined by
\begin{equation}
\Lambda(L)=\{ v\in V\,|\,E(v,\lambda)\subseteq\Zh, \forall
\lambda\in \Lambda  \}.
\end{equation}

The line bundle $L$ is called \textit{symmetric} if $(-1)^*_TL\simeq L$ where
$(-1)^*_T$ is the automorphism of the torus $T$ coming from the map $v\mapsto -v$ on $V_{\Rh}$.
It is easy to see that line bundle $L(H,\chi)$ is symmetric if and only if
$\chi\subseteq\{\pm 1\}$. Thus for a symmetric line bundle the characteristic
$c\in \frac12\Lambda(L)/\Lambda(L)$.

\paragraph{Theta function.} In this paragraph we assume that $H$ is positive
definite.
Since any line bundle over $V$ is trivial we can identify
$H^0(L)$ with $H^0(\Oc_V)^{\Lambda}$, the subspace
of holomorphic functions on $V$ invariant under the action of $\Lambda$.
Recall that a line bundle $L=L(H,\chi)$ is described in terms of
canonical factor of automorphy $a_{(H,\chi)}$. Thus $H^0(L)$ can be
identified with the set of holomorphic functions $\vartheta:V\to\Ch$
satisfying
\begin{equation}
\vartheta(v^++\lambda^+)=a_{(H,\chi)}(\lambda^+,v^+)\,\vartheta(v^+)
\label{B:theta}
\end{equation}
for all $v\in V$ and $\lambda\in\Lambda$.

To construct a theta function it is convenient to introduce a
classical factor of automorphy. This differs from the canonical
factor of automorphy and depends on the choice of Lagrangian
decomposition $V=V_1\oplus V_2$.

Let us recall the discussion from section~\ref{subsec:part} above. A
choice of Lagrangian decomposition defines two natural real
subspaces inside $V^+$: $V_1^+$ and $V^+_2$ respectively. Notice
that each of them generates $V^+$ as a complex vector space. The
hermitian form $H$ restricted to $V^+_2\times V_2^+$ defines a real
symmetric form. Denote by $B$ its $\Ch$-linear extension. The
following properties of $H$ and $B$ are easy to verify
\begin{equation}
B|_{V^+\times V_2^+}=H|_{V^+\times V_2^+}
\quad\text{and}\quad
B|_{V_2^+\times V^+}=H|_{V_2^+\times V^+}-2iE|_{V_2^+\times V^+}.
\label{restr}
\end{equation}
Note that the $\Ch$-bilinear form $B$ is completely determined by
the Lagrangian subspace $V_2$ and does not depend on a choice of
Lagrangian subspace $V_1$. Moreover $\Re(H-B)$ is positive definite
on $V_1^+$. This follows since $V^+=V_2^++iV_2^+$, a vector $v_1\in
V_1$ can uniquely written as $v_1=v_2^{+\,\prime}+i
v_2^{+\,\prime\prime}$:
\begin{equation*}
\Re(H-B)(v_1,v_1)=\Re(2iE(v_2^{+\,\prime},v_1^+)-2E(v_2^{+\,\prime\prime},v_1^+))
=2E(v_1^+,v_2^{+\,\prime\prime})
=2E(iv_2^{+\,\prime\prime},v_2^{+\,\prime\prime})=2H(v_2^{+\prime\prime},v_2^{+\,\prime\prime}).
\end{equation*}

The bilinear form $B$ enables us to introduce the classical
factor of automorphy for $L(H,\chi)$. Define $e_{(H,\chi)}:\Lambda\times V^+\to \Ch^*$ by
\begin{equation}
e_{(H,\chi)}(\lambda,v^+):=\chi(\lambda)\,e^{\pi(H-B)(v^+,\lambda^+)
+\frac{\pi}{2}\,(H-B)(\lambda^+,\lambda^+)}.
\end{equation}
A simple calculation shows that
\begin{equation*}
e_{(H,\chi)}(\lambda,v^+)= a_{(H,\chi)}(\lambda,v^+)\,{ e^{\frac{\pi}{2}B(v^+,v^+)}
\over e^{\frac{\pi}{2}B(v^++\lambda^+,v^++\lambda^+)}}
\end{equation*}
Therefore the classical factor of automorphy differs from the
canonical factor of automorphy by a coboundary and hence defines an
equivalent line bundle.

The reason for introducing the classical factor of automorphy is
that $e_{(H,\chi)}$ is invariant under a shift of lattice vector
from $\Lambda_2$, while $a_{(H,\chi)}$ is not. Thus, with the
classical factor of automorphy the  functional equation
$\tilde{\vartheta}(v^++\lambda^+)
=e_{(H,\chi)}(\lambda,v^+)\,\tilde{\vartheta}(v^+)$ can be solved
using Fourier transform. The solutions of this functional equation
and \eqref{B:theta} are related by the cocycle
$\vartheta(v^+)=e^{\frac{\pi}{2}B(v^+,v^+)}\tilde{\vartheta}(v^+)$,
so
\begin{subequations}
\begin{equation}
\vartheta^c(v^+) :=e^{-\pi H(v^+,c^+)-\frac{\pi}{2}\,H(c^+,c^+)
+\frac{\pi}{2}\,B(v^++c^+,v^++c^+)} \sum_{\lambda\in
\Lambda_1}e^{-\frac{\pi}{2}\,(H-B)(\lambda^+,\lambda^+)
+\pi(H-B)(v^++c^+,\lambda^+)}.
\end{equation}
solves \eqref{B:theta}. Note that this expression can also be
rewritten as
\begin{equation}
\vartheta^c(v^+)=e^{\frac{\pi}{2}\,B(v^+,v^+)
+\frac{\pi}{2}(H-B)(c_2^+,c_1^+)}
\sum_{\lambda\in \Lambda_1-c_1}e^{-\frac{\pi}{2}\,(H-B)(\lambda^+,\lambda^+)
+\pi(H-B)(v^++c_2^+,\lambda^+)}.
\end{equation}
\label{theta-cpx}
\end{subequations}

\paragraph{Classical theta function.}
Let $(X_{\mathbf{T}}=\Ch^g/\Lambda_{\mathbf{T}},H=H_{\mathbf{T}})$
denote the principally polarized abelian variety corresponding to
$\mathbf{T}$ in the Siegel upper half space. Here
$\Lambda_{\mathbf{T}}=\Lambda_1\oplus \Lambda_2$ with
$\Lambda_1=\mathbf{T}\Zh^g$ and $\Lambda_2=\Zh^g$. This
decomposition induces a decomposition $\Ch^g=V_1\oplus V_2$ where
$V_1=\mathbf{T} \Rh^g$ and $V_2=\Rh^g$ and we can write every $v\in
\Ch^g$ uniquely as $v=\mathbf{T} v^1+v^2$.

Note that for all $v,w\in \Ch^g$ we have $B(v,w)=v^t(\Im\mathbf{T})^{-1}w$
and $(H-B)(v,w)=-2iv^tw^1$. The classical \textit{Riemann
theta function} with the characteristics
$\!\left[\begin{smallmatrix}
c^1
\\
c^2
\end{smallmatrix}
\right]
$
is defined by
\begin{equation}
\vartheta\!\left[\begin{smallmatrix}
c^1
\\
c^2
\end{smallmatrix}
\right](v,\mathbf{T})
=\sum_{\ell\in \Zh^g}\exp\bigl[
i\pi(\ell+c^1)^t\mathbf{T}(\ell+c^1)+2\pi i (v+c^2)^t(\ell+c^1)
\bigr].
\label{Riemann_theta}
\end{equation}
{}From eq.~\eqref{theta-cpx} it follows that the functions
$\vartheta^c_{\mathbf{T}}$ and $\vartheta\!\left[\begin{smallmatrix}
c^1
\\
c^2
\end{smallmatrix}
\right]
$
are related by
\begin{equation}
\vartheta^{\mathbf{T}c^1+c^2}_{\mathbf{T}}(v)
=e^{\frac{\pi}{2}\,B(v,v)-i\pi c^1c^2}\,
\vartheta\!\left[\begin{smallmatrix}
c^1
\\
c^2
\end{smallmatrix}
\right](v,\mathbf{T}).
\end{equation}

The classical Riemann theta function has the following properties:
\begin{enumerate}
\item For every $c^1,\,c^2\in \Rh^g$ it satisfies the
functional equation
\begin{equation}
\vartheta\!\left[\begin{smallmatrix}
c^1
\\
c^2
\end{smallmatrix}
\right](v+\mathbf{T}\lambda^1+\lambda^2,\mathbf{T})
=
e^{2\pi i (c^1\lambda^2-c^2\lambda^1)-i\pi \lambda^1\mathbf{T}\lambda^1
-2\pi i v\lambda^1}
\vartheta\!\left[\begin{smallmatrix}
c^1
\\
c^2
\end{smallmatrix}
\right](v,\mathbf{T})
\end{equation}
for all $v\in \Ch^g$ and $\lambda^1,\lambda^2\in \Zh^g$.

\item For all $c^1,c^2\in\Rh^g$
\begin{equation}
\vartheta\!\bigl[\begin{smallmatrix}
c^1+\ell^1
\\
c^2+\ell^2
\end{smallmatrix}
\bigr]
=e^{2\pi i (\ell^2)^tc^1}\vartheta\!\left[\begin{smallmatrix}
c^1
\\
c^2
\end{smallmatrix}
\right]
\end{equation}
if and only if $\ell^1,\ell^2\in \Zh^g$.

\item Change of characteristic translates to
\begin{equation}
\vartheta\!\left[\begin{smallmatrix}
c^1
\\
c^2
\end{smallmatrix}
\right](v,\mathbf{T})
=e^{i\pi\, c^1\mathbf{T}c^1+2\pi i\, c^1(v+c^2)}\,
\vartheta\!\left[\begin{smallmatrix}
0
\\
0
\end{smallmatrix}
\right](v+\mathbf{T}c^1+c^2,\mathbf{T}).
\end{equation}

\item It satisfies the following heat equation:
\begin{equation}
\frac{\pd^2\vartheta\!\left[\begin{smallmatrix}
c^1
\\
c^2
\end{smallmatrix}
\right]}{\pd v_I\pd v_J}(v,\mathbf{T})
=2\pi i (1+\delta_{IJ})\,\frac{\pd \vartheta\!\left[\begin{smallmatrix}
c^1
\\
c^2
\end{smallmatrix}
\right]}{\pd \mathbf{T}_{IJ}}(v,\mathbf{T})
\end{equation}
\end{enumerate}

\paragraph{Dependence on a choice of Lagrangian decomposition.}
Suppose we have chosen another Lagrangian decomposition $\Lambda=\Lambda_1'\oplus
\Lambda_2$ with the same $\Lambda_2$. Since the $\Ch$-bilinear form
$B$ is completely determined by $V_2$ that theta function is
\begin{equation}
\vartheta^c_{\Lambda_1'}(v)=e^{-\pi H(v,c)-\frac{\pi}{2}\,H(c,c)
+\frac{\pi}{2}\,B(v+c,v+c)}
\sum_{\lambda'\in \Lambda_1'}e^{-\frac{\pi}{2}\,(H-B)(\lambda',\lambda')
+\pi(H-B)(v+c,\lambda')}.
\end{equation}
The Lagrangian decompositions $\Lambda_1'\oplus\Lambda_2$ and $\Lambda_1\oplus\Lambda_2$ are
related by a linear transformation $f:\Lambda_1\to \Lambda_2$ satisfying
\begin{equation}
E(f(\lambda_1),\mu_1)+E(\lambda_1,f(\mu_1))=0.
\end{equation}
Explicitly, a lattice element $\lambda$ can be written
in two ways $\lambda=\lambda_1+\lambda_2$ and $\lambda=\lambda_1'+\lambda_2'$
where
\begin{equation}
\lambda_1'=\lambda_1+f(\lambda_1)\quad\text{and}\quad
\lambda_2'=\lambda_2-f(\lambda_1).
\end{equation}
Further we will denote $\Lambda_1'$ by $\Lambda_1^f$.
Using this representation we express the sum over
$\Lambda_1^f$ in terms $\Lambda_1$
\begin{equation*}
\vartheta^c_{\Lambda_1'}(v)=e^{-\pi H(v,c)-\frac{\pi}{2}\,H(c,c)
+\frac{\pi}{2}\,B(v+c,v+c)}
\sum_{\lambda\in \Lambda_1}e^{-\frac{\pi}{2}\,(H-B)(\lambda,\lambda)
+\pi(H-B)(v+c,\lambda)}\,e^{-i\pi E(f(\lambda),\lambda)}.
\end{equation*}
Consider the function $F(\lambda)=e^{-i\pi E(f(\lambda),\lambda)}$
defined on $\Lambda_1$. It is easy to see that this function
satisfies the equation
\begin{equation*}
F(\lambda_1+\mu_1)=F(\lambda_1)F(\mu_1)
\end{equation*}
and thus there exist an element $w_f\in\frac12 \Lambda_2$ such that
\begin{equation}
F(\lambda_1)=e^{2\pi i E(w_f,\lambda_1)}.
\end{equation}
In fact,  if $\{E^j\}$ is a basis  for the  Lagrangian subspace
$V_1$ and $\{F_j\}$ is a dual basis for $V_2$, then the map $f$ is
represented by an integral symmetric matrix $f^{jk}$ and
$w_f=\frac12 f^{jj}F_j$. Thus we obtain that
\begin{equation}
\theta^c_{\Lambda_1^f}(v)=e^{i\pi E(w_f,c)}\,\theta^{c+w_f}_{\Lambda_1}(v).
\label{L1dept}
\end{equation}
One sees that the change $\Lambda_1\mapsto\Lambda_1^f$ corresponds
to a change
of the characteristic $c_2\mapsto c_2+w_f$\footnote{Sometimes
it is incorrectly said that it corresponds to a change
of spin structure. }.

\section{Path integral measure}
\label{sec:appC}
\setcounter{equation}{0}

Let $X$ be an $n$-manifold. A metric $g_X$ on $X$ defines a Hodge
metric on the vector space of differential forms. In our conventions
all forms are \textit{dimensionless}. However the Hodge $*$
operation is dimensionfull: dimension of $*\omega_p$ is
$[L]^{n-2p}$. It is convenient to introduce a dimensionless norm
\begin{equation}
\|\omega_p\|^2_{L}:=L^{2p-n}\int_X \omega_p\wedge *\omega_p
\end{equation}
where $L$ is some parameter of dimension of length, $[L]$. The
operator $d^{\dag}|_{\Omega^p}=(-1)^{np+n+1}*d*$ has dimension
$[L]^{-2}$. We also introduce the dimensionless Laplace operator
\begin{equation}
\Delta_p^{L}:=L^{2}(dd^{\dag}+d^{\dag}d)|_{\Omega^p}.
\end{equation}

\paragraph{Gauge fields.} Consider a gauge potential $a\in\Omega^{g+1}(X)$.
Denote by $\mathcal{G}_{g+1}$ the group of gauge transformations
$a\mapsto a+\omega_{g+1}$ where $\omega_{g+1}\in\Omega^{g+1}_{\Zh}(X)$.
In this paragraph we want to obtain a formula for
\begin{equation*}
\int_{\Omega^{g+1}(X)}\frac{\Ds a}{\mathrm{Vol}(\mathcal{G}_{g+1})}
\end{equation*}
Using the Hodge decomposition we can write $a$ uniquely as
\begin{equation*}
a=a^h+a^T+d\alpha_{g}^T
\end{equation*}
Here $a^h\in\mathscr{H}^{g+1}(X)$ is a harmonic form, $a^T\in
\mathrm{im}\, d^{\dag}\cap \Omega^{g+1}(X)$ and $\alpha^T_g\in
\mathrm{im}\,d^{\dag}\cap \Omega^g(X)$. This implies
\begin{equation}
  \|\delta a\|^2_{L}=\|\delta a^h\|^2_{L}
+\|\delta a^T\|^2_{L} +\langle\delta\alpha^T,
(L^2d^{\dag}d)\delta\alpha^T\rangle_{L} \end{equation}
 Thus
\begin{equation}
\int_{\Omega^{g+1}(X)}\Ds a=\int_{\mathscr{H}^{g+1}(X)}\Ds a^h\int\Ds a^T \Ds\alpha_g^{T}\,
\bigl[\det{}'(L^2\,d^{\dag}d\bigr|_{\Omega^g(X)\cap \mathrm{im}\, d^{\dag}})\bigr]^{1/2}.
\label{Da}
\end{equation}
It is convenient to introduce the notation
\begin{equation}
L_p:=\det{}'(L^2 d^{\dag}d\bigr|_{\Omega^p(X)\cap \mathrm{im}\, d^{\dag}}).
\end{equation}
The gauge group $\mathcal{G}_{g+1}$ has several connected
components labelled by the harmonic forms
with integral periods $\mathscr{H}^{g+1}_{\Zh}(X)$. Using the Hodge
decomposition we can write
\begin{equation}
\int_{\Omega^{g+1}(X)}\frac{\Ds a}{\mathrm{Vol}(\mathcal{G}_{g+1})}
=\int_{\mathscr{H}^{g+1}(X)/\mathscr{H}^{g+1}_{\Zh}(X)}\Ds a^h\int
\frac{\Ds a^T \Ds\alpha_g^{T}}{\mathrm{Vol}(\mathcal{G}^0_{g+1})}\,
L_g^{1/2}
\label{Da2}
\end{equation}
where $\mathcal{G}^0_{g+1}
\cong\Omega^{g+1}_{\text{exact}}(X)/\mathcal{G}_g$ is the connected
component of the identity of the gauge group $\mathcal{G}_{g+1}$.

\paragraph{Volume of the gauge algebra.}
To calculate the volume $\mathrm{Vol}(\mathcal{G}_{g+1}^0)$ we
notice that
\begin{equation}
\mathrm{Vol}(\mathcal{G}_{g+1}^0)=
\int_{\Omega^{g+1}_{\text{exact}}/\mathcal{G}_g}\Ds \alpha_{g+1}=
\int_{\Omega^g} \frac{\Ds \alpha_{g}}{\mathrm{Vol}(\mathcal{G}_g)}
\label{int1}
\end{equation}
Using the Hodge decomposition we can write $\alpha_g=\alpha_g^{h}+
\alpha_g^T+d\alpha_{g-1}^T$,
and thus
\begin{equation}
\mathrm{Vol}(\mathcal{G}_{g+1}^0)=
\int_{\mathscr{H}^{g}(X)/\mathscr{H}^g_{\Zh}(X)}\Ds \alpha_{g}^h\,
\int\frac{\Ds \alpha^T_{g}\,\Ds\alpha_{g-1}^T}{\mathrm{Vol}(\mathcal{G}_{g}^0)}\,
L_{g-1}^{1/2}.
\label{int2}
\end{equation}
The integral over the harmonic forms is a finite dimensional integral which
yields the volume $V_p:=\mathrm{Vol}_L(\mathscr{H}^p/\mathscr{H}^p_{\Zh})$ of
the harmonic torus.
The volume of the harmonic torus is
\begin{equation}
V_p=L^{b_p(2p-n)/2}\left[\det \int_X \omega_{\alpha}\wedge *\omega_{\beta}\right]^{1/2}
\label{C:Vp}
\end{equation}
where $b_p=\dim \mathscr{H}^p$, and  $\{\omega_{\alpha}\}$ is an
\textit{integral} basis of harmonic $p$-forms
($\alpha=1,\dots,b_p$). Notice that $V_p$ does not depend on a
choice of the integral basis $\{\omega_{\alpha}\}$.

Successively applying the formula \eqref{int2} one finds
that all terms $\int\Ds \alpha_p^T$ but one cancel, and
\begin{equation}
\mathrm{Vol}(\mathcal{G}_{g+1}^0)=
\int\Ds\alpha_g^T\,\prod_{p=0}^{g}\left[
V_p\,L_{p-1}^{1/2}
\right]^{(-1)^{g-p}}.
\end{equation}
Combining this result with \eqref{Da2} one finds
\begin{equation}
\int_{\Omega^{g+1}}\frac{\Ds a}{\mathrm{Vol}(\mathcal{G}_{g+1})}
=\int_{\mathscr{H}^{g+1}/\mathscr{H}^{g+1}_{\Zh}}\Ds a^h\int_{\Omega^{g+1}\cap\mathrm{im}\, d^{\dag}}\Ds a^T\,
\left\{\prod_{p=0}^{g}
\left[\frac{\det{}'(L^2d^{\dag}d)_p}{V_p^2}\right]^{(-1)^{g-p}}\right\}^{1/2}
\label{integral}
\end{equation}


\section{Metric variation}
\label{app:metric} \setcounter{equation}{0} In this appendix we
present the  derivation of some well known facts about the
stress-energy tensor of the self-dual field, using the trick
\eqref{metric-variat}.

\paragraph{Formula for $T^{\pm}_{\mu\nu}$.}
{}From equation \eqref{Tmunu} one finds
\begin{equation}
\delta g^{\mu\nu}T_{\mu\nu}(F^{\pm})\vol(g)
=\frac{\pi}{8}\bigl[
F\wedge *\xi_g F-F\wedge \xi_g*F
\pm F\wedge *\xi_g*F\mp F\wedge\xi_g F
\bigr].
\label{Tpm}
\end{equation}
Now from the identity $0=\delta_g(*(*F))=0$ we obtain the following
equation
\begin{equation}
0=-\tr(\delta g^{-1}g)\, F+*(\xi_g*F)+\xi_gF
\end{equation}
expressing $\xi_g*F$ through $\xi_g F$. Thus
\begin{equation}
F\wedge *(\xi_g*F)=-F\wedge \xi_g F\quad
\text{and}\quad
F\wedge (\xi_g*F)=\tr(\delta g^{-1} g)F\wedge *F-F\wedge *\xi_g F.
\end{equation}
Substituting these equations into \eqref{Tpm} one finds
\begin{equation}
\delta g^{\mu\nu}T_{\mu\nu}(F^{\pm})\vol(g)
=\frac{\pi}{4}\Bigl[\xi_gF\wedge(*F\pm F)-\frac12\tr(\delta g^{-1}g)F\wedge *F\Bigr]
\end{equation}
{}From this equation it follows that the stress-energy tensor for a
$2\ell+1$-form $F$ is the sum of the stress-energy tensor for the
self-dual form $F^+$ and the anti-self-dual form $F^-$:
\begin{equation}
T_{\mu\nu}(F^++F^-)=T_{\mu\nu}(F^+)+T_{\mu\nu}(F^-).
\end{equation}

\section{Splitting of the sum over instantons}
\label{app:sum}
\setcounter{equation}{0}

In this appendix we spell out the splitting theorems relevant to
splitting the sum over instantons in a theory such as that described
in section 2. The splitting is in terms of a sum over ``conformal
blocks.'' In the present case the ``conformal blocks'' are theta
functions of level $k$.  The main splitting theorem is Theorem E.1
below. We then show how the failure to include subtle phases such as
the quadratic refinement $\Omega$ can change the set of conformal
blocks deduced from this splitting technique. This is exhibited
explicitly in Theorem E.2. In several papers in the literature on
M5-branes the sum over instantons is incorrectly written as the
untwisted sum, rather than as the twisted sum.

\subsection{Symplectic structure, complex structure, and metric}

Let $V_{\Rh}$ be a real vector space with symplectic form $\omega$
which is $\Zh$-valued on a lattice $\Gamma \subset V_{\Rh}$ of full
rank $2g$.  Now we choose a Lagrangian decomposition $V_1\oplus V_2$
of the vector space $V_{\Rh}$ such that
$\Gamma_1\oplus\Gamma_2=(V_1\cap \Gamma)\oplus (V_2\cap \Gamma)$ is
the decomposition of the lattice $\Gamma$. Choose an integral basis
$\{\alpha_I\}$  ($I=1,\dots,g$) for $\Gamma_1$ and a complementary
basis $\{\beta^I\}$ for $\Gamma_2$:
\begin{equation}
\omega(\alpha_I,\alpha_J)=0=\omega(\beta^I,\beta^J)
\quad\text{and}\quad
\omega(\alpha_I,\beta^J)=\delta_I{}^J.
\end{equation}

We now assume there is also  a complex structure $J$ (e.g. in the
geometrical setting $J=-*$ is defined by the Hodge star) which is
  compatible with the symplectic form $\omega$, so we have a metric
\begin{equation}
g(u,v):=\omega(J\cdot u, v).
\label{metric}
\end{equation}

Now choose the basis $\zeta_I$ of type $(1,0)$. By definition
$\zeta^I$ is a basis of solutions of the equation
$J\cdot\zeta_I=i\zeta_I$. One can express the complex
structure $J$ in terms of the components of the
complex period matrix $\mathbf{T}_{IJ}$.
To this end we choose a basis $\zeta_I$ of the form
\begin{equation}
\zeta_I=\alpha_I+\mathbf{T}_{IJ}\beta^J.
\end{equation}
{}From $g(\zeta_I,\zeta_J)=g(\zeta_J,\zeta_I)$ we learn that
$\mathbf{T}_{IJ}$ is symmetric, and $g$ is of type $(1,1)$.
Note that
\begin{equation}
g(\zeta_I,\bar{\zeta}_J)=2\Im\mathbf{T}_{IJ}
\end{equation}
which implies that $\Im\mathbf{T}_{IJ}$ is a positive definite
matrix. If we write $\mathbf{T}=X+iY$ then the complex structure can
be written as
\begin{equation}
J\cdot\begin{pmatrix}
\alpha_I
\\
\beta^I
\end{pmatrix}
=
\begin{pmatrix}
-(XY^{-1})_I{}^J & -(Y+X
Y^{-1}X)_{IJ}
\\
(Y^{-1})^{IJ} &
(Y^{-1}X)^I{}_J
\end{pmatrix}
\begin{pmatrix}
\alpha_J
\\
\beta^J
\end{pmatrix}
.\label{mTJ}
\end{equation}

Any element $v\in V_{\Rh}$ can be written as $v=v^++v^-$ where
$J\cdot v^{\pm}=\pm i v^{\pm}$
and $(v^+)^*=v^-$. Suppose we are given $v=v^I_1\alpha_I+v^2_I\beta^I$
then
\begin{subequations}
\begin{align}
v^+&=\frac{1}{2}(1-iJ)v=+\frac{1}{2i}(v^2-v_1\cdot\bar{\mathbf{T}})\cdot Y^{-1}\cdot\zeta;
\\
v^-&=\frac{1}{2}(1+iJ)v=-\frac{1}{2i}(v^2-v_1\cdot\mathbf{T})\cdot Y^{-1}\cdot\bar{\zeta}.
\end{align}
\end{subequations}
This implies
\begin{equation}
g(v,v)=2g(v^+,v^-)=(v^2-v_1\cdot\bar{\mathbf{T}})\cdot Y^{-1}\cdot(v^2-\mathbf{T}\cdot v_1).
\end{equation}

Let $\nu=n^I\alpha_I+m^I\beta_I$
then the metric \eqref{metric} takes a form:
\begin{equation}
g(\nu,\nu)=
\begin{pmatrix}
n^I
&
m_I
\end{pmatrix}
\begin{pmatrix}
(Y+X
Y^{-1}X)_{IJ} &
-(XY^{-1})_I{}^J
\\
-(Y^{-1}X)^I{}_J &
(Y^{-1})^{IJ}
\end{pmatrix}
\begin{pmatrix}
n^J
\\
m_J
\end{pmatrix}
\end{equation}

\subsection{Theta function}

Let us define the level $k/2$  theta function (by convention the
level can be half integral) with   characteristics
$\theta_I,\,\phi^I$ by the series
\begin{equation}
\Theta_{k/2,\gamma}
\!\left[\begin{smallmatrix}
\theta
\\
\phi
\end{smallmatrix}
\right]
\!(\mathbf{T};a_1,a^2)
:=e^{i\pi k a_1\cdot\mathbf{T}\cdot a_1 -i\pi k a_1\cdot a^2+i\pi k \theta\cdot \phi}
\sum_{\{p_L^I\}\in\Zh^g+\gamma+\theta}
e^{i\pi k\, p_L\cdot \mathbf{T}
\cdot p_L+2\pi i k\, p_L\cdot(a^2-\mathbf{T}\cdot a_1-\phi)}
\label{thetaell}
\end{equation}
where $\gamma^I\in\{0,\frac{1}{k},\dots,\frac{k-1}{k}\}$.
Here we assume that $\Im\mathbf{T}_{IJ}$ is a positive definite matrix,
and thus the series \eqref{thetaell} converges absolutely.
The characteristics $\phi_I$ and $\theta^I$ take values in $\Rh$.
The  theta function with \textit{zero} characteristics is denoted by
$\Theta_{k/2,\gamma} (\mathbf{T};a_1,a^2) $. Notice that the series
defining the theta function depends on a choice of Lagrangian
decomposition $\Gamma_1\oplus \Gamma_2$. More precisely, the sum
goes over $\Gamma_1+\gamma+\theta$ where $\gamma\in
\frac{1}{k}\Gamma_1/\Gamma_1$, $\theta\in V_1$; the second
characteristic $\phi$ take values in $V_2$ and the complex period
matrix depends on a choice of both $\Gamma_1$ and $\Gamma_2$.

The theta function \eqref{thetaell} satisfies the following
functional equation
\begin{equation}
\Theta_{k/2,\gamma}
\!\left[\begin{smallmatrix}
\theta
\\
\phi
\end{smallmatrix}
\right]
\!(\mathbf{T};a_1+\lambda_1,a^2+\lambda^2)
=\Omega_0(\lambda)\,e^{i\pi k\omega(\lambda,a) }\,
\Theta_{k/2,\gamma}
\!\left[\begin{smallmatrix}
\theta
\\
\phi
\end{smallmatrix}
\right]
\!(\mathbf{T};a_1,a^2)
\end{equation}
where
\begin{equation}
\Omega_0(\lambda):=e^{-i\pi k \lambda_1\cdot\lambda^2 +
2\pi i k(\theta\cdot\lambda^2-\phi\cdot\lambda_1)}.
\end{equation}
This equation means that the theta function is a section of the line
bundle $\boldsymbol{\mathcal{L}}^{\otimes k}$ over the torus
$V_{\Rh}/\Gamma$. If $\theta\in \frac12\Gamma_1$ and $\phi \in
\frac12\Gamma_2$ then the line bundle $\Lb^{\otimes k}$ is a
symmetric line bundle. Note that different values of $\gamma$ lead
to different sections of the same line bundle, but different values
of $\theta, \phi$ lead to different line bundles.

\paragraph{Properties.}
The complex conjugate is
\begin{equation}
\overline{\Theta_{k/2,\gamma}
\!\left[\begin{smallmatrix}
\theta
\\
\phi
\end{smallmatrix}
\right]
\!(\mathbf{T};a_1,a^2)
}
=\Theta_{k/2,\gamma}
\!\left[\begin{smallmatrix}
\theta
\\
-\phi
\end{smallmatrix}
\right]
\!(-\bar{\mathbf{T}};a_1,-a^2)
.
\label{theta-cc}
\end{equation}
The theta function with the shifted characteristics is related
to the original theta function by
\begin{equation}
\Theta_{k/2,\gamma}
\!\left[\begin{smallmatrix}
\theta+m
\\
\phi+n
\end{smallmatrix}
\right]
\!(\mathbf{T};a_1,a^2)
=
e^{i\pi k(m\cdot n+m\cdot\theta-n\cdot\phi)}
\Theta_{k/2,\gamma}
\!\left[\begin{smallmatrix}
\theta
\\
\phi
\end{smallmatrix}
\right]
\!(\mathbf{T};a_1,a^2)
.
\end{equation}

\paragraph{Modular transformations.}
The change of symplectic basis $\{\alpha_I,\beta^I\}$ and Lagrangian decomposition
is described by the group $Sp(2g,\Zh)$. It consists of the matrices of the form
\begin{equation}
g=\begin{pmatrix}
A & B
\\
C & D
\end{pmatrix}
\quad\text{and}\quad
D^tA-B^tC=\mathbbmss{1}_g,\quad
D^tB=B^tD,\quad C^tA=A^tC.
\end{equation}
The generators can be chosen to be
\begin{enumerate}
\item $\begin{pmatrix} A & 0
\\
0 & A^{-1,t}
\end{pmatrix}$, $A\in GL(g,\Zh)$ i.e. $\det A=\pm 1$.
This transformation describes change of integral basis in Lagrangian subspaces
$V_1$ and $V_2$.

\item $\begin{pmatrix} \mathbbmss{1}_g & B \\ 0 & \mathbbmss{1}_g\end{pmatrix}$ where
$B$ is symmetric $g\times g$ matrix.
This transformation describes a change of Lagrangian subspace $V_1$:
$V_1\oplus V_2\to V_1'\oplus V_2$.

\item $S=\begin{pmatrix} 0 & -\mathbbmss{1}_g\\ \mathbbmss{1}_g & 0\end{pmatrix}$.
This transformation exchanges the Lagrangian subspaces:
$V_1\oplus V_2\to V_2\oplus V_1$.

\end{enumerate}

These generators act as follows on the theta function \eqref{thetaell}:

\noindent1. A-transform:
\begin{subequations}
\begin{equation}
\Theta_{k/2,\gamma}
\!\left[\begin{smallmatrix}
\theta
\\
\phi
\end{smallmatrix}
\right]
\!(A\mathbf{T}A^t;a_1,a^2)
=
\Theta_{k/2,\gamma}
\!\left[\begin{smallmatrix}
A^t\theta
\\
A^{-1}\phi
\end{smallmatrix}
\right]
\!(\mathbf{T};A^t a_1,A^{-1}a^2)
\end{equation}
2. Generalization of the T-transform:
\begin{multline}
\Theta_{k/2,\gamma}
\!\left[\begin{smallmatrix}
\theta
\\
\phi
\end{smallmatrix}
\right]
\!(\mathbf{T}+B;a_1,a^2)
=e^{-\frac{i\pi}{2} k\,B_{II}\theta^I}\,
e^{i\pi k B_{II}\gamma_I(\gamma_I-1)
+2\pi i k\sum_{I<J}B_{IJ}\gamma_I\gamma_J}\,
\\
\times
\Theta_{k/2,\gamma}
\!\left[\begin{smallmatrix}
\theta
\\
\phi-B\theta-\frac12 B_{II}
\end{smallmatrix}
\right]
\!(\mathbf{T};a_1,a^2-Ba_1).
\end{multline}

\noindent 3. $S$-transform:
\begin{equation}
\Theta_{k/2,\gamma}
\!\left[\begin{smallmatrix}
\theta
\\
\phi
\end{smallmatrix}
\right]
\!(-\mathbf{T}^{-1};a_1,a^2)
\\
=\det(-i\mathbf{T})^{1/2}\,k^{-g/2}
\sum_{\gamma'\in(\frac{1}{k}\Zh/\Zh)^{\otimes g}}e^{-2\pi i k\, \gamma_I'\gamma_I}
\Theta_{k/2,\gamma'}
\!\left[\begin{smallmatrix}
-\phi
\\
\theta
\end{smallmatrix}
\right]
\!(\mathbf{T};-a^2,a_1)
.
\end{equation}
\label{Theta-mod}
\end{subequations}

\subsection{Splitting the twisted sum}
We want to express
\begin{equation}
\mathcal{Z}_{p,q}(a):=\sum_{R\in \Gamma}\Omega_{pq}(R)\,
e^{-\frac{\pi p}{2q}
g(R-qa,R-qa)+i\pi p\, \omega(a,R)}
\label{sum1}
\end{equation}
in terms of theta functions for the complex tori $V_{\Rh}/\Gamma$.
Here $a=a^I_1\alpha_I+a^2_I\beta^I$. $\Omega_{pq}(R)$ take values
in $\{\pm 1\}$ and is a quadratic
refinement of $(-1)^{pq\omega}$:
\begin{equation}
\Omega_{pq}(R+R')=\Omega_{pq}(R)\Omega_{pq}(R') \,e^{i\pi
qp\,\omega(R,R')}.
\end{equation}

The results of this subsection can be summarized follows:
\begin{lemma}
If $pq=p\mod 2$ then the twisted sum $\mathcal{Z}_{p,q}(a)$ defines
a section of a line bundle
$\boldsymbol{\mathcal{L}}^{\otimes pq}$ over the finite dimensional tori $V_{\Rh}/\Gamma$.
\end{lemma}

\begin{thm}
If $pq=p\mod 2$ and $\mathrm{gcd}(p,q)=1$ then the twisted sum
$\mathcal{Z}_{p,q}(a)$ splits
\begin{equation}
\mathcal{Z}_{p,q}(a)=(\det\tfrac{2q}{p}\Im \mathbf{T})^{1/2}\,
\mathop{\sum_{
\gamma_p\in (\frac{1}{p}\Zh/\Zh)^g}}_{
\gamma_q\in (\frac{1}{q}\Zh/\Zh)^g}
\Theta_{pq/2,\gamma_p+\gamma_q}
\!\left[\begin{smallmatrix}
\theta
\\
\phi
\end{smallmatrix}
\right]
\!(\mathbf{T};0,0)
\overline{\Theta_{pq/2,\gamma_p-\gamma_q}
\!\left[\begin{smallmatrix}
\theta
\\
\phi
\end{smallmatrix}
\right]
\!(\mathbf{T};a_1,a^2)
}
\label{splitting}
\end{equation}
where $\Theta_{pq/2,\gamma}
\!\left[\begin{smallmatrix}
\theta
\\
\phi
\end{smallmatrix}
\right]
\!(\mathbf{T};a_1,a^2)$
is a theta function with characteristics defined by series \eqref{thetaell},
$a=a_1^I\alpha_I+a^2_I\beta^I$.
\end{thm}

In particular, if $p=q=1$ then the twisted sum \eqref{sum1} factorizes
as the square of a \textit{single} theta function.

\begin{rem}
If $\gcd(p,q)=m\ne 1$ then the twisted sum also splits, however
it splits in terms of level $pq/m^2$ theta functions on the torus
$V_{\Rh}/mV_{\Zh}$.
\end{rem}

\begin{proof}[Proof of the lemma.]
The function $\psi(a)$ descends to a  section of a line bundle $\boldsymbol{\mathcal{L}}^{pq}$
if satisfies the equation
\begin{equation}
\psi(a+\lambda)=\Omega_{pq}(\lambda)\,e^{i\pi pq
\omega(a,\lambda)}\psi(a). \label{secdsd}
\end{equation}
It is matter of simple algebra to show that the sum \eqref{sum1}
satisfies equation \eqref{secdsd} iff $pq=p\mod 2$, where we use the
fact that $\Omega_{pq}(q\lambda) = \Omega_{pq}(\lambda)$.
\end{proof}

\begin{proof}[Proof of the theorem.]
We write $R=n^I\alpha_I+m_I \beta^I$. Then we can write
  the quadratic
refinement $\Omega_{pq}$ in the form
\begin{equation}
\Omega_{k}(R)=e^{-i\pi k n\cdot m+2\pi i k(m\cdot
\theta-n\cdot\phi)}.
\end{equation}
Here $\theta$ and $\phi$ are half integrally quantized, so that
$\Omega_k(R)\in \{\pm 1\}$.

To split the sum we need to do Poisson resummation over $m_I$. The
result is:
\begin{multline}
\mathcal{Z}_{p,q}(a)=
(\det\tfrac{2q}{p}\Im \mathbf{T})^{1/2}
e^{-i\pi k a_1\cdot\bar{\mathbf{T}}\cdot a_1+i\pi k a_1\cdot a^2}
\sum_{n^I,w^I}
\exp\Bigl[
i\pi k\,p_L\cdot\mathbf{T}\cdot p_L
-i\pi k\,p_R\cdot\mathbf{\bar{T}}\cdot p_R
\Bigr]
\\
\times
\exp\Bigl[
-2\pi i k\, p_R\cdot(a^2-\bar{T}\cdot a_1)
-2\pi ik\,q\phi\cdot(p_L+p_R)
\Bigr]
\label{S1'}
\end{multline}
where
\begin{equation}
p_L^I=\frac{1}{2q}\, n^I+\frac{1}{2}\,\Bigl[qn^I+2q\theta^I+\frac{2}{p} w^I\Bigr]
\quad\text{and}\quad
p_R^I=\frac{1}{2q}\, n^I-\frac{1}{2}\,\Bigl[qn^I+2q\theta^I+\frac{2}{p} w^I\Bigr].
\label{psiell}
\end{equation}

{}From the conditions $pq=p\mod 2$ and $\mathrm{gcd}(p,q)=1$ it
follows that for $q=\mathrm{odd}$, $p$ an  arbitrary integer s.t.
$\mathrm{gcd}(p,q)=1$, we can  write $q=2r+1$ and shift $w^I\mapsto
w^I-prn^I-pr2\theta^I$:
\begin{equation}
p_L^I=\frac{1}{2q}\, n^I+\frac{1}{2}\,\Bigl[n^I+2\theta^I+\frac{2}{p} w^I\Bigr]
\quad\text{and}\quad
p_R^I=\frac{1}{2q}\, n^I-\frac{1}{q}\,\Bigl[qn^I+2q\theta^I+\frac{2}{p} w^I\Bigr].
\end{equation}
Now introduce
\begin{equation*}
n^I=qt^I+2q\gamma_q^I\quad\text{and}\quad
w^I=\frac{1-q}{2}\,pt^I-pq\gamma_q^I+ps^I+p\gamma_p^I.
\end{equation*}
where $s^I,t^I\in\Zh$, $\gamma_q^I\in\{0,\frac{1}{q},\dots,\frac{q-1}{q}\}$
and $\gamma_p^I\in\{0,\frac{1}{p},\dots,\frac{p-1}{p}\}$. Thus
\begin{equation}
p_L^I=\underbrace{t^I+s^I}_{n_L}+\gamma_q^I+\gamma^I_p+\theta^I
\quad\text{and}\quad
p_R^I=\underbrace{-s^I}_{n_R}+\gamma_q^I-\gamma^I_p+\theta^I.
\end{equation}
One sees that $n_L$ and $n_R$ are independent
summation variables.
Thus the sum \eqref{S1'} splits as in \eqref{splitting}.
\end{proof}

\subsection{Splitting the untwisted sum}
We want to express
\begin{equation}
\mathcal{Z}_{p,q}(a):=\sum_{R\in V_{\Zh}}e^{-\frac{\pi p}{2q}
g(R-qa,R-qa)+i\pi p\, \omega(a,R)}
\label{sumutw}
\end{equation}
in terms of theta functions for the complex tori $V_{\Rh}/V_{\Zh}$.

\begin{thm}
If $pq=p\mod 2$ then the untwisted sum \eqref{sumutw} defines a
section of a vector bundle (not a line bundle) over
$V_{\Rh}/V_{\Zh}$. If in addition $\mathrm{gcd}(p,q)=1$ then the
untwisted sum $\mathcal{Z}_{p,q}(a)$ splits:
\\
\begin{subequations}
(a) if $p=\mathrm{odd}$ and $q=\mathrm{odd}$
\begin{equation}
\mathcal{Z}_{p,q}(a)=(\det\tfrac{2q}{p}\Im\mathbf{T})^{1/2}\,
\frac{1}{2^g}\sum_{\theta,\phi\in\{0,\frac12\}^g}
\mathop{\sum_{
\gamma_p\in (\frac{1}{p}\Zh/\Zh)^g}}_{
\gamma_q\in (\frac{1}{q}\Zh/\Zh)^g}
\Theta_{pq/2,\gamma_p+\gamma_q}
\!\left[\begin{smallmatrix}
\theta
\\
\phi
\end{smallmatrix}
\right]
\!(\mathbf{T};0,0)
\overline{\Theta_{pq/2,\gamma_p-\gamma_q}
\!\left[\begin{smallmatrix}
\theta
\\
\phi
\end{smallmatrix}
\right]
\!(\mathbf{T};a_1,a^2)
}
\label{splitting:u}
\end{equation}
\\
(b) if $p=\mathrm{even}$ and $q=\mathrm{odd}$
\begin{equation}
\mathcal{Z}_{p,q}(a)=(\det\tfrac{2q}{p}\Im\mathbf{T})^{1/2}\,
\mathop{\sum_{
\gamma_p\in (\frac{1}{p}\Zh/\Zh)^g}}_{
\gamma_q\in (\frac{1}{q}\Zh/\Zh)^g}
\Theta_{pq/2,\gamma_p+\gamma_q}
(\mathbf{T};0,0)
\overline{\Theta_{pq/2,\gamma_p-\gamma_q}
(\mathbf{T};a_1,a^2)
}
\label{splitting:uev}
\end{equation}
\end{subequations}
Here $\Theta_{pq/2,\gamma}
\!\left[\begin{smallmatrix}
\theta
\\
\phi
\end{smallmatrix}
\right] \!(\mathbf{T};a_1,a^2)$ is the theta function with
characteristics defined by the series \eqref{thetaell}, and
$a=a_1^I\alpha_I+a^2_I\beta^I$.
\end{thm}

\begin{rem}
If $\gcd(p,q)=m\ne 1$ then the untwisted sum also splits, however
it splits in terms of the level $pq/m^2$ theta functions on the torus
$V_{\Rh}/mV_{\Zh}$.
\end{rem}

\begin{proof}[Proof of the theorem.]
Decompose $a=a_1^I\alpha_I+a^2_I\beta^I$ and $R=n^I\alpha_I+m_I\beta^I$.
To split the sum \eqref{sumutw} we need to do Poisson resummation over $m_I$:
\begin{equation}
\mathcal{Z}_{p,q}(a)=(\det\tfrac{2q}{p}Y)^{1/2}
e^{-i\pi k a_1\cdot\bar{\mathbf{T}}\cdot a_1+i\pi k a_1\cdot a^2}
\sum_{n^I,w^I}
\exp\Bigl[
i\pi k\,p_L\cdot\mathbf{T}\cdot p_L
-i\pi k\,p_R\cdot\mathbf{\bar{T}}\cdot p_R
+2\pi i k p_R\cdot(\bar{T}\cdot a_1-a^2)
\Bigr]
\label{S0'}
\end{equation}
where $Y=\Im\mathbf{T}$, $k=qp$ and
\begin{equation}
p_L^I=\frac{1}{2q}\, n^I+\frac{1}{p}\,w^I
\quad\text{and}\quad
p_R^I=\frac{1}{2q}\, n^I-\frac{1}{p}\,w^I.
\label{pLpR0}
\end{equation}
Now we have to consider two cases: $(p=\text{odd},q=\text{odd})$ and
$(p=\text{even},q=\text{odd})$
 separately.

\paragraph{$\mathbf{q=odd,\,p=odd}$:}
We split $n^I$ and $w^I$ in \eqref{pLpR0} as follows
\begin{equation*}
w^I=p\,t^I+p\gamma^I_p\quad\text{and}\quad n^I=2qs^I+2q\gamma_q^I+2q\theta^I
\end{equation*}
where $s^I,t^I\in\Zh$, $\theta^I\in\{0,\frac12\}$,
$\gamma^I_p\in\{0,\frac{1}{p},\dots,\frac{p-1}{p}\}$
and $\gamma^I_q\in\{0,\frac{1}{q},\dots,\frac{q-1}{q}\}$.
Thus we have
\begin{equation*}
p_L^I=\underbrace{s^I+t^I}_{n_L}+\theta^I+\gamma^I_q+\gamma_p^I
\quad\text{and}\quad
p_R^I=\underbrace{s^I-t^I}_{n_R}+\theta^I+\gamma_q^I-\gamma^I_p.
\end{equation*}
We certainly want to consider $n_L=s+t$ and $n_R=s-t$ as independent summation
variables however they are not independent: $n_L^I-n_R^I$ are even integers
for all $I$.
This difficulty can be overcome by inserting
the following function  into the sum \eqref{S0'}:
\begin{equation*}
\frac{1}{2^g}\sum_{\phi_I\in\{0,\frac12\}}e^{2\pi i k \phi_I(n_R-n_L)^I}
=\left\{%
\begin{array}{ll}
    1, & \hbox{iff the integers $(n_L-n_R)^I$ are even for \textit{all} $I$;} \\
    0, & \hbox{otherwise.} \\
\end{array}%
\right.
\end{equation*}
Thus for $k=qp=\text{odd}$ and $\gcd(p,q)=1$ the sum \eqref{S0'} can be written in term of
the level $k$ theta functions with characteristics as in \eqref{splitting:u}.

\paragraph{$\mathbf{q=odd}$, $\mathbf{p=even}$:} Now we do the following change
of variable in \eqref{pLpR0}:
$n^I=qt^I+2q\gamma_q^I$ and
$w^I=(p/2)t^I+p s^I+p\gamma^I_p$
where $t^I\in\Zh$.
In this case equations \eqref{pLpR0} take the form
\begin{equation*}
p_L^I=\underbrace{t^I+s^I}_{n_L}+\gamma_p^I+\gamma_q^I\quad\text{and}\quad
p_R^I=\underbrace{-s^I}_{n_R}+\gamma_p^I-\gamma_q^I.
\end{equation*}
The variables $n_L$ and $n_R$ are independent summation
variables, and thus we obtain \eqref{splitting:uev}.
\end{proof}



{\small
\begin{thebibliography}{99}
\bibitem{6} N. Marcus and J. H. Schwarz, ``Field Theories That Have No Manifestly Lorentz
Invariant Formulation,'' Phys. Lett. B115 (1982) 111.

\bibitem{7} W. Siegel, ``Manifest Lorentz Invariance Sometimes Requires Nonlinearity,'' Nucl.
Phys. B238 (184) 307.

\bibitem{8} P. Goddard and D. Olive, ``Algebras, Lattices, and Strings'' (preprint, 1983), Phys.
Scripta T15 (1987).

\bibitem{9}
C. Imbimbo and A. Schwimmer, ``The Lagrangian Formulation Of Chiral Scalars,''
Phys. Lett. B193 (1987) 455.

\bibitem{Floreanini:1987as}
  R.~Floreanini and R.~Jackiw,
  ``Selfdual Fields As Charge Density Solitons,''
  Phys.\ Rev.\ Lett.\  {\bf 59}, 1873 (1987).

\bibitem{10} C. M. Hull, ``Covariant Quantization Of Chiral Bosons And Anomaly Cancellation,''
Phys. Lett. B206 (1988) 234.

\bibitem{11} J. M. F. Labastida and M. Pernici, ``Lagrangians For Chiral Bosons And The Heterotic
String,'' Nucl. Phys. B306 (1988) 516.

\bibitem{12} L. Mezincescu and R.I. Nepomechie, ``Critical Dimensions For Chiral Bosons,'' Phys.
Rev. D37 (1988) 3067.

\bibitem{Henneaux:1988gg}
  M.~Henneaux and C.~Teitelboim,
  ``Dynamics Of Chiral (Selfdual) P Forms,''
  Phys.\ Lett.\ B {\bf 206}, 650 (1988).

\bibitem{13}
F. P. Devecchi and M. Henneaux, ``Covariant Path Integral
For Chiral p-Forms,'' Phys. Rev. D54 (1996) 1606.

\bibitem{14} P. P. Srivastava, ``On A Gauge Theory Of Selfdual Field And Its Quantization,'' Phys.
Lett. B234 (1990) 93.

\bibitem{15} B. McClain, Y. S. Wu, and F. Yu, ``Covariant Quantization Of Chiral Bosons and
OSp(1, 1|2) Symmetry'', Nucl. Phys. B343 (1990) 689.

\bibitem{16} C. Wotzase, ``The Wess-Zumino Term For Chiral Bosons,'' Phys. Rev. Lett. 66 (1991)
129.

\bibitem{17} I. Martin and A. Restuccia, ``Duality Symmetric Actions And Canonical Quantiza-
tion,'' Phys. Lett. B323 (1994) 311.

\bibitem{18} J. H. Schwarz and A. Sen, ``Duality Symmetric Actions,'' Nucl. Phys. B411 (1994)
35, hep-th/9304154.
32

\bibitem{19} E. Verlinde, ``Global Aspects Of Electric-Magnetic Duality,'' Nucl. Phys. B455 (1995)
211, hep-th/9506011.

\bibitem{20} M. Perry and J. H. Schwarz, ``Interacting Chiral Gauge Fields In Six Dimensional
Born-Infeld Theory,'' hep-th/9611065.

\bibitem{21} J. H. Schwarz, ``Coupling A Self-Dual Tensor To Gravity In Six Dimensions,'' hep-
th/9701008.

\bibitem{22} N. Berkovits, ``Local Actions With Electric and Magnetic Sources,'' hep-th/9610134,
``Super-Maxwell Actions With Manifest Duality,'' hep-th/9612174.

\bibitem{23} I. Bengtsson and A. Kleppe, ``On Chiral p-Forms,'' hep-th/9609102; I. Bengtsson,
``Manifest Duality In Born-Infeld Theory,'' hep-th/9612174.

\bibitem{24} P. Pasti, D. Sorokin, and M. Tonin, Phys. Lett. B352 (1995) 59, Phys. Rev. D52
(1995) R4277, ``On Lorentz Invariant Actions For Chiral P-Forms,'' hep-th/9611100,
``Covariant Action For A D = 11 Five-Brane With The Chiral Field,'' Phys. Lett.
B398 (1997) 41; I. Bandos, K. Lechner, A. Nurmagambetov, P. Pasti, D. Sorokin,
``Covariant Action For The Super-Five-Brane Of M-Theory,'' hep-th/9701149; G.
Dall'Agata, K. Lechner, and M. Tonin, ``Action For IIB SUpergravity In Ten
Dimensions,'' hep-th/9812170.

\bibitem{25} M. Aganagic, J. Park, C. Popescu, and J. Schwarz, ``Worldvolume Action For The
M-Theory Fivebrane,'' Nucl. Phys. B496 (1997) 191, hep-th/9701166.

\bibitem{26} P. S. Howe, E. Sezgin, and P. C. West, ``Covariant Field Equations Of The M Theory
Five-Brane,'' Phys. Lett. B399 (1997) 49, hep-th/9702008, ``The Six-Dimensional Self-
Dual Tensor,'' Phys. Lett. B400 (1997) 255, hep-th/9702111.

\bibitem{27} Y.-G. Miao, J.-G. Zhou, and Y.-Y. Liu, ``New Way Of The Derivation Of First Order
Wess-Zumino Terms,'' Phys. Lett. B323 (1994) 169; Y.-G. Miao and H. J. W. Muller-
Kirsten, ``Self-Duality Of Various Chiral Boson Actions,'' hep-th/9912066.

\bibitem{28} A. Mazyntsia, C. R. Preitschopf, and D. Sorokin,
``Dual Actions For Chiral Bosons,''
hep-th/9808049.

\bibitem{29}
X.~Bekaert, M.~Henneaux and A.~Sevrin,
``Deformations of chiral two-forms in six dimensions,''
Phys.\ Lett.\ B {\bf 468} (1999) 228
[arXiv:hep-th/9909094],
\\
``Symmetry-deforming interactions of chiral p-forms,''
Nucl.\ Phys.\ Proc.\ Suppl.\  {\bf 88} (2000) 27 [arXiv:hep-th/9912077].



\bibitem{Witten:1996hc}
  E.~Witten,
  ``Five-brane effective action in M-theory,''
  J.\ Geom.\ Phys.\  {\bf 22}, 103 (1997)
  [arXiv:hep-th/9610234].

\bibitem{Witten:1999vg}
  E.~Witten,
  ``Duality relations among topological effects in string theory,''
  JHEP {\bf 0005}, 031 (2000)
  [arXiv:hep-th/9912086].

\bibitem{Hopkins:2002rd}
  M.J.~Hopkins and I.M.~Singer,
  ``Quadratic functions in geometry, topology, and M-theory,''
  arXiv:math.at/0211216.

\bibitem{Henningson:1999dm}
  M.~Henningson, B.~E.~W.~Nilsson and P.~Salomonson,
  ``Holomorphic factorization of correlation functions in  (4k+2)-dimensional
  (2k)-form gauge theory,''
  JHEP {\bf 9909}, 008 (1999)
  [arXiv:hep-th/9908107].

\bibitem{Dolan:1998qk}
  L.~Dolan and C.~R.~Nappi,
  ``A modular invariant partition function for the fivebrane,''
  Nucl.\ Phys.\ B {\bf 530}, 683 (1998)
  [arXiv:hep-th/9806016].

\bibitem{Gerasimov:2004yx}
  A.~A.~Gerasimov and S.~L.~Shatashvili,
  ``Towards integrability of topological strings. I: Three-forms on  Calabi-Yau
  manifolds,''
  JHEP {\bf 0411}, 074 (2004)
  [arXiv:hep-th/0409238].

\bibitem{Moore:2004jv}
  G.~W.~Moore,
  ``Anomalies, Gauss laws, and page charges in M-theory,''
  Comptes Rendus Physique {\bf 6}, 251 (2005)
  [arXiv:hep-th/0409158].

\bibitem{Diaconescu:2000wy}
  D.~E.~Diaconescu, G.~W.~Moore and E.~Witten,
  ``E(8) gauge theory, and a derivation of K-theory from M-theory,''
  Adv.\ Theor.\ Math.\ Phys.\  {\bf 6}, 1031 (2003)
  [arXiv:hep-th/0005090].

\bibitem{Maldacena:2001ss}
  J.~M.~Maldacena, G.~W.~Moore and N.~Seiberg,
  ``D-brane charges in five-brane backgrounds,''
  JHEP {\bf 0110}, 005 (2001)
  [arXiv:hep-th/0108152].

\bibitem{Avis:1979de}
  S.~J.~Avis and C.~J.~Isham,
  ``Generalized Spin Structures On Four-Dimensional Space-Times,''
  Commun.\ Math.\ Phys.\  {\bf 72}, 103 (1980).

\bibitem{Belov:2005ze}
  D.~Belov and G.~W.~Moore,
  ``Classification of abelian spin Chern-Simons theories,''
  arXiv:hep-th/0505235.

\bibitem{Alvarez-Gaume:1986mi}
L.~Alvarez-Gaume, J.~B.~Bost, G.~W.~Moore, P.~C.~Nelson and C.~Vafa,
  ``Bosonization In Arbitrary Genus,''
  Phys.\ Lett.\ B {\bf 178} (1986) 41.

\bibitem{Alvarez-Gaume:1987vm}
L.~Alvarez-Gaume, J.~B.~Bost, G.~W.~Moore, P.~C.~Nelson and C.~Vafa,
``Bosonization On Higher Genus Riemann Surfaces,'' Commun.\ Math.\
Phys.\  {\bf 112}, 503 (1987).

\bibitem{Moore:1999gb}
  G.~W.~Moore and E.~Witten,
  ``Self-duality, Ramond-Ramond fields, and K-theory,''
  JHEP {\bf 0005}, 032 (2000)
  [arXiv:hep-th/9912279].

\bibitem{Freed:2000tt}
  D.~S.~Freed and M.~J.~Hopkins,
  ``On Ramond-Ramond fields and K-theory,''
  JHEP {\bf 0005}, 044 (2000)
  [arXiv:hep-th/0002027].

\bibitem{Birkenhake} C.~Birkenhake and H.~Lange, \textit{Complex abelian varieties},
second edition, Springer, ISBN 3-540-20488-1

\bibitem{FMS} D.~Freed, G.~Moore and G.~Segal,
\textit{Heisenberg Groups and Nincommutative Fluxes},
to appear;

\bibitem{Bekaert:1998yp}
  X.~Bekaert and M.~Henneaux,
  ``Comments on chiral p-forms,''
  Int.\ J.\ Theor.\ Phys.\  {\bf 38}, 1161 (1999)
  [arXiv:hep-th/9806062].

\bibitem{dfreed} D.S.~Freed, ``K-Theory in Quantum Field Theory'',
math-ph/0206031

\bibitem{qftstr}
P. Deligne, P. Etingof, D. S. Freed, L. C. Jeffrey, D. Kazhdan, J. W. Morgan, D.
R. Morrison, E. Witten (eds.), Quantum Fields and Strings: A Course for Mathematicians,
Volume 2, American Mathematical Society, Providence, RI, 1999.

\bibitem{Diaconescu:2003bm}
  E.~Diaconescu, G.~W.~Moore and D.~S.~Freed,
  ``The M-theory 3-form and E(8) gauge theory,''
  arXiv:hep-th/0312069.




\bibitem{Alvarez-Gaume:1983ig}
  L.~Alvarez-Gaume and E.~Witten,
  ``Gravitational Anomalies,''
  Nucl.\ Phys.\ B {\bf 234}, 269 (1984).

\bibitem{Dijkgraaf:2002ac}
  R.~Dijkgraaf, E.~Verlinde and M.~Vonk,
  ``On the partition sum of the NS five-brane,''
  arXiv:hep-th/0205281.

\bibitem{Witten:1993ed}
  E.~Witten,
  ``Quantum background independence in string theory,''
  arXiv:hep-th/9306122.

\bibitem{Axelrod:1989xt}
  S.~Axelrod, S.~Della Pietra and E.~Witten,
  ``Geometric Quantization Of Chern-Simons Gauge Theory,''
  J.\ Diff.\ Geom.\  {\bf 33}, 787 (1991).


\bibitem{MS}
J.W.~Milnor and J.D.~Stashef, \textit{Characteristic classes}, Annals of
Mathematics Study, Number 76, Princeton University Press and University of
Tokyo Press.
\\
D.B.A.~Epstein, Lectures by N.E.~Steenrod, \textit{Cohomology operations},
Annals of
Mathematics Study, Number 50.

\bibitem{Stong} R.~Stong and T.~Yoshida, ``Wu classes'',
Proccedings of American Mathematical Society, Vol. 100, No. 2 (Jun. 1987),
352--354.

\bibitem{Spacetime} M.~Kriele, \textit{Spacetime. Foundations of General
Relativety and Differential Geometry}, Springer 1999.


\end{thebibliography}
}
\end{document}